\documentclass{aa}  
\usepackage{hyperref}
\usepackage{graphicx}
\usepackage{tabularx}
\usepackage{txfonts}
\usepackage[switch]{lineno}
\usepackage{xcolor}
\hypersetup{
    colorlinks,
    linkcolor={blue!50!black},
    citecolor={blue!50!black},
    urlcolor={blue!80!black}
}

\begin{document} 

\title{A census of galactic spider binary millisecond pulsars with the Nan\c{c}ay Radio Telescope}

\author{
C.~Blanchard\inst{1}
\and
L.~Guillemot\inst{1,2}
\and 
G.~Voisin\inst{3}
\and 
I.~Cognard\inst{1,2}
\and
G.~Theureau\inst{1,2,3}
}

\institute{
LPC2E, OSUC, Univ Orleans, CNRS, CNES, Observatoire de Paris, F-45071 Orleans, France\\
\email{clara.blanchard@cnrs-orleans.fr}
\and
Observatoire Radioastronomique de Nan\c{c}ay, Observatoire de Paris, Universit\'e PSL, Université d'Orl\'eans, CNRS, 18330 Nan\c{c}ay, France
\and
LUTH, Observatoire de Paris, PSL Research University, CNRS, Universit\'e Paris Diderot, Sorbonne Paris Cit\'e, 92195 Meudon, France
}

\date{Received December 18, 2024; accepted April 12, 2025}

\authorrunning{C. Blanchard et al.}
\titlerunning{A census of spider binary pulsars with the NRT}

\abstract
   {Spider binary pulsars are systems in which a millisecond pulsar (MSP) tightly orbits ($P_b \lesssim$ 1 day) a low mass ($m_c \lesssim$ 0.5 $M_\odot$) non-degenerate or semi-degenerate star. Spider systems often display eclipses around superior conjunction, and their orbital periods often exhibit rapid time variations. The eclipse phenomenon is currently poorly understood. However, eclipses are excellent probes of plasma physics, intra-binary shocks, and they can also be used to study MSP formation processes.}
   {In this work we present Nan\c{c}ay Radio Telescope (NRT) observations of a sample of 19 spider pulsars conducted over several years. The sample includes 11 eclipsing and 8 non-eclipsing systems. We aim to provide a homogeneous phenomenological study of the eclipses, in order to compare spiders and study their group properties, as those eclipsing systems are often studied individually. We then confront our results with those derived from other studies, mainly optical observations and analyses of the pulsar companions. }
  {We analyzed eclipses via pulsar timing, using a 2D template matching technique allowing to simultaneously determine radio pulse Times Of Arrival (TOAs) and Dispersion Measures (DMs) along the lines of sight to the pulsars. The eclipses were then fit with a phenomenological model which gives a measurement of the duration and asymmetry of the eclipses. These parameters were then compared to other eclipse and system measurements, mainly derived from optical observations of the companion, to discuss the potential link between the presence of eclipses and orbital inclination, eclipsing systems being known to have higher mass functions than non-eclipsing ones. Finally, we formed polarisation-calibrated profiles for the pulsars in our sample, and derived some of their main polarization properties.}
   {We present here a comprehensive review of the NRT NUPPI backend spider pulsars dataset. We also present the first review and systematic analysis of a large sample of eclipsers, monitored with the NRT over several years. The phenomenological fit allowed us to derive a number of parameters to compare the eclipsers with each other, which led to the categorization of eclipsers depending on the shape of their eclipses. We present the polarimetric properties of the 19 spiders in the sample alongside their profiles, which were previously unpublished in some cases. We compared the mass function distributions of the eclipsing and non-eclipsing systems in our sample, and found, in agreement with previous studies, that eclipsing systems have higher mass functions than their non-eclipsing counterparts, suggesting that the latter have lower orbital inclinations. For the eclipsing systems, we found evidence for a positive correlation between eclipse duration and mass function, as expected if more eclipsing material crosses the line-of-sight in higher inclination systems. For the entire sample, we found marginal evidence for increasing pulse profile width with decreasing mass function, possibly indicating that the low mass function spiders are indeed those seen under low inclinations. We finally conducted a comprehensive literature review of the published inclination measurements for the pulsars in the sample and compared the inclinations to eclipse parameters, finding no clear correlations between orbital inclination and eclipse properties, unexpectedly. Nevertheless, the small number of available orbital inclination constraints, contradicting each other in some cases, hinders such searches for correlations.
   }  
   {}


\maketitle


\section{Introduction}

Since the discovery of the first spider pulsar B1957+20, the so-called original black widow pulsar \citep{B1957}, these extreme binary systems have been used to study a broad range of physics. Spider binaries are usually defined as millisecond pulsars (MSPs) in tight orbits ($P_b \lesssim 1$ day) with light and non-degenerate or semi-degenerate stars. Depending on the companion mass, the spiders are divided into two sub-populations: Black Widows (BWs) that have the lightest companions ($m_c \lesssim 0.07 M_\odot $), and Redbacks (RBs) that have sub-solar companions ($0.07 M_\odot \lesssim m_c \lesssim 0.5 M_\odot $) \citep{Roberts_2012}. In these systems, the powerful pulsar winds ablate the outer layers of the tidally locked companion. Neutron stars (NSs) are the cornerstone of several research fields, such as general relativity tests, astronomical transients study, gravitational wave astronomy, or the Equation of State (EoS) of neutron star mass.  Different EoSs allow for different mass-radius relationships and different maximum masses. Therefore, measuring as many NSs masses as possible is key to constraining EoSs. MSPs have an average mass higher than that of young pulsars \citep{O&F_2016}, and spiders are believed to have high masses among MSPs \citep{Linares_2020}. However, as noted in \citet{Clark_2023}, the mass measurements for spider pulsars, often obtained from optical observations, may be overestimated because of the underestimated inclinations coming from oversimplified heating models, especially in the case of BWs. Recent measurements from the same team bring back some spider masses closer to the average mass value for MSPs. 

MSPs are rotation-powered neutron stars with short rotational periods ($P \lesssim 10$ ms) that are believed to have been spun up by accretion and thus by the transfer of angular momentum from a binary companion \citep{Bisnovatyi_1974, Alpar_1982}. Spiders are thought to be a possible evolutionary channel for isolated MSPs \citep{Ruderman_1989}. The binary stage would therefore be a temporary phase during which the pulsar accretes and ablates its companion, spinning up the pulsar rotation via the transfer of angular momentum. This idea is supported by the transitioning systems that have been seen to switch back and forth from an accretion state (X-ray emitting, radio-quiet) to an RB state during the last two decades \citep[see e.g.][]{Papitto_2013,Bassa_2014,Roy_2015}

Spiders often undergo ``eclipses'' around superior conjunction, during a fraction of the orbital period. It is not the companion that is occulting the beam but instead matter around it, corresponding to the ablated material outflowing from the companion. This fog is not gravitationally bound to the companion, as it is very often larger than its Roche-lobe. The exact physical mechanism is currently unclear, but the eclipses are usually explained by dispersion, scattering, and absorption of radio emission by the diffuse intrabinary material. A review of the different eclipses mechanisms can be found in \citet{Thompson_1994}. Some works have linked individual systems to specific mechanisms: for instance, scattering and cyclotron absorption are considered the primary mechanisms producing the eclipses of PSR~J2051$-$0827, while  for PSR~J1544+4937, cyclotron-synchrotron absorption is believed to be the dominant eclipse mechanism at play \citep{Kansabanik_2021}. Note that this study used observations conducted around 400 MHz where PSR~J1544+4937 display eclipses but this pulsar does not display eclipses at 1.4 GHz. Most processes at stake depend on the observation frequency, so that eclipse properties also depend on frequency, and studying these phenomena at different radio frequencies is thus important. Dispersive effects being strongest at low frequencies, eclipses of spider systems have often been observed and studied below $\sim$~1~GHz. Studies of eclipse properties of pulsars at frequencies above 1~GHz have been conducted \citep[see e.g.][]{Polzin_2019}; however, to our knowledge, no systematic studies of the properties of ensembles of eclipsing pulsars have been published at these high frequencies.

The Nan\c{c}ay Radio Telescope (NRT) is a meridian telescope equivalent to a 94 m parabolic dish, located near Orléans (France). The NRT can track objects with declinations above $\sim -39^\circ$, for approximately 1~hr around transit. Detailed descriptions of the telescope and of pulsar observations with the NRT can be found in \citet{Guillemot_2023}. The NRT has been observing spiders for more than two decades, monitoring known systems and searching for new ones, with one BW discovered: PSR~J2055+3829 \citep{Guillemot_2019}, found during the SPAN512 survey \citep{SPAN512, Desvignes_2022}. This paper constitutes the first effort to systematically analyze the L-band data on the sample of spider systems observed with the NRT, providing an opportunity to look into their group properties.

In this paper we present the analysis of NRT data on 19 spiders observed over several years. In Sect.~\ref{sect:sample} we describe the sources included in this census, and their properties derived from timing and optical observations. In Sect.~\ref{sect:analysis} we present the analysis conducted for this work: the data preparation and timing method (Sect.~\ref{sect:timing}), the eclipse characterization (Sect.~\ref{sect:eclipses}) the morphological analysis of their eclipses (for pulsars exhibiting eclipses; see Sect.~\ref{sect:study}). Sect.~\ref{sect:polar} presents an overview of their polarimetric properties. Results are presented in Sect.~\ref{sect:results}: the correlations between the eclipse parameters (Sect.~\ref{subsect:eclipse_fit_results}), the dependence of eclipse parameters on the mass function (Sect~\ref{subsect:mass_f}) as well as on orbital inclination (Sect.~\ref{sect:optical}). Sect.~\ref{subsect:width} presents a search for correlations pulse profile widths and the mass functions of the systems. We finally summarise our work in Sect.~\ref{sect:conclusion}. 


\section{Pulsar sample and dataset properties}
\label{sect:sample}

Since the beginning of pulsar monitoring programs at Nan\c{c}ay, a total of 19 spider pulsars have been followed regularly with the NRT. Before mid-2011, pulsars were observed with the Berkeley-Orl\'eans-Nan\c{c}ay (BON) backend, and in August 2011 BON was replaced by the NUPPI backend as the primary pulsar instrumentation in operation at Nan\c{c}ay. 
The NUPPI pulsar observation backend is a version of the Green Bank Ultimate Pulsar Processing Instrument\footnote{\url{https://safe.nrao.edu/wiki/bin/view/CICADA/NGNPP}} designed for Nan\c{c}ay. It records 512~MHz of frequency bandwidth, in the form of 128 channels of 4~MHz each, that are coherently dedispersed in real time. In this article we only considered data taken with the NUPPI backend, since NUPPI observations of spider pulsars are much more numerous and of higher quality than BON observations. The majority of NUPPI observations have been conducted at L-band, at a central frequency of 1484~MHz.

\begin{figure}[ht]
\centering
\includegraphics[width=\hsize]{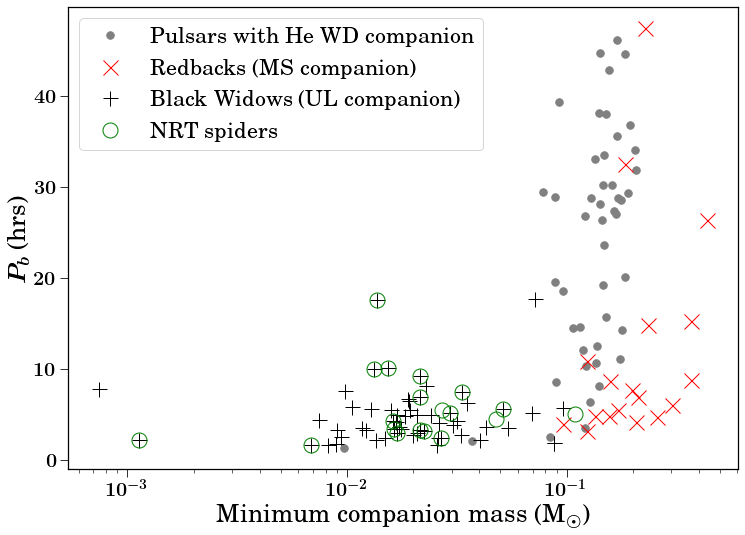}
\caption{Orbital periods of known spider systems against minimum companion masses. In this figure, spider systems are categorised as BW or RB depending on the companion type listed in the ATNF catalog: Main Sequence (MS) stars for RBs, or Ultra-low mass objects (UL) in the case of BWs. Some NRT spiders have no information about the companion type in the catalogue hence some circles have no cross inside.}
\label{fig:atnf}
\end{figure}

\begin{figure*}[ht]
\begin{center}
\includegraphics[width=\textwidth]{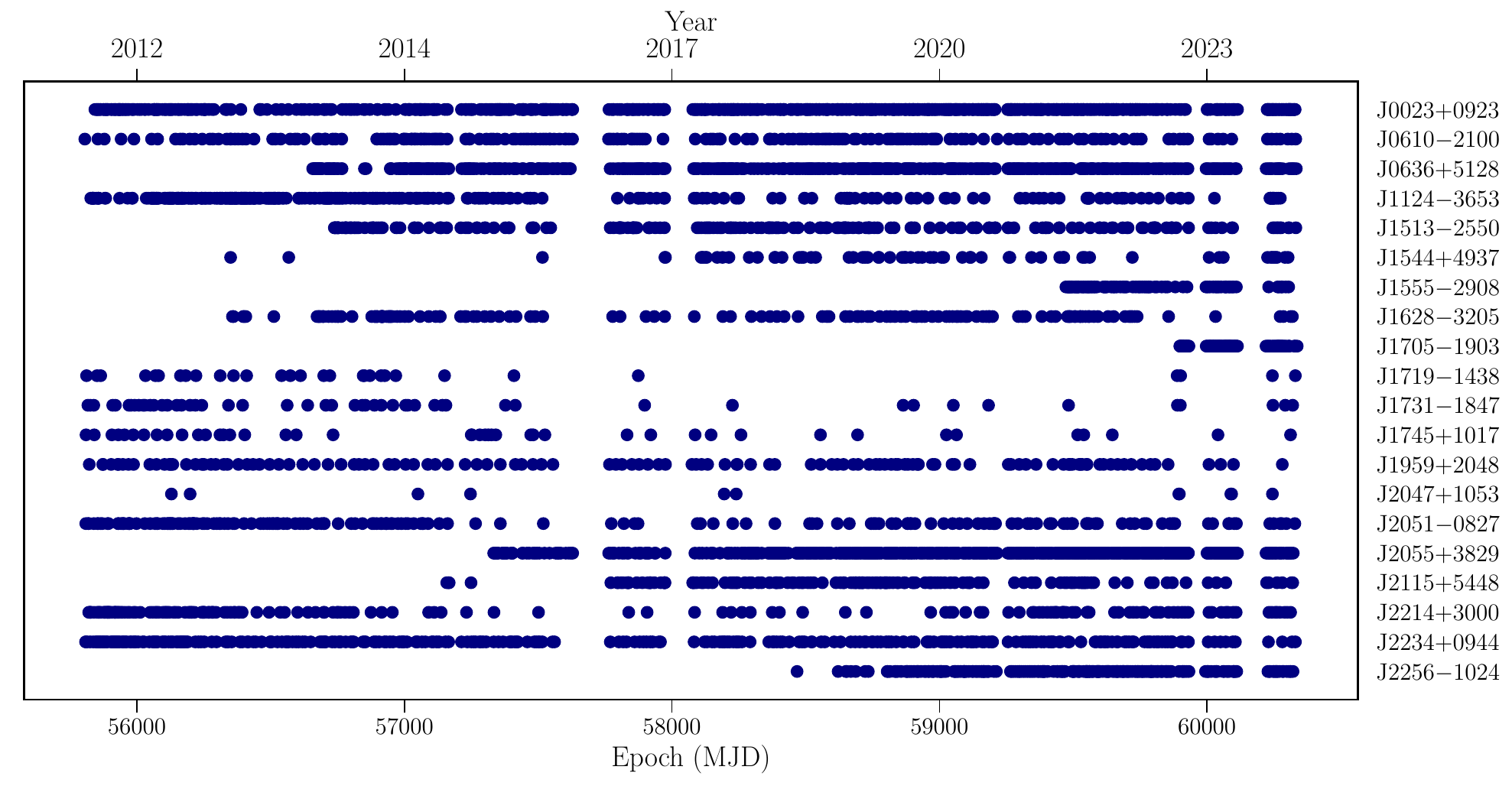}
\caption{Time span of the NRT dataset considered in this study. Each dot represents an observation made with the NUPPI backend.}
\label{fig:data_span}
\end{center}
\end{figure*}

Fig.~\ref{fig:atnf} puts the pulsar sample into the context of the known spider pulsar population as listed in the ATNF catalog \citep{Manchester_2005}\footnote{Available at: \url{https://www.atnf.csiro.au/research/pulsar/psrcat/}}.
The names of the 19 spider pulsars in our sample, as well as an overview of the NRT dataset considered in this study for these pulsars can be seen in Fig.~\ref{fig:data_span}. In Table~\ref{tab:obs_sample} we list the total time spans of the observations and the number of observations for each pulsar.  We note that there is only one RB pulsar in our sample: PSR~J1628$-$3205. All the other objects are BW pulsars. In spite of the under-representation of RB-type objects in our sample, orbital periods and companion masses are fairly representative of those of the global, currently-known spider population, as can be seen from Fig.~\ref{fig:atnf}. The scientific rationale behind the monitoring of these spider pulsars with the NRT is the follow-up of pulsars discovered by radio surveys or radio searches at the locations of \textit{Fermi} Large Area Telescope unassociated sources, long-term timing and orbital variability studies, or the characterization of their eclipses.

\begin{table}
\caption[]{Properties of the NRT dataset used in this study. }
\label{tab:obs_sample}
\centering
\begin{tabular}{cccc}
\hline
\hline
Pulsar & Time span (d) & $N_{obs}$ & Median S/N \\
\hline
\hline
J0023+0923 & 4487 & 688 & 27.0 \\
J0610$-$2100 & 4527 & 211 & 40.1 \\
J0636+5128 & 3677 & 416 & 55.6 \\
J1124$-$3653 & 4447 & 348 & 6.5 \\
J1513$-$2550 & 3594 & 207 & 27.0 \\
J1544+4937 & 3953 & 67 & 15.8 \\
J1555$-$2908 & 833 & 79 & 20.6 \\
J1628$-$3205 & 3962 & 145 & 29.5 \\
J1705$-$1903 & 439 & 69 & 141.7 \\
J1719$-$1438 & 4519 & 30 & 31.1 \\
J1731$-$1847 & 4504 & 52 & 26.8 \\
J1745+1017 & 4503 & 43 & 38.9 \\
J1959+2048 & 4460 & 125 & 15.4 \\
J2051$-$0827 & 4520 & 181 & 96.3 \\
J2055+3829 & 2989 & 523 & 17.7 \\
J2115+5448 & 3162 & 143 & 28.9 \\
J2214+3000 & 4493 & 168 & 21.5 \\
J2234+0944 & 4523 & 303 & 55.0 \\
J2256$-$1024 & 1854 & 217 & 52.4 \\
\hline
\end{tabular}
\tablefoot{For each pulsar, we list the total time span of the observations, the number of observations, and the median signal-to-noise ratio (S/N) of the frequency- and time-averaged observations. In this study we considered all available NRT observations recorded using the NUPPI backend until MJD~60369 (29 February 2024).}
\end{table}

Astrometric, spin and dispersion measure (DM) properties for the 19 pulsars in our sample are listed in Table~\ref{tab:spin_sample}, and basic orbital properties are listed in Table~\ref{tab:binary_sample}. Companion masses in the latter table were derived from the Keplerian parameters, using:\\

\begin{equation}
m_\mathrm{f} = \frac{(m_c \sin(i) )^3}{(m_p +m_c)^2} = \frac{4 \pi ^2}{G} \frac{(a_p \sin (i))^3}{P_b^2}, \\
\label{eq:comp_mass}
\end{equation}

\noindent
where $m_\mathrm{f}$ is the mass function, $m_p$ (resp. $m_c$) is the mass of the pulsar (resp. companion), $i$ is the orbital inclination, $G$ is the gravitation constant, $a_p$ is the semi-major of the pulsar orbit relative to the common center of mass, and $P_b$ is the orbital period. The companion masses in Table~\ref{tab:binary_sample} were determined assuming a pulsar mass of $1.5\ M_\odot$. Minimum companion masses were determined using $i = 90^\circ$, while median values were calculated assuming $i = 60^\circ$. 

In the following subsections we describe some of the main properties of the pulsars in our sample.

\begin{table*}
\caption[]{Astrometric and spin properties of the 19 pulsars constituting our spider pulsar sample.}
\label{tab:spin_sample}
\centering

\begin{tabular}{cccccccc}
\hline
\hline
Pulsar & Discovery article & Gal. lon. & Gal. lat.  & P &  DM & $d$  \\
 & & (J2000) &  (J2000)  & (ms) & (pc cm$^{-3}$) & (kpc) \\
\hline
J0023+0923 & \citet{Bangale_2024} & 111.38 & $-$52.85 & 3.05 & 14.3 & 1.24 \\
J0610$-$2100 & \citet{Burgay_06} & 227.75 & $-$18.18 & 3.86 & 60.7 & 3.26 \\
J0636+5128 & \citet{Stovall_14} & 163.91 & 18.64 & 2.87 & 11.1 & 0.21 \\
J1124$-$3653 & \citet{Bangale_2024} & 284.09 & 22.76 & 2.41 & 44.9 & 0.99 \\
J1513$-$2550 & \citet{SanpaArsa_PhD} & 338.82 & 26.96 & 2.12 & 46.9 & 3.96 \\
J1544+4937 & \citet{Bhattacharyya_2013} & 79.17 & 50.17 & 2.16 & 23.2 & 2.98 \\
J1555$-$2908 & \citet{Ray_2022} & 344.48 & 18.50 & 1.79 & 75.9 & 7.55 \\
J1628$-$3205 & \citet{Ray_2012} & 347.43 & 11.48 & 3.21 & 42.1 & 1.22 \\
J1705$-$1903 & \citet{Morello_2018} & 3.25 & 13.03 & 2.48 & 57.5 & 2.34 \\
J1719$-$1438 & \citet{Bailes_2011} & 8.86 & 12.84 & 5.79 & 36.8 & 0.34 \\
J1731$-$1847 & \citet{Keith_2010} & 6.89 & 8.15 & 2.35 & 106.5 & 4.77 \\
J1745+1017 & \citet{Barr_2013} & 34.87 & 19.25 & 2.65 & 24.0 & 1.22 \\
J1959+2048 & \citet{B1957} & 59.20 & -4.70 & 1.61 & 29.1 & 1.73 \\
J2051$-$0827 & \citet{Stappers_1996} & 39.19 & $-$30.41 & 4.51 & 20.7 & 1.47 \\
J2055+3829 & \citet{Guillemot_2019} & 80.62 & $-$4.26 & 2.09 & 91.8 & 4.59 \\
J2115+5448 & \citet{SanpaArsa_PhD} & 95.04 & 4.11 & 2.60 & 77.4 & 3.11 \\
J2214+3000 & \citet{Ransom_2011} & 86.86 & $-$21.67 & 3.12 & 22.6 & 1.68 \\
J2234+0944 & \citet{Ray_2012} & 76.28 & $-$40.44 & 3.63 & 17.8 & 1.58 \\
J2256$-$1024 & \citet{Crowter_2020} & 59.23 & $-$58.29 & 2.30 & 13.8 & 1.33 \\
\hline
\end{tabular}
\tablefoot{For each pulsar we list the discovery article, the galactic coordinates, the spin period P, the dispersion measure DM, and the distance $d$. The latter distance values are based on the DMs and were estimated using the YMW16 model for the distribution of free electrons in the Galaxy \citep{Yao_2017}.}
\end{table*}

\begin{table*}
\caption[]{Binary properties of the 19 pulsars constituting our spider pulsar sample.}
\label{tab:binary_sample}
\centering

\begin{tabular}{ccccccc}
\hline
\hline
Pulsar & $P_b$ & $a_1$ & $m_\mathrm{f}$ & $m_{c,\ \mathrm{min}}$ & $m_{c,\ \mathrm{med}}$ & Eclipsing at 1.4 GHz? \\
 & (hrs) & ($10^{-2}$ lt-s) & ($10^{-16} M_\odot$) & ($10^{-3} M_\odot$) & ($10^{-3} M_\odot$) & \\
\hline
J0023+0923 & 3.33 & 3.48 & 1.56 & 17.54 & 15.25 & - \\
J0610$-$2100 & 6.86 & 7.35 & 3.46 & 22.90 & 19.90 & - \\
J0636+5128 & 1.60 & 0.90 & 0.12 & 7.35 & 6.39 & - \\
J1124$-$3653 & 5.45 & 7.96 & 6.98 & 29.03 & 25.21 & Yes \\
J1513$-$2550 & 4.29 & 4.08 & 1.52 & 17.36 & 15.09 & Yes \\
J1544+4937 & 2.90 & 3.29 & 1.73 & 18.16 & 15.78 & - \\
J1555$-$2908 & 5.60 & 15.14 & $4.54 \times 10^{-2}$ & 54.78 & 47.51 & Yes \\
J1628$-$3205 & 5.00 & 41.03 & $1.14\times 10^3$ & 167.90 & 144.72 & Yes \\
J1705$-$1903 & 4.41 & 10.44 & $2.39\times 10^{-2}$ & 44.05 & 38.23 & Yes \\
J1719$-$1438 & 2.18 & 0.18 & $5.24 \times 10^{-4}$ & 1.21 & 1.05 & - \\
J1731$-$1847 & 7.47 & 12.02 & $1.28 \times 10^{-2}$ & 35.60 & 30.91 & Yes \\
J1745+1017 & 17.53 & 8.82 & 0.92 & 14.66 & 12.74 & - \\
J1959+2048 & 9.17 & 8.92 & 3.47 & 22.93 & 19.92 & Yes \\
J2051$-$0827 & 2.38 & 4.51 & 6.64 & 28.54 & 24.79 & Yes \\
J2055+3829 & 3.11 & 4.53 & 3.93 & 23.92 & 20.78 & Yes \\
J2115+5448 & 3.25 & 4.48 & 3.51 & 23.01 & 20.00 & Yes \\
J2214+3000 & 10.00 & 5.91 & 0.85 & 14.27 & 12.41 & - \\
J2234+0944 & 10.07 & 6.84 & 1.30 & 16.47 & 14.31 & - \\
J2256$-$1024 & 5.11 & 8.30 & 8.98 & 31.60 & 27.44 & Yes \\
\hline
\end{tabular}
\tablefoot{For each pulsar, we list the orbital period $P_b$, the projected semi-major axis of the pulsar orbit, $a_1 = a_p\ \sin(i)$, the mass function $m_\mathrm{f}$ and estimates of the minimum and median companion masses, $m_\mathrm{c,\ min}$ and $m_\mathrm{c,\ med}$. Companion masses were derived using Equation~\ref{eq:comp_mass}, assuming a pulsar mass $m_p = 1.5\ M_{\odot}$. Minimum and median masses were respectively derived assuming orbital inclinations $i$ of $90^\circ$ and $60^\circ$. In the last column we indicate whether the pulsar is seen to eclipse in the 1.4~GHz NRT data or not, with eclipsing pulsars defined as those displaying excess delays in their timing residuals around superior conjunction (see Sect.~\ref{sect:analysis}).}
\end{table*}

\subsection{PSR J0023+0923}

PSR J0023+0923 was discovered as part of a survey of unassociated \textit{Fermi} Large Area Telescope (LAT) sources with the Green Bank Telescope (GBT) at 350~MHz \citep{Bangale_2024}. It has a period of 3.05~ms, and a DM of 14.32 pc cm$^{-3}$ putting it at a distance of 1.24~kpc according to the YMW16 model for the distribution of free electrons in the Milky Way \citep{Yao_2017}. It has been observed in X-rays \citep{Gentile_2014}, and has been detected with the LAT as a pulsed source of $\gamma$-ray emission \citep[see e.g.][]{Smith_3PC}. 

It is classified as a BW due to its very typical $0.016\ M_\odot$ minimum companion mass, and short orbital period of 3.3~hrs. The companion has been observed at optical wavelengths with multiple instruments \citep{Breton_2013,Draghis_2019,MataSanchez_2023}. The system is not displaying eclipses in the L-band, and is not reported to display eclipses at 350~MHz or 2~GHz \citep{Bangale_2024}.

\subsection{PSR J0610$-$2100}

PSR J0610$-$2100 was discovered during the Parkes High-Latitude pulsar survey \citep{Burgay_06}. It has a period of 3.86~ms and a DM of 60.7~pc cm$^{-3}$ placing it at a distance of 3.26~kpc according to the YMW16 model. It was first reported to emit pulsed $\gamma$ rays by \citet{Espinoza_13}. 

It is classified as a BW due to its $0.021\ M_\odot$ minimum companion mass, and short orbital period of 6.9~hrs. The companion has been detected at optical wavelengths in archival observations obtained with FORS2 \citep{Pallanca_2012}. This system is not known to display eclipses.

\subsection{PSR J0636+5128}

PSR J0636+5128 was found in the Green Bank Northern Celestial Cap Pulsar Survey \citep{Stovall_14}. It has a period of 2.87~ms, and a DM of 11.1 pc cm$^{-3}$. The YMW16 model predicts a very small distance of 0.21 kpc for this DM and set of sky coordinates. Pulsed $\gamma$-ray emission was first detected and reported by \citet{Smith_2019}. With its ultra-light companion (minimum mass of $0.007\ M_\odot \simeq 7.3 M_J$) and short orbital period of 1.6~hrs, it is classified as a BW pulsar. Its orbital period is among of shortest currently known among the binary pulsar population. 

The companion has been detected in 2014 Gemini archival data \citep{Draghis_18}. This system is not known to display eclipses. 

Unlike PSRs~J0023+0923 and J0610$-$2100 that have been observed with NUPPI soon after it was installed in August 2011, NRT observations of J0636+5128 began in late 2013.

\subsection{PSR J1124$-$3653}

PSR~J1124$-$3653 was found during a 350~MHz survey of unassociated \textit{Fermi} LAT sources with the GBT \citep{Bangale_2024}. It has a period of 2.41~ms, and a DM of 44.9~pc cm$^{-3}$ putting it at a distance of 0.99~kpc according to the YMW16 model. It has been observed in X-rays \citep{Gentile_2014}, and is detected to emit pulsed $\gamma$-ray emission \citep{Smith_3PC}. 

With its low minimum companion mass of $0.027\ M_\odot$ and short orbital period of 5.4~hrs, it is a BW pulsar. It is extremely faint in 1.4~GHz NRT observations: as can be seen from Table~\ref{tab:obs_sample} the median S/N of our observations of this one is only 6.5. Although the pulsar is seen to display eclipses at 1.4~GHz in the NRT data, the very low S/N values prevented us from obtaining satisfactory results for the eclipse analysis using the same analysis method as for the others.

\subsection{PSR J1513$-$2550}

PSR~J1513$-$2550 was found in a survey for new pulsars at the locations of \textit{Fermi} LAT unassociated sources at 800~MHz with the GBT\citep{SanpaArsa_PhD}. It has a period of 2.12~ms, and a DM of 46.9~pc cm$^{-3}$. The YMW16 model of free electrons in the Galaxy places the pulsar at a distance of 3.96~kpc. It has been detected as a pulsed source of $\gamma$ rays with the LAT \citep{Smith_3PC}. 

It is classified as a BW due to its $0.016\ M_\odot$ minimum companion mass and short orbital period of 4.3~hrs. At 1.4~GHz this system is observed to eclipse, with short-duration (duty cycles of less than 10\% of the orbital period) total eclipses. NRT observations of J1513$-$2550 began in March 2014.

\subsection{PSR J1544+4937}

This 2.16~ms pulsar with a DM of 23.2 pc.cm$^{-3}$ was discovered as part of a search for pulsars in LAT unassociated sources with the GMRT\citep{Bhattacharyya_2013}. The YMW16-predicted distance for this DM value is 2.98~kpc. The low minimum companion mass ($ 0.017\ M_\odot$) and short orbital period ($P_b \sim 2.9$ hrs) make it a BW pulsar. The analysis of \textit{Fermi} LAT data revealed that J1544+4937 emits pulsed $\gamma$-ray emission \citep{Bhattacharyya_2013,Smith_3PC}.

The companion has been detected in optical \citep{Tang_2014} and is thought to be nearly Roche-lobe filling \citep{MataSanchez_2023}. This system displays total eclipses at 322~MHz, and the TOAs present significant delays around superior conjunction at 607~MHz, with delays reaching more than 10\% of the pulsar's spin period \citep{Bhattacharyya_2013}. The 1.4~GHz NRT data do not appear to reveal any significant delays around superior conjunction. 

NRT monitoring of this object began in early 2013, \textit{i.e.}, soon after the discovery of this pulsar.

\subsection{PSR J1555$-$2908}

PSR~J1555$-$2908 was found in a search of \textit{Fermi} LAT unassociated sources with the GBT at 800~MHz\citep{Ray_2022}. LAT data analysis revealed that the pulsar also emits pulsed $\gamma$-ray emission. This pulsar has a period of 1.79~ms, and a DM of 75.9 pc~cm$^{-3}$. For this DM and line-of-sight the YMW16 model predicts a distance of 7.55~kpc, making this pulsar the most distant in our sample.

The minimum companion mass of $0.051\ M_\odot$ and short orbital period of 5.6~hrs make PSR~J1555$-$2908 a BW pulsar. Optical observations constrained the minimum companion mass to be $0.060 \pm0.005\ M_\odot$, placing the system at the upper limit of what is considered a BW \citep{Kennedy_2022}. The companion is severely bloated, with a Roche-lobe filling factor ("measured as the ratio of the stars radius to the L1 point along the line joining both") of $f=0.98 \pm 0.02$.

\citet{Clark_2023} showed by analyzing \textit{Fermi} LAT data that PSR~J1555$-$2908 displays eclipses in $\gamma$ rays. Eclipses at such short wavelengths indicate that the system is seen nearly edge-on, the derived orbital inclination is greater than 83$^\circ$. Moreover, analysis of the $\gamma$-ray eclipses enabled model-independent constraints on the pulsar mass, estimated to be $m_p = 1.645 \pm 0.065 M_\odot$. 

NRT monitoring of PSR~J1555$-$2908 began relatively recently, with first observations conducted in mid-2021. 

\subsection{PSR J1628$-$3205}

PSR~J1628$-$3205 was discovered in a survey of \textit{Fermi} LAT unassociated sources at 350 MHz with the GBT \citep{Ray_2012}. Pulsations were later detected in the $\gamma$-ray data recorded by the LAT, and reported in \citet{Smith_3PC}. This pulsar has a rotational period of 3.21~ms, and a DM of 42.1~pc cm$^{-3}$, placing the pulsar at a distance of 1.22~kpc according to the YMW16 model. 

Its comparatively larger minimum companion mass of $0.109\ M_\odot$ and short orbital period of 5 hrs make it a RB pulsar. It is the only RB in our sample. In fact, optical observations by \citet{Li_2014} constrained the companion mass to be larger than $ 0.161 M_\odot$. Analysis of the 1.4~GHz NRT data does not reveal eclipses around superior conjunction. 

Routine monitoring of this pulsar with the NRT began in early 2013.

\subsection{PSR J1705$-$1903}

PSR~J1705$-$1903 was discovered with the Parkes radio telescope as part of the HTRU survey \citep{Morello_2018}. It has a period of 2.48 ms, and a DM of 57.5 pc.cm$^{-3}$. The YMW16 distance for this DM and line-of-sight is 2.34~kpc. This pulsar is one of the few in this sample that are not detected in $\gamma$ rays yet. It also does not appear to have any $\gamma$-ray counterpart: the nearest source from the \textit{Fermi} LAT 14-Year Point Source Catalog \citep[hereafter 4FGL;][]{4FGL} is 11' away, and is associated with PSR~J1705$-$1906 \citep{Smith_3PC}.

It is classified as a BW due to its low minimum companion mass of $0.047 M_\odot$ and short orbital period of 4.4~hrs. This system does not display total eclipses in the NRT data at 1.4 GHz, but does display significant delays at superior conjunction. We therefore categorize it as an eclipser. 

NRT monitoring on this one started in November 2022. The NRT dataset on this pulsar has the shortest time span among the datasets considered in this study.

\subsection{PSR J1719$-$1438}
\label{sect:1719}

PSR~J1719$-$1438 was discovered with the Parkes radio telescope as part of the HTRU survey \citep{Bailes_2011}. It has a period of 5.79~ms, and a DM of 36.8 pc cm$^{-3}$, placing it at a very close distance of 0.34~kpc according to the YMW16 model. Like PSR~J1705$-$1903, it is currently undetected in $\gamma$ rays, and there are no 4FGL sources within $1^\circ$ of the pulsar position.

This system has a peculiar status: it is classified as a BW by default, but has an extremely light companion, with a minimum companion mass of $0.001\ M_\odot$. The companion is believed to be Jupiter-sized but up to 10 times more dense, around 23 g~cm$^{-3}$. As the companion is likely closer to a planet than it is to a star, it has been suggested that the system could be part of a sub-class of BWs that have a different evolutionary path \citep{Guo_2022}, or that it is not a BW at all, but could be the missing link between X-ray binaries and isolated pulsars \citep{Haaften_2013}. It has been searched for in optical \citep{Bailes_2011} but was not detected at the expected magnitude and location. 

Computational evolutionary models constrain the companion to be Helium or Carbon rich \citep{Bailes_2011}. The exact nature of the companion is yet unknown, but potential progenitors could be a He or C white dwarf \citep{Bailes_2011} or a He star \citep{Guo_2022}. This could make PSR~J1719$-$1438 a member of a separate category, different from BWs, in which the companions are usually considered non-degenerate. This system is not known to display eclipses.

\subsection{PSR J1731$-$1847}

PSR~J1731$-$1847 was discovered with the Parkes radio telescope during the HTRU survey \citep{Bates_2011}. It has a spin period of 2.34~ms, and a DM of 106.6 pc~cm$^{-3}$, the largest in our pulsar sample. The YMW16 distance for this object is 4.77~kpc. Interestingly, the pulsar is located only 6.5' away from a 4FGL source, 4FGL~J1731.7$-$1850, with no known association. Nevertheless, no $\gamma$-ray pulsations have been reported from PSR~J1731$-$1847 yet.

The orbital period is 7.5~hrs, and the minimum companion mass is $ 0.033\ M_\odot$, making this object a BW. This system displays eclipses from 1.4~GHz down to at least 700~MHz \citep{Bates_2011}.

\subsection{PSR J1745+1017}

PSR~J1745+1017 was discovered with the Effelsberg radio telescope in a search for pulsars at the locations of unassociated \textit{Fermi} LAT sources \citep{Barr_2013}. It has a rotational period of 2.65~ms and a DM of 24.0~pc cm$^{-3}$. The YMW16 model places this pulsar at a distance of 1.22~kpc. Unsurprisingly, given the fact that the pulsar was discovered within the confidence region of a LAT source with no known counterpart, the pulsar was detected to emit pulsed $\gamma$-ray emission soon after it was discovered \citep{Barr_2013}.

With an orbital period of 17.5~hrs and a minimum companion mass of $0.014\ M_\odot$, it is classified as a BW pulsar. This system is not known to display eclipses.

\subsection{PSR J1959+2048}

The discovery of the so-called original BW PSR~J1959+2048 (also known as B1957+20), was reported in \citet{B1957} after a detection with the Arecibo telescope in a survey for millisecond pulsars. It has a period of 1.61~ms, and a DM of 21.9 pc~cm$^{-3}$. The YMW16 distance for this pulsar is 1.73~kpc. Pulsed $\gamma$ rays from this pulsar were detected using LAT data \citep{Guillemot_2012}. The latter study also reported on the detection of pulsed X-ray emission using XMM-Newton data, at the 4-$\sigma$ confidence level. However, new pulsation searches in X-rays using NICER data did not confirm this detection \citep{Ng_2022}. 

It is classified as a BW, with its $0.022\ M_\odot$ minimum companion mass and 9.2~hr orbital period. This system displays eclipses from 430 MHz \citep{B1957} up to the L-band frequency range. The detection of eclipses shed new light upon the mechanism spinning up young pulsars to millisecond rotation periods. While radio eclipses of J1959+2048 have been studied thoroughly \citep[see][]{Main_2018,Lin_2023}, the pulsar was also observed to eclipse in $\gamma$ rays \citep{Clark_2023}, indicating that the system is seen nearly edge-on: $i > 84.1^\circ$. \citet{Clark_2023} were also able to put constraints on the mass of the pulsar, finding $m_p = 1.805 \pm 0.135\ M_\odot$, a higher value than the canonical $1.4\ M_\odot$, as for J1555+2908, albeit much lower that the initially inferred 2.4 $M_\odot$ \citep{Kerkwijk_2011}.

\subsection{PSR J2051$-$0827}

PSR~J2051$-$0827 was the second spider discovered. It was reported in \citet{Stappers_1996} after a detection with the Parkes telescope during a survey of the southern sky. It has a spin period of 4.51~ms, and a DM of 20.7 pc~cm$^{-3}$ placing it at a distance of 1.47~kpc according to the YMW16 model. 

It is classified as a BW, with its minimum companion mass of $0.027\ M_\odot$ and 2.4~hr orbital period. This system displays eclipses from 436~MHz \citep{Stappers_1996} up to the L-band. \citet{Polzin_2019} showed that the eclipses are not seen at every orbit, and are variable, in duration and in occurrence. Additionally, the centroid of the eclipse is seen to vary in orbital phase over a year-long period \citep[see Fig.~4 of][for a comprehensive plot of Effelsberg observations of J2051$-$0827 eclipses]{Polzin_2019}. Although these properties are unique among spider pulsars, this uniqueness is to be taken with caution, given the fact that PSR~J2051$-$0827 has been studied much more extensively than most other spider pulsars. PSR~J2051$-$0827 is also known to display rapid variations of its orbital properties \citep{Shaifullah_2016} that might be linked to the above-mentioned variability of the eclipses. Changes in rotation measure during the eclipse \citep{Wang_2023} and intra-system plasma lensing \citep{Lin_2021} have also been reported for this pulsar. This is also the first (and to this date only) system for which the quadrupole moment of the companion has been measured \citep{Voisin_2020}.

PSR~J2051$-$0827 has been detected to emit pulsed $\gamma$-ray emission \citep{Wu_2012,Smith_3PC}, and its companion has been detected at optical wavelengths \citep{Dhillon_2022}.

\subsection{PSR J2055+3829}

PSR~J2055+3829 was discovered at 1.4~GHz as part of the SPAN512 survey conducted with the NRT \citep{Guillemot_2019}. It has a spin period of 2.09~ms, and a DM of 91.9~pc cm$^{-3}$. The YMW16 distance for this DM and line-of-sight is 4.59~kpc. The pulsar has not been detected as a pulsed source of $\gamma$-ray emission, and there are no $\gamma$-ray sources within $1^\circ$ of the pulsar position in the 4FGL catalog. 

It is classified as a BW, with a minimum companion mass of $0.027\ M_\odot$ and an orbital period of 2.4~hrs. As can be seen from Fig.~4 of \citet{Guillemot_2019}, PSR~J2055+3829 is observed to display total eclipses at 1.4~GHz. 

It has been observed with the NRT since late-2015 with an average observation cadence of one observation every 5.5 days.

\subsection{PSR J2115+5448}

PSR~J2115+5448 was discovered with the GBT at the location of a \textit{Fermi} LAT unassociated source \citep{SanpaArsa_PhD}. It has a spin period of 2.60~ms, and a DM of 77.4~pc cm$^{-3}$, for which the YMW16 model predicts a distance of 3.11~kpc. The pulsar was detected to produce pulsed $\gamma$-ray emission by analyzing LAT photons \citep{Smith_3PC}.

It is classified as a BW, with its minimum companion mass of $0.022\ M_\odot$ and short orbital period of 3.2~hrs. Analysis of the NRT data reveals that the pulsar displays variable eclipses around superior conjunction, with the radio signal being totally eclipsed during some orbits, and only delayed during others. 

Routine monitoring of this pulsar with the NRT began in mid-2015.

\subsection{PSR J2214+3000}

This pulsar was discovered by \citet{Ransom_2011} at the location of a \textit{Fermi} LAT source with no known counterparts. It was quickly detected as a pulsed source of $\gamma$-ray emission, confirming that it was indeed powering the LAT unassociated source. It has a spin period of 3.12~ms, and a DM of 22.6~pc cm$^{-3}$ placing the system at a distance of 1.68~kpc according to the YMW16 model. 

Its orbital period of 10~hrs and minimum companion mass of $0.007 M_\odot$ make it a BW system. First detection of the companion has been reported by \citet{Schroeder_2014}. It has not been reported to display eclipses, and the analysis of the 1.4~GHz NRT indeed did not yield any eclipse detection.

\subsection{PSR J2234+0944}

PSR~J2234+0944 was discovered at the location of a \textit{Fermi} LAT unassociated source with the Parkes radio telescope \citep{Ray_2012}. It was later confirmed to emit pulsed $\gamma$-ray emission \citep[see e.g.][]{Smith_3PC}. 

It has a spin period of 3.63~ms, a DM of 17.8~pc cm$^{-3}$, an orbital period of 10~hrs and a minimum companion mass of $0.008 M_\odot$. The YMW16 model places this pulsar at a distance of 1.58~kpc. This pulsar's orbital properties make it a BW-type object. 

Investigation of the 1.4~GHz NRT data did not reveal the presence of eclipses, and the pulsar has not yet been reported to display eclipses at other frequencies.

\subsection{PSR J2256$-$1024}

PSR~J2256$-$1024 was discovered with the GBT as part of a drift scan survey at 350~MHz \citep{Crowter_2020}. It has a spin period of 2.30~ms, and a DM of 13.8~pc cm$^{-3}$. The YMW16 model of free electrons in the Galaxy predicts a distance of 1.33~kpc for this direction and DM value. 

The pulsar has been observed in X-rays \citep{Gentile_2014} and has been detected to emit pulsed $\gamma$-ray emission \citep{Smith_3PC}. Optical observations of the system yielded a detection of the companion object \citep{Breton_2013}. The pulsar is seen to exhibit total eclipses in the 1.4~GHz NRT data.

PSR~J2256$-$1024 is monitored at the NRT since late-2018.


\section{Eclipse analysis}
\label{sect:analysis}

\subsection{Data preparation and timing analysis}
\label{sect:timing}

In order to characterize the properties of the eclipses detected in the 1.4~GHz NRT data, we started by conducting timing analyses of all the selected pulsars, to construct timing solutions enabling us to determine pulsar orbital phases with good accuracy. 

The NUPPI data on the selected pulsars were cleaned of radio-frequency interference (RFI) using the \textsc{surgical} method of the \textsc{CoastGuard} pulsar analysis library \citep{Lazarus_2016}. The PSRCHIVE software library \citep{Hotan_2004} was used to calibrate the polarization information, and to perform all subsequent data reduction steps. 

The timing analysis was done in two steps. We first constructed reference profiles for each of the considered pulsars by summing up eight time- and frequency-averaged observations with high S/N values, and smoothing the obtained integrated profiles. We then extracted TOAs from the individual observations by cross-correlating them with the reference profile for the corresponding pulsar. In this first step, the cross-correlations were performed using the Fourier domain with Markov chain Monte Carlo algorithm implemented in the \texttt{pat} routine of PSRCHIVE. We generated one TOA per 64~MHz of frequency bandwidth in order to track time variations of the DM, and per 5~min, for the TOAs to cover less than a few percents of the pulsar's orbital period. TOA data were analyzed using the \textsc{Tempo2} pulsar timing suite \citep{Hobbs_2006}. We discarded TOAs around superior conjunction in the case of pulsars observed to display eclipses in the NRT data. We then obtained timing solutions for each pulsar by fitting the timing residuals (\textit{i.e.}, the differences between measured arrival times and those predicted by \textsc{Tempo2}) for their spin, astrometric, DM and orbital parameters. DM variations were modeled by fitting for up to two time derivatives, and similarly variations of the orbital period, which are commonly observed in these systems, were fitted for by including up to 17 derivatives (in the case of PSR J2051$-$0827). 

In the second step of the timing analysis, we used the timing solutions obtained from the previous analysis step to refold all individual observations, and formed 2D template profiles from archives with 8 frequency sub-bands of 64~MHz each for each pulsar. For pulsars with observations that have S/N values larger than 100, we used the highest-S/N observation as the 2D template. For the other pulsars we summed up the three highest-S/N observations to form the 2D template. With these standard profiles with frequency resolution we could follow the same analysis procedure as in \citet{Guillemot_2019} and use the wide-band template matching technique implemented in the \texttt{PulsePortraiture} library \citep{Pennucci_2014} to extract one TOA per 10~min of observation, for the entire bandwidth. Extracting TOAs using the 2D template-matching method implemented in the \texttt{PulsePortraiture} toolkit has several advantages: first of all, as can be seen from Table~\ref{tab:obs_sample}, many of the pulsars in our sample are detected with low S/N values at 1.4~GHz with the NRT. For observations near the detection limit, using the usual \texttt{pat} routine to determine one TOA per frequency-band of 64~MHz can result in a large number of outlier TOAs. In Table~\ref{tab:1Dvs2D} we list the fraction of TOA outliers as obtained with the 1D and 2D TOA extraction methods. Extracting TOAs with the wideband-matching technique results in lower outlier rates in all cases, and in some cases (see e.g. PSR~J1513$-$2550 or J2055+3829) the outlier rate decreases dramatically when using the 2D template matching method.

\begin{table}
\caption{Fraction of outliers in the TOA datasets.}
\label{tab:1Dvs2D}
\centering

\begin{tabular}{cccc}
\hline
\hline
Pulsar & 1D F8 & 2D & 1D fscrunch \\
\hline
J0023+0923 & 53.8\% & 5.3\% & 3.8\%  \\
J0610$-$2100 & 9.1\% & 2.1\% & 1.1\% \\
J0636+5128 & 12.2\% & 3.4\% & 1.5\%  \\
J1124$-$3653 & 82.1\% & 60.3\% & 56.3\% \\
J1513$-$2550 & 18.8\% & 5.4\% & 6.8\% \\
J1544+4937  & 34.8\% & 2.9\% & 2.6\% \\
J1555$-$2908  & 13.9\% & 6.4\% & 6.4\% \\
J1705$-$1903 & 0.6\% & 0.0\% & 0.0\% \\
J1719$-$1438 & 68.8\% & 2.0\% & 1.0\% \\
J1731$-$1847 & 41.4\% & 28.1\% & 77.4\% \\
J1745+1017 & 16.4\% & 1.2\% & 2.4\% \\
J1959+2048 & 43.4\% & 6.6\% & 8.9\% \\
J2051$-$0827 & 9.0\% & 3.1\% & 3.4\% \\
J2055+3829 & 17.4\% & 11.5\% & 11.7\% \\
J2115+5448 & 36.2\% & 23.5\% & 40.4\% \\
J2214+3000 & 90.2\% & 5.3\% & 4.0\% \\
J2234+0944 & 14.4\% & 6.5\% & 6.4\% \\
J2256$-$1024 & 48.2\% & 5.4\% & 5.1\% \\
\hline
\end{tabular}
\tablefoot{For each pulsars we compared TOAs extracted from eight frequency sub-bands of 64~MHz each using the standard template-matching method (``1D F8''), and TOAs extracted using the entire available bandwidth, with the wideband template-matching technique as implemented in the \texttt{PulsePortraiture} library (``2D''). For reference we also list the outlier rates as obtained when extracting TOAs from fully frequency-scrunched observations (``1D fscrunch''). Outliers are defined here are those with residuals larger than 1\% of $P_{spin}$ considering the uncertainty.}
\end{table}

\begin{figure}[ht]
   \centering
   \includegraphics[width=\hsize]{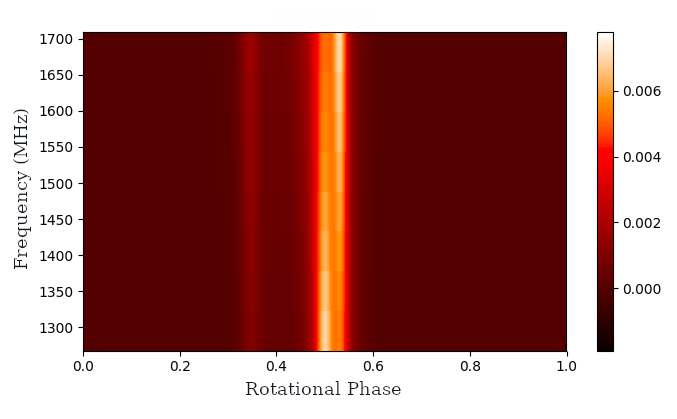}
      \caption{The 2D template profile of PSR~J0023+0923 as determined using the \texttt{PulsePortraiture} software suite. The sub-components of the main pulse of this pulsar are observed to evolve with frequency, with the first one being brighter than the second at lower frequencies, and fainter at the top of the observed bandwidth. When collapsed in frequency, this profile displays two equally bright peaks.}
         \label{fig:J0023}
   \end{figure}

Second, in some pulsars the pulse profile is observed to vary significantly with frequency, as is illustrated in Fig.~\ref{fig:J0023} for PSR~J0023+0923. For these pulsars, TOAs determined using the 2D template-matching method are more accurate than TOAs extracted using a single frequency-averaged profile for the entire band. Finally, as mentioned in \citet{Guillemot_2019}, in addition to determining a single TOA for the entire available bandwidth in a given time sample, the \texttt{PulsePortraiture} library also determines a DM value by comparing the reference profile and the 2D profile in the considered time sample, theoretically enabling us to investigate short-term variations of the DM around superior conjunction. However, in our sample the majority of the pulsars are too faint to give satisfactory DM measurements, hence we decided on conducting the following study on the TOAs only.
After the latter analysis step, we obtained TOA and DM datasets for each pulsar in the sample, with one measurement per 10 min of observation. In the next subsection we present a phenomenological analysis of the eclipses properties of the pulsar in our sample, using these TOA datasets.

\subsection{Eclipse characterization}
\label{sect:eclipses}

\begin{figure*}[ht]
   \centering
   \includegraphics[width=\hsize]{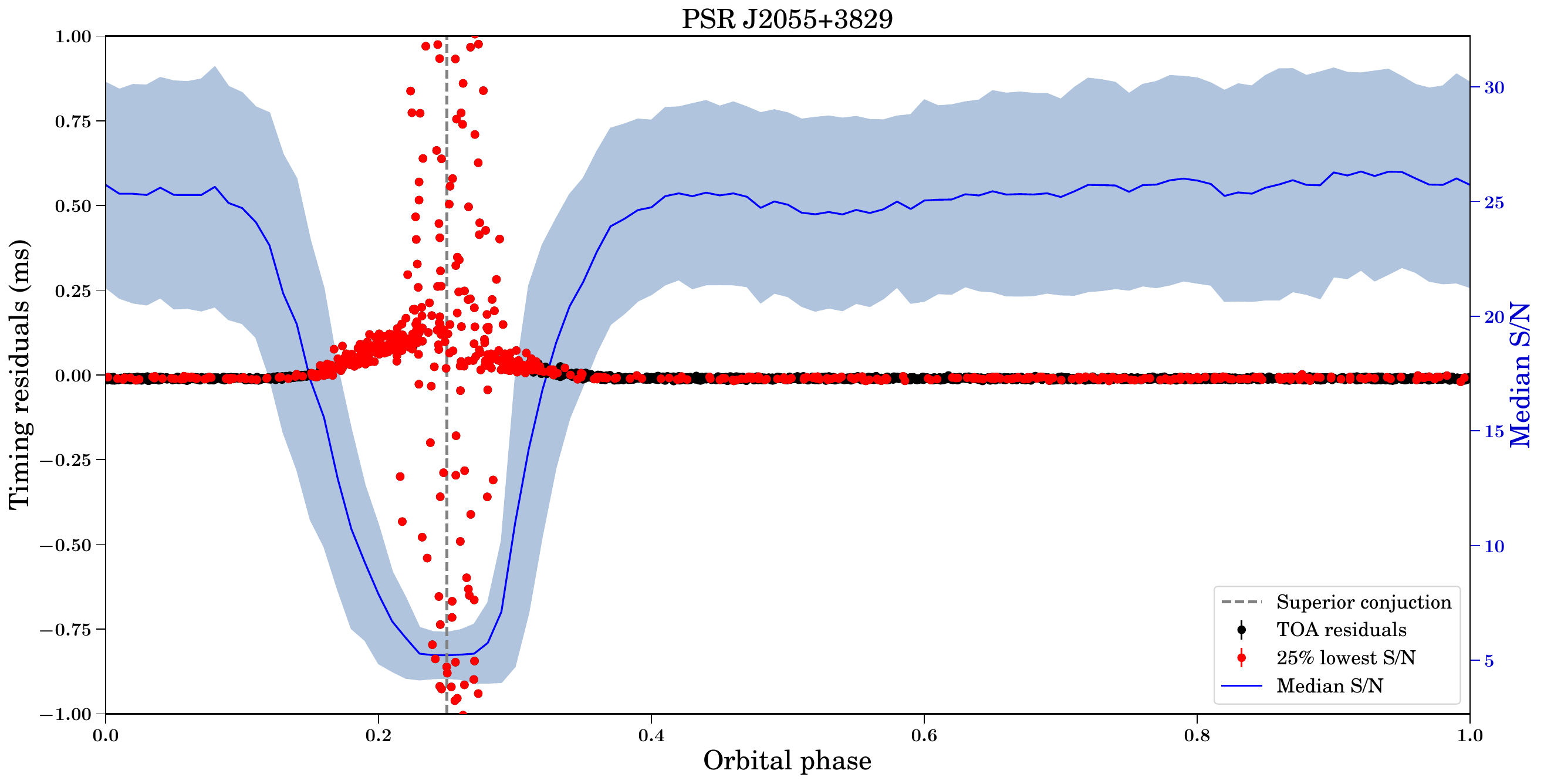}
      \caption{Timing residuals and median S/N values as a function of orbital phase for PSR~J2055+3829. TOA residuals are shown as red and black dots, with red dots corresponding to observations with S/N values among the 25\% lowest. Error bars, which are much smaller than the vertical scale, are not visible in this figure. The blue curve is the median S/N of the TOAs computed over sliding windows of width 0.1 in orbital phase, and the blue zone shows the median absolute deviation around the median values. The dashed vertical line corresponds to superior conjunction, at orbital phase 0.25.}
        \label{fig:SNR_drop}
   \end{figure*}

Depending on the precise geometry of the systems, the radio signal from eclipsing spider pulsars can disappear or get attenuated over a certain phase interval around superior conjunction. The TOA extraction procedure described in Sect.~\ref{sect:analysis} determined TOAs in every 10-min sample of the pulsar observations, \textit{i.e.}, independently of whether the pulsar is detected or not. The procedure also determines a S/N associated which each TOA, therefore providing a way of discriminating eclipsing and non-eclipsing spiders: around superior conjunction, pulsars exhibiting eclipses are expected to be detected with lower S/N values than well away from superior conjunction, whereas pulsars that do not have eclipses should be detected with similar S/N values throughout their orbit. Figure~\ref{fig:SNR_drop} shows timing residuals and S/N values as a function of orbital phase, for PSR~J2055+3829.

It should be noted that the data points shown in Fig.~\ref{fig:SNR_drop} were recorded over a very large number of pulsar orbits, and highly-sensitive observations of spider pulsars have shown that the eclipse properties of some objects vary from one orbit to another \citep{Polzin_2019}. Fig.~\ref{fig:SNR_drop} therefore shows the ``average'' eclipse properties of PSR~J2055+3829 in our dataset. Nevertheless, median S/N values for this pulsar are clearly seen to drop around superior conjunction. We thus define PSR~J2055+3829 as being an eclipsing pulsar in the 1.4~GHz NRT data, and generally define eclipsing pulsars as those displaying S/N drops around superior conjunction. In Figs.~\ref{fig:SNR_drop_appendix1} to \ref{fig:SNR_drop_appendix3} we show plots of the S/N values as a function of orbital phase for all the pulsars in our sample. 

From these plots we classify eclipsers and non-eclipsers as follows: a pulsar is considered an eclipser when the median S/N at superior conjunction is lower than the median S/N of $\phi$ > 0.5 minus the median absolute deviation. With this definition, eight out of 19 spiders of the sample are eclipsers. Those eight pulsars display what we call ``total'' eclipses, during which the beam is not detectable for a few sub-integrations, leading to very low S/N TOAs. In addition, around superior conjunction the residuals of the eclipsers display a significant delay, typically around 5 to 8\% of the spin period. These TOAs are usually associated with S/N values lower than those of TOAs well away from the eclipse region, but still higher than the threshold we defined to determine non detections. The additional delay in TOAs that correspond to detections is due to the additional plasma crossing the line-of-sight.

Nevertheless, three pulsars in our sample display eclipsing behavior in our data, without strictly respecting the criteria mentioned above for defining eclipsing pulsars. PSR~J1124$-$3653, whose TOA residuals around superior conjunction are not distributed around zero unlike in other orbital phase ranges, appears to display eclipses. However, it is too faint for our eclipse analysis to yield satisfactory results. PSR~J1705$-$1903 displays ``shallow'' eclipses: the S/N drop for this pulsar is less pronounced than in other pulsars due to the fact that its radio beam is detectable throughout the orbit. Yet, residuals around superior conjunction clearly depart from zero up to approximately 0.5\% of its spin period, \textit{i.e.}, much less strongly than in other eclipsers. Because the beam is eclipsed by plasma, the eclipsing mechanisms are frequency dependent, and eclipses are more pronounced at lower frequencies compared to higher frequencies. This also means that the frequency below which the pulsed signal disappears, called the cutoff frequency, varies from one pulsar to another \citep[it can also vary with time, see e.g.][for a detailed study of the variation of the eclipse cutoff frequency in PSR~J1544+4937]{Kumari_2024}. The cutoff frequency in PSR~J1705$-$1903 appears to be lower than that of most known eclipsers, at a frequency below 1.2~GHz. No observations of this pulsar at these low frequencies have been reported, to the best of our knowledge. The third pulsar, PSR~J2051$-$0827, does not display any drop in S/N around superior conjunction. However, its TOA residuals depart from zero, indicating that extra material is traversed by the radio beam. PSR~J2051$-$0827's eclipses therefore also appear to have a cutoff frequency below 1.2~GHz. This pulsar has been observed thoroughly and displays total eclipses at 660~MHz \citep{Stappers_1996}. In conclusion, PSRs~J1124$-$3653, J1705$-$1903 and J2051$-$0827 will still be considered eclipsers in the rest of the paper, and are flagged as such in Table~\ref{tab:binary_sample}. However, due to the above-mentioned peculiarities, the phenomenological fits presented in the following section did not yield satisfactory results for the very faint pulsar J1124$-$3653, and did not produce optimal results for PSRs~J1705$-$1903 and J2051$-$0827, due to the lack of clear, total eclipse phases in the data for these objects.

As mentioned above, the TOA extraction procedure determined TOAs for every 10 min sample, regardless of whether the observed pulsar is indeed detected in the considered time sample or not. TOAs in these time samples corresponding to non-detections therefore needed to be discarded from our datasets. We again made selections in our timing datasets based on the S/N values associated with the TOAs: as can be seen from Fig.~\ref{fig:SNR_drop}, TOAs with S/N values among the 25\% lowest are mostly located in the orbital phase interval corresponding to the eclipse. For the 10 eclipsing pulsars we discarded TOAs associated with S/N values within the first quartile. This threshold of 25\% was chosen as a compromise, for the 10 considered objects, between the need to eliminate data points corresponding to non-detections while keeping large enough numbers of data points. 

We finally modelled the timing residuals of the pulsars in our sample to determine their average eclipse properties, following a phenomenological approach. This analysis is presented in the following subsection.

\begin{table*}
\caption[]{Parameters of the functions describing the eclipse ingress and egress phases.}
\label{tab:results_fit}
\centering

\begin{tabular}{ccccccc}
\hline
\hline
Pulsar & \multicolumn{3}{c}{Ingress} & \multicolumn{3}{c}{Egress} \\
\hline
 & $\mu_i$ &$\tau_i$ ($\phi^{-1}$) & $\Phi_\mathrm{start}$ & $\mu_e$ & $\tau_e$ ($\phi^{-1}$) & $\Phi_\mathrm{end}$ \\
\hline
*J1513$-$2550 & 0.37(9) & > 0.01 & 0.23 & 0.21(1) & 0.019(2) & 0.31(1) \\
*J1555$-$2908 & 0.37(9) & > 0.01 & 0.21 & 0.242(4) & 0.0191(9) & 0.37(2) \\
J1628$-$3205 & 0.224(8) & 0.024(2) & 0.102(4) & 0.290(1) & 0.0355(6) & 0.471(2) \\
J1705$-$1903 & 0.440(3) & 0.0461(7) & 0.092(2) & 0.072(6) & 0.041(1) & 0.381(3) \\
J1731$-$1847 & 0.20(7) & 0.01(1) & 0.134(7) & 0.3430(5) & 0.0123(2) & 0.42(2) \\
J1959+2048 & 0.260(3) & 0.0078(6) & 0.21(1) & 0.205(3) & 0.0231(8) & 0.345(7) \\
J2051$-$0827 & 0.346(1) & 0.0356(3) & 0.1509(7) & 0.119(2) & 0.0440(5) & 0.361(1) \\
J2055+3829 & 0.2342(9) & 0.0179(2) & 0.127(2) & 0.236(1) & 0.0227(3) & 0.3713(5) \\
*J2115+5448 & 0.36(9) & > 0.01 & 0.20 & 0.236(1) & 0.0258(4) & 0.352(1) \\
J2256$-$1024 & 0.273(3) & 0.0114(6) & 0.205(1) & 0.211(1) & 0.0200(2) & 0.3297(5) \\
\hline
\end{tabular}
\tablefoot{The parameters are determined from the fit presented in Sect.~\ref{sect:study}. $\mu$ and $\phi$ are orbital phases and therefore have no unit. The 3 ``abrupt'' eclipsers (starred names) have no proper ingress phase, hence the beginning of the eclipse is taken at the last TOA above S/N threshold before the superior conjunction and the $\tau_i$ is only an upper limit.}
\end{table*}

\begin{table*}
\caption[]{Parameters derived from the results of the fit of pulsar eclipse properties. }
\label{tab:params_fit}
\centering

\begin{tabular}{ccccccccc}
\hline
\hline
Pulsar & \multicolumn{2}{c}{$T_\mathrm{delay}$} & \multicolumn{2}{c}{$T_\mathrm{obsc}$} & $T_\mathrm{obsc}$ / $T_\mathrm{delay}$ & $S$ & $D$ & Category \\
\hline
 & (\% $P_B$) & (min) &(\% $P_B$) & (min) & &  &  \\
\hline
J1513$-$2550 & 8.3(5) & 21(1) & 6.4 & 16.6 & 0.78(2) & > 0.53 & 0.38(3) & A \\
J1555$-$2908 & 15.7(2) & 52.7(7) & 10.5 & 35.1 & 0.67(1) & > 0.52 & 0.331(6) & A \\
J1628$-$3205 & 37.0(4) & 111(3) & 16.8 & 50.5 & 0.456(5) & 0.67(6) & 0.67(3)& P \\
J1705$-$1903 & 28.9(3) & 76.6(9) & 0* & 0.8 & 0* & 1.13(3) & 1.21(3) & S \\
J1731$-$1847 & 29(3) & 1.3(1)e+02 & 21.5 & 96.4 & 0.75(2) & 1(1) & 0.7(2) & A/P \\
J1959+2048 & 13.1(2) & 72(8) & 4.6 & 25.5 & 0.353(5) & 0.34(3) & 0.39(6) & A/P \\
J2051$-$0827 & 21.0(1) & 29.9(2) & 0* & 0.0 & 0* & 0.81(1) & 0.90(1) & S \\
J2055+3829 & 24.4(2) & 45.6(3) & 12.3 & 23.0 & 0.504(2) & 0.79(1) & 1.02(2) & P \\
J2115+5448 & 15.1(1) & 29.5(2) & 7.7 & 15.0 & 0.508(3) & > 0.39 & 0.485(5) & A \\
J2256$-$1024 & 12.5(1) & 38.2(4) & 3.2 & 9.8 & 0.258(4) & 0.57(3) & 0.56(1) & P \\
\hline
\end{tabular}
\tablefoot{For each pulsar we list $T_\mathrm{delay}$, the total duration of the eclipse, \textit{i.e.}, the time over which the residuals display an excess delay, $T_\mathrm{obsc}$ (``obsc'' for obscuration), the time over which the pulsar is completely eclipsed, the $T_\mathrm{obsc} / T_\mathrm{delay}$ ratio, and the ingress/egress slope ratio $S$ and duration ratio $D$, as defined in Sect.~\ref{sect:study}. Values of $T_\mathrm{delay}$ and $T_\mathrm{obsc}$ are expressed in fraction of the orbital period and in minutes. Values of $S$ and $D$ were determined using the values of $\tau$ and $\Phi$ listed in Table~\ref{tab:results_fit}. In the column category, A stands for ``abrupt'', \textit{i.e.}, eclipsers for which the S/N cut-off left no TOAs forming an ingress slope, P stands for ``progressive'', meaning eclipsers that display a proper ingress phase, and S stands for ``shallow'', meaning eclipsers for which the beam is detectable through the whole orbit.}
\end{table*}

\subsection{Phenomenological modelling}
\label{sect:study}

For each of the 10 pulsars found to display eclipses in the NRT data, we fitted the timing residuals cleaned of the 25\% lowest S/N detections to a symmetrical function over a background level: 

\begin{equation}
F(\phi) = f_i(\phi) H(-(\phi-0.25)) + f_e(\phi) H(\phi-0.25) + b,
\end{equation}

\noindent
with:

\begin{equation}
f_i(\phi) = \exp \left( \frac{\phi - \mu_i}{\tau_i} \right) \\ 
f_e(\phi) = \exp \left( - \frac{\phi - \mu_e}{\tau_e} \right).
\end{equation}

\noindent
and $H$ the Heaviside step function: $H(x) = \begin{cases} 1, & x \geq 0 \\ 0, & x < 0 \end{cases}$
\newline

\begin{figure}[h]
   \centering
   \includegraphics[width=\hsize]{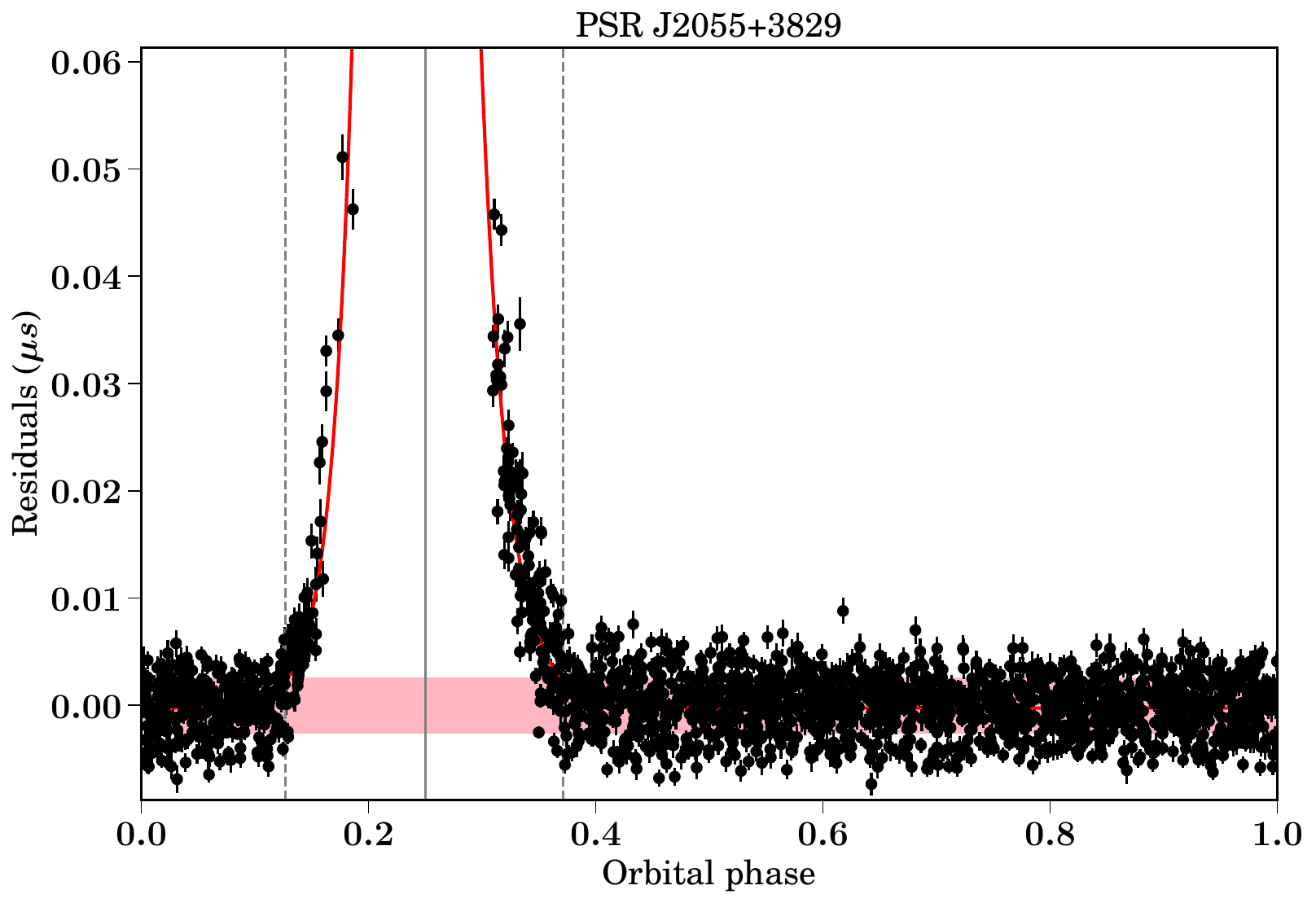}
      \caption{Timing residuals as a function of orbital phase around superior conjunction, indicated with a solid line, for PSR~J2256$-$1024. The best-fit function obtained from the analysis presented in Sect.~\ref{sect:study} is shown in red. The light red zone shows the RMS of the timing residuals for orbital phases $\phi > 0.5 $, and the dashed vertical lines correspond to the values of $\Phi_\mathrm{start}$ and $\Phi_\mathrm{end}$ as inferred from the fit.}
         \label{fig:J2256_fit}
   \end{figure}

In the above expressions, $b$ is the background level, corresponding to the average timing residual value away from the eclipse, and the $f_i$ and $f_e$ are the functions used to fit the timing residuals in the ingress and egress phases. Each of the two functions has two parameters: a centroid $\mu$ and a characteristic phase $\tau$. Five parameters are therefore fit for, for each pulsar: $b$, and the $\mu$ and $\tau$ parameters. The function for the ingress part is fitted to timing residuals at orbital phases from 0 to 0.25, and that for the egress part is fitted to residuals at orbital phases from 0.25 to 0.5. The parameter space was explored using a parallelized affine-invariant Monte Carlo sampling algorithm implemented in the \textsc{EMCEE} library \citep{emcee}. Fig.~\ref{fig:J2256_fit} shows an example of timing residuals as a function of orbital phase around superior conjunction for PSR~J2256$-$1024, along with the best-fit function found from this analysis. Figures showing the fit results for the other analyzed pulsars can be found in Appendix~B. Results of the fits are summarized in Table~\ref{tab:results_fit}.

From this fit we determined two additional parameters, the values of which are collected in Table~\ref{tab:results_fit}: $\Phi_\mathrm{start}$, the orbital phase at which the ingress phase begins, and $\Phi_\mathrm{end}$, the orbital phase at which the egress phase ends. Between $\Phi_\mathrm{start}$ and $\Phi_\mathrm{end}$, the best-fit function is above $b + \sigma$, with $\sigma$ denoting the standard deviation of the timing residuals away from the eclipse phase (in this case, we considered orbital phases larger than 0.5). For ``abrupt'' eclipses, no data points could be fit in the ingress phase. In these cases, $\Phi_\mathrm{start}$ was defined as the orbital phase of the last valid data point before superior conjunction.

Measuring $\Phi_\mathrm{start}$ and $\Phi_\mathrm{end}$ enabled us to determine eclipse duty cycles, $T_\mathrm{delay}$, as $T_\mathrm{delay} = \Phi_\mathrm{end} - \Phi_\mathrm{start}$. For each pulsar we also determined $T_\mathrm{obsc}$, corresponding to the phase interval over which the pulsar is completely eclipsed in our data, \textit{i.e.}, the duration between the leftmost point with $\phi > 0.25$ and the rightmost point with $\phi < 0.25 $. Values of $T_\mathrm{eclipse}$ and $T_\mathrm{obsc}$ from our analysis are listed in Table~\ref{tab:params_fit}. In this Table, the values of $T_\mathrm{delay}$ and $T_\mathrm{obsc}$ are expressed in fraction of the orbital period and in time. In Table~\ref{tab:params_fit} we also list slope ratios $S = \tau_i / \tau_e$, and ingress/egress duration ratios $D$, calculated as $D = (0.25 - \Phi_\mathrm{start}) / (\Phi_\mathrm{end} - 0.25)$. The closer $D$ is to 1, the more symmetric the eclipse is, and conversely. 

The parameter $T_\mathrm{obsc} / T_\mathrm{delay}$ is the fraction of the eclipse phase over which the beam is not detectable. The closer to 1 this parameter is, the shorter the ingress and/or egress phases are before and/or after the flux drops below the detection level. Conversely, a value of 0 indicates that the beam is detectable throughout the eclipse.


\section{Polarimetry}
\label{sect:polar}

In addition to determining the eclipse properties of the pulsars in our sample that display eclipses in the 1.4~GHz NRT data, we also characterized the polarimetric properties of these pulsars. To construct polarimetric profiles for the 19 pulsars, we selected 1.4~GHz NUPPI observations conducted after MJD~58800 (13 November 2019), \textit{i.e.}, observations that can be polarization calibrated using the improved calibration scheme presented in \citet{Guillemot_2023}. For pulsars displaying eclipses we only considered data taken away from the eclipse regions. Individual observations, prepared with the same procedure as described in Sect.~\ref{sect:timing}, were summed in time using the \texttt{psradd} tool of PSRCHIVE, to form high S/N polarimetric profiles. For pulsars with Rotation Measure (RM) information available in the ATNF pulsar catalog, we corrected the polarimetric profiles from Faraday rotation of the polarization position angle across the bandwidth. For the other five pulsars (namely, PSRs~J1124$-$3653, J1513$-$2550, J1555$-$2908, J1628$-$3205, J2115+5448), no RM information was available in the ATNF catalog at the time of writing. We attempted to determine RM values from the summed NUPPI polarimetric profiles, but the low degrees of linear polarization in these pulsars and very low S/N values in some cases prevented us from determining the RMs. The summed observations were finally scrunched in frequency.

The resulting polarimetric profiles at 1.4~GHz for the 19 pulsars in our sample are shown in Figs.~\ref{fig:profile_appendix1} to \ref{fig:profile_appendix3}. Pulsar profiles that were not corrected for Faraday rotation are marked with a star. We note that the 1.4~GHz polarimetric profiles of PSRs~J1124$-$3653, J1555$-$2908, J1628$-$3205, and J2115+5448 were previously unpublished, to the best of our knowledge. Polarimetric profiles for the other pulsars are consistent with the previously published ones. In Table~\ref{tab:polar} we list the pulsar RMs, and values of the polarized intensities and linear, circular and absolute circular intensities, as determined using the \texttt{psrstat} tool of PSRCHIVE. For pulsars for which we lacked an RM value and thus could not correct the data for Faraday rotation, we do not quote the polarized intensities and linear intensities. Polarization fractions vary broadly across the sample, with values ranging from 13\% to 66\%.

\begin{table*}
\caption[]{Polarization parameters for the 19 pulsars in our sample. }
\label{tab:polar}
\centering

\begin{tabular}{ccccccc}
\hline
\hline
Pulsar & RM & RM Reference & $I_p/I$  & $\langle L \rangle / I$ & $\langle V \rangle / I$ & $\langle \| V \| \rangle / I$ \\
 & (rad m$^{-2}$) & & (\%) & (\%) & (\%) & (\%) \\
\hline
J0023+0923 & $-6$ & \citet{Wang_2023} & $30$ & $29$ & $3$ & $4$ \\
J0610$-$2100 & $33.7$ & \citet{Spiewak_2022} & $66$ & $65$ & $-6$ & $5$ \\
J0636+5128 & $-7.0$ & \citet{Wahl_2022} & $35$ & $33$ & $-5$ & $6$ \\
J1124$-$3653 & / & / & / & / & $3$ & $8$ \\
J1513$-$2550 & / & / & / & / & $-8$ & $2$ \\
J1544+4937 & $10.3$ & \citet{Wang_2023} & $15$ & $12$ & $1$ & $2$ \\
J1555$-$2908 & / & / & / & / & $5$ & $4$ \\
J1628$-$3205 & / & / & / & / & $-2$ & $3$ \\
J1705$-$1903 & $-5.6$ & \citet{Spiewak_2022} & $19$ & $18$ & $1$ & $1$ \\
J1719$-$1438 & $17.3$ & \citet{Spiewak_2022} & $38$ & $24$ & $-27$ & $12$ \\
J1731$-$1847 & $21.5$ & \citet{Spiewak_2022} & $51$ & $47$ & $-20$ & $17$ \\
J1745+1017 & $24.2$ & \citet{Wang_2023} & $48$ & $40$ & $-23$ & $17$ \\
J1959+2048 & $-70.1$ & \citet{Wang_2023} & $27$ & $20$ & $15$ & $10$ \\
J2051$-$0827 & $-46$ & \citet{Han_2018} & $13$ & $7$ & $-1$ & $8$ \\
J2055+3829 & $-65.8$ & \citet{Wang_2023} & $17$ & $17$ & $> 0.1$ & $> 0.1$ \\
J2115+5448 & / & / & / & / & $-3$ & $4$ \\
J2214+3000 & $-44.5$ & \citet{OSullivan_2023} & $29$ & $29$ & $> 0.1$ & $> 0.1$ \\
J2234+0944 & $-10.4$ & \citet{Wang_2023} & $15$ & $12$ & $4$ & $4$ \\
J2256$-$1024 & $15.0$ & \citet{Crowter_2020} & $18$ & $13$ & $2$ & $10$ \\
\hline
\end{tabular}
\tablefoot{Rotation Measure (RM) values were taken from the articles listed in the ``RM Reference'' column. Polarized intensities ($I_p$), linear ($\langle L \rangle$), circular ($\langle V \rangle$) and absolute circular ($\langle \| V \| \rangle$) intensities are quoted in percentages of the total intensity ($I$). See Sect.~\ref{sect:polar} for details on the analysis.}
\end{table*}

   \begin{figure}[h]
   \centering
   \includegraphics[width=\hsize]{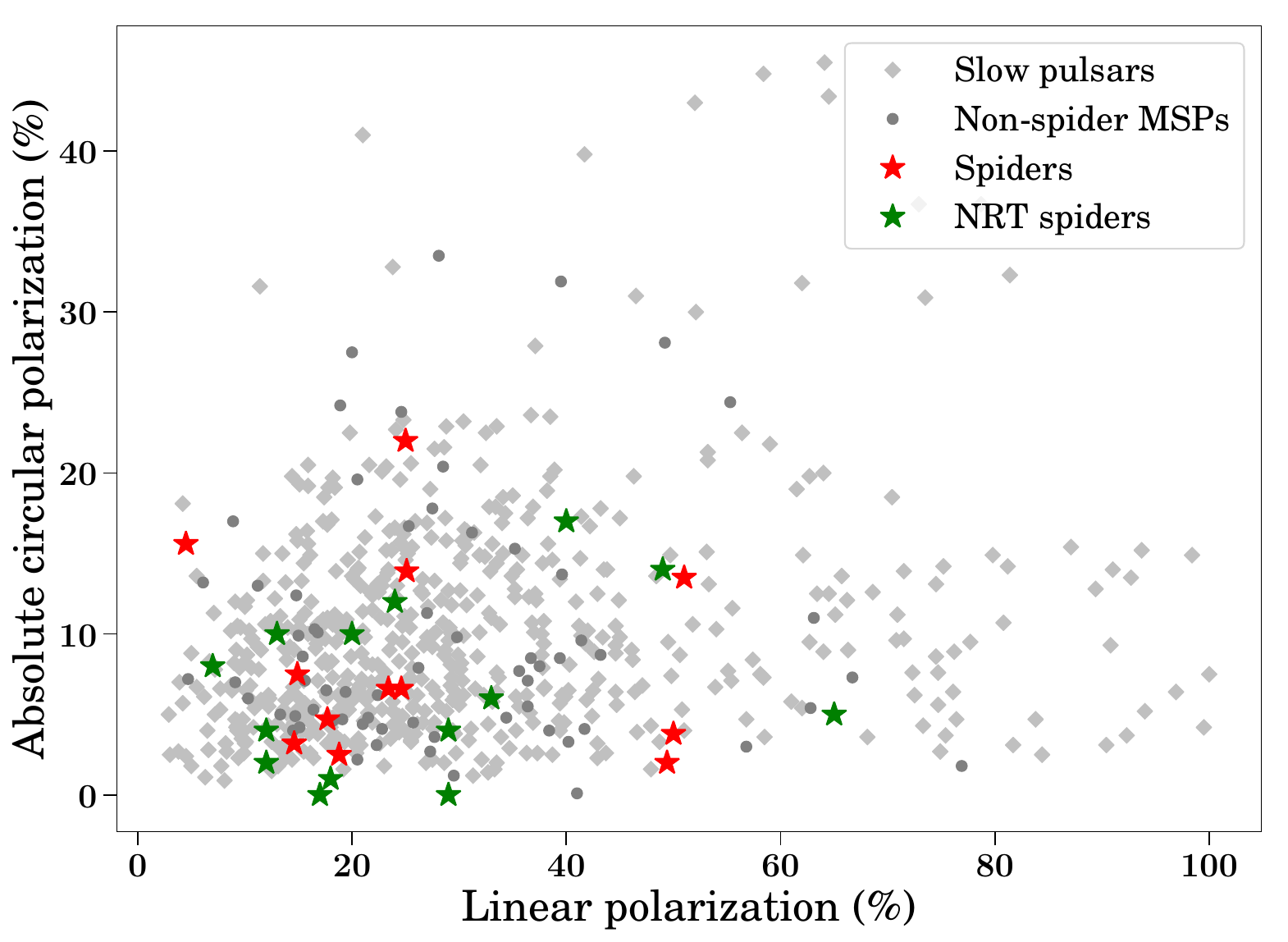}
      \caption{Linear polarization fractions and absolute circular polarization fractions for the 682 pulsars from \citet{Wang_2023}. Grey diamonds are slow pulsars, grey dot represent isolated MSPs or MSPs in non-spider systems (pulsars orbitting white dwarf companions or main sequence stars). Red stars are spiders that are not in the NRT sample, and green stars are the spiders analyzed in this articles. The values for the latter objects can be found in Table~\ref{tab:polar}.}
         \label{fig:FAST_polar}
   \end{figure}

In order to put our polarization fraction measurements in a broader context, we plotted in Fig~\ref{fig:FAST_polar} the linear and absolute circular polarization fractions listed in Table~\ref{tab:polar} along with the measurements for 682 pulsars published in \citet{Wang_2023}, obtained using the Five-hundred-meter Aperture Spherical radio Telescope (FAST). One can see on this figure that most spider pulsars appear to have low degrees of polarization, similarly to non-spider MSPs. It can also be noted that NRT measurements are complementary to those of \citet{Wang_2023} and display similar trends. We performed 2D Kolmogorov-Smirnov tests using the publicly-available software \textsc{ndtest}\footnote{Written by Zhaozhou Li, \url{https://github.com/syrte/ndtest}}, to compare the polarization fractions of spider pulsars with those of other MSPs and slow pulsars. For spider pulsars we combined NRT and FAST polarization measurements, leading to a sample of 26 objets with linear and absolute circular polarization fractions. The other samples included 66 non-spider MSPs, and 582 slow pulsars. We found a probability that polarization data for spider MSPs and slow pulsars originate from the same parent distribution of approximately 8.5\%, suggesting that the two populations have distinct polarization properties. On the other hand, the test found a larger probability of $\sim 20$\% that the data for spider and non-spider MSPs originate from the same parent distribution, thus preventing us from concluding.


\section{Analysis of eclipse and pulse profile properties}
\label{sect:results}

\subsection{Correlations between  eclipse parameters}
\label{subsect:eclipse_fit_results}

\begin{figure}[h]
   \centering
   \includegraphics[width=\hsize]{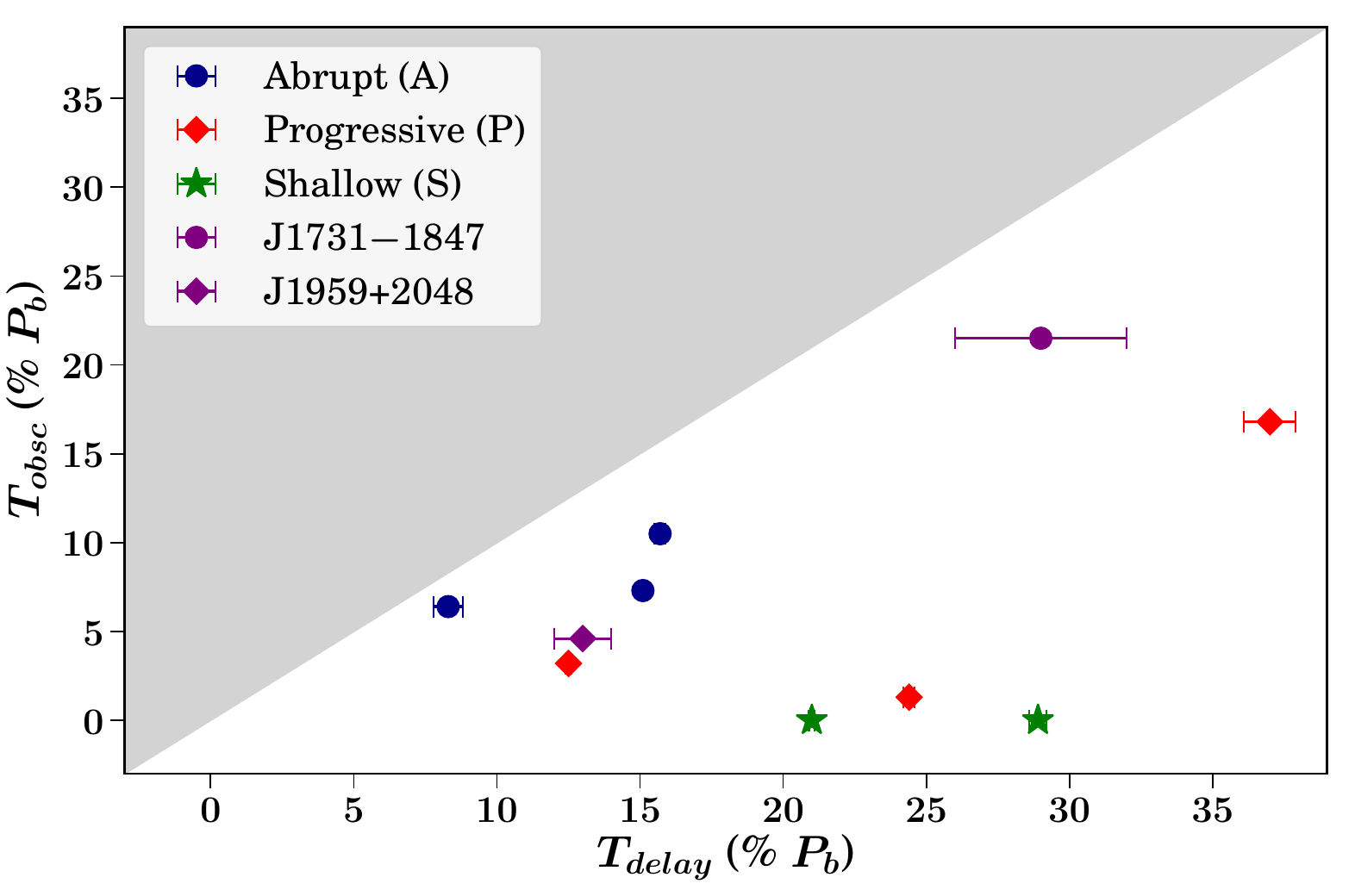}
      \caption{Obscuration duration ($T_\mathrm{obsc}$) against delay duration ($T_\mathrm{delay}$). The grey zone line is the 1:1 line. An eclipsing pulsar on this line would have no ingress and egress phases. The further a pulsar is from this line, the smaller is the fraction of the eclipse over which the beam is not detectable. Marker colors and shapes depend on the classifications, as given in Table~\ref{tab:params_fit}: blue circles correspond to the ``abrupt'' eclipsers, PSRs~J1513$-$2550, J1555$-$2908 and J2115+5448, red diamonds are the ``progressive'' eclipsers, PSRs~J1628$-$3205, J2055+3829 and J2256-1024 and green stars are the ``shallow'' eclipsers, namely PSRs J1705$-$1903 and J2051$-$0827.}
         \label{fig:TeVSTt}
   \end{figure}

\begin{figure}[h]
   \centering
   \includegraphics[width=\hsize]{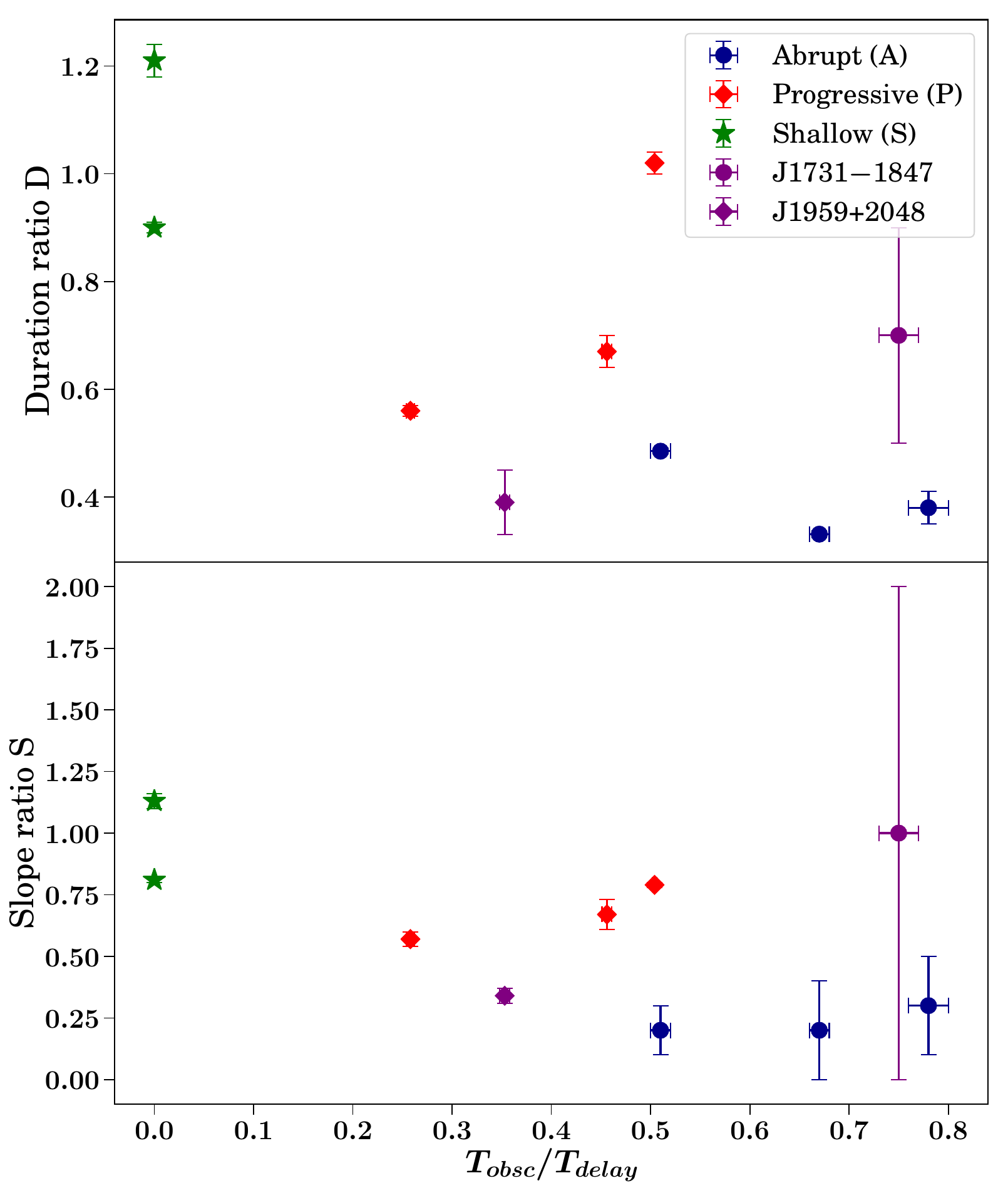}
      \caption{Ingress/egress duration ratio $D$ and ingress/egress slope ratio $S$ as a function of the $T_\mathrm{obsc} / T_\mathrm{delay}$ ratio, \textit{i.e.}, the fraction of the orbit over which the radio beam of the pulsar is not detectable. Values of $S$ and $D$ can be found in Table~\ref{tab:params_fit}. Colors and symbols are the same as in Fig.~\ref{fig:TeVSTt}.}
         \label{fig:ratioVSparams}
   \end{figure}

Two categories of eclipsers can be distinguished in Figs.~\ref{fig:fit_appendix1} and \ref{fig:fit_appendix2}, in addition to the ``shallow'' eclipsers PSRs J1705$-$1703 and J2051$-$0827. On the one hand, eclipsing pulsars displaying no ingress phases, for which the S/N cut-off left no TOAs forming an ingress slope outside the red zone (RMS of TOA residuals far from the eclipse), namely PSRs J1513$-$2550, J1555$-$2908 and J2115+5448, and on the other hand, pulsars with a progressive ingress phase, namely PSRs J1628$-$3205, J2055+3829 and J2256$-$1024. It is worth noting that two pulsars lie in a grey area between the earlier two categories: PSRs J1731$-$1847 and J1959+2048. PSR~J1731$-$1847 is the least observed eclipser in our sample, and very few TOAs remained after data selection, making it difficult to probe the ingress region and search for a proper slope in that region. Unlike PSR~J1731$-$1847, the data selection left numerous TOAs in the ingress phase of PSR~J1959+2048. However, this one displays a very steep ingress and is seen to eclipse in gamma rays \citep{Clark_2023}, similarly to PSR~J1555$-$2908 which also displays an ``abrupt'' ingress phase. For these reasons, the two pulsars are represented with different colors than others in Figs.~\ref{fig:TeVSTt} and \ref{fig:ratioVSparams}.

Looking at Table~\ref{tab:params_fit}, we can notice that eclipse durations ($T_\mathrm{delay}$) vary widely, ranging from 8.3\% up to 37\% of the respective orbital periods, as do the obscuration durations ($T_\mathrm{obsc}$), that range from 0\% to 21.5\% of the orbital period. These two parameters do not appear to be strongly correlated, as can be seen from Fig.~\ref{fig:TeVSTt}. Nevertheless, it appears from this figure that $T_\mathrm{obsc}$ seems to be closer to $T_\mathrm{delay}$ for ``abrupt'' eclipsers, represented as blue circles, than for those classified as being ``progressive'' eclipsers, shown with red diamonds. In this regard, PSR~J1731$-$1847 bears resemblance to ``abrupt'' eclipsers, while PSR~J1959+2048 shows similarities with ``progressive'' eclipsers.

In addition, it can be seen from Table~\ref{tab:params_fit} that most duration ratios $D$ are found to be smaller than 1 for this pulsar sample, meaning that these objects have longer egress phases than ingress phases. This is not the case for PSRs~J1705$-$1903 and J2055+3829, as can be seen from Fig~\ref{fig:fit_appendix1}: the eclipses of these pulsars are asymmetric but with longer ingress phases.

From the upper panel of Fig.~\ref{fig:ratioVSparams} we can notice a very rough trend of negative correlation between the symmetry parameter $D$ and the fraction of the eclipse over which the beam is not detectable, $T_\mathrm{obsc} / T_\mathrm{delay}$, with abrupt eclipsers displaying larger $T_\mathrm{obsc} / T_\mathrm{delay}$ and lower $D$ values, progressive eclipsers having intermediate $T_\mathrm{obsc} / T_\mathrm{delay}$ and $D$ values, and shallow eclipsers having high $D$ values and null $T_\mathrm{obsc} / T_\mathrm{delay}$ ratios. The difficulty to assign PSRs~J1731$-$1847 and J1959+2048 one of the classifications given in Table~\ref{tab:params_fit} is clear from this figure, since PSR~J1731$-$1847 has a high $T_\mathrm{obsc} / T_\mathrm{delay}$ value and an intermediate $D$ value, while PSR~J1959+2048 has an intermediate $T_\mathrm{obsc} / T_\mathrm{delay}$ value and a low $D$ ratio. This porosity between the two categories is not surprising, because of the strong frequency dependence of eclipsing behaviour: for instance, some pulsars classified as ``progressive'' eclipsers at 1.4~GHz could appear as ``abrupt'' eclipsers at lower frequencies, where eclipses become more abrupt and obscuring phases become longer. It could be that the three categories we have distinguished in fact belong to a spectrum of eclipse behaviours. 

Finally, slope ratios $S$ are found to be smaller than 1 in almost all cases, corresponding to ingress phases that are generally steeper than the egress phases. In the case of the ``abrupt'' eclipses, values of the $S$ parameter have large uncertainties. For these objects, the S/N cut left no TOAs in the ingress region, resulting in ingress slopes being difficult to measure accurately. The fit found slope values of $0.01 \pm 0.01$, leading to uncertain $S$ parameter values. Otherwise, we note from the lower panel of Fig.~\ref{fig:ratioVSparams} that the ingress/egress slope ratio $S$ shows, similarly to the $D$ parameter, a trend of negative correlation with the fraction of the eclipse over which the beam is not detected, $T_\mathrm{obsc} / T_\mathrm{delay}$; however the sample is too limited to be conclusive.

\subsection{Relationship between eclipsing behaviour and mass function}
\label{subsect:mass_f}

Comparing properties of BW systems, \citet{Freire_2005} found evidence for eclipsing binaries having higher mass functions than non-eclipsing ones. Since the mass function scales as $\sin(i)^3$, this could be interpreted as eclipsing systems being seen with higher orbital inclinations.
After the numerous discoveries of new BW systems over the last 15 years, \citet{Guillemot_2019} re-investigated the mass functions of these systems (see Fig.~6 in the latter article and discussions therein), finding that the mass function distributions of eclipsing and non-eclipsing objects only have $\sim 0.007$\% probability of originating from the same parent distributions. We here investigate whether the same trend holds for the pulsars in our sample.

\subsubsection{Mass function distribution for eclipsers and non-eclipsers}
\label{subsec:mf_eclipsers_noneclipsers}

\citet{Freire_2005} and \citet{Guillemot_2019} focused on BW pulsars in their studies. Similarly, our pulsar sample comprises BW pulsars only, with two notable exceptions: first, PSR~J1628$-$3205, which is a RB pulsar and therefore has the heaviest mass function among the sample (see Table~\ref{tab:binary_sample}). Given that the mass function scales as $m_c^3$, and since the masses of RB companions are roughly an order of magnitude larger than those of BW companion objects (by definition of the two categories, see Fig.~\ref{fig:atnf}), the mass functions of RBs are roughly three orders of magnitude heavier than those of BWs. On the other hand, the median companion mass of PSR~J1719$-$1438 is about an order of magnitude lighter than those of other BWs (see Table ~\ref{tab:binary_sample} and Sec~\ref{sect:1719} for more details), so that this object has the lowest mass function among the pulsars in our sample, by two orders of magnitude.

   \begin{figure}[htbp!]
   \centering
   \includegraphics[width=\hsize]{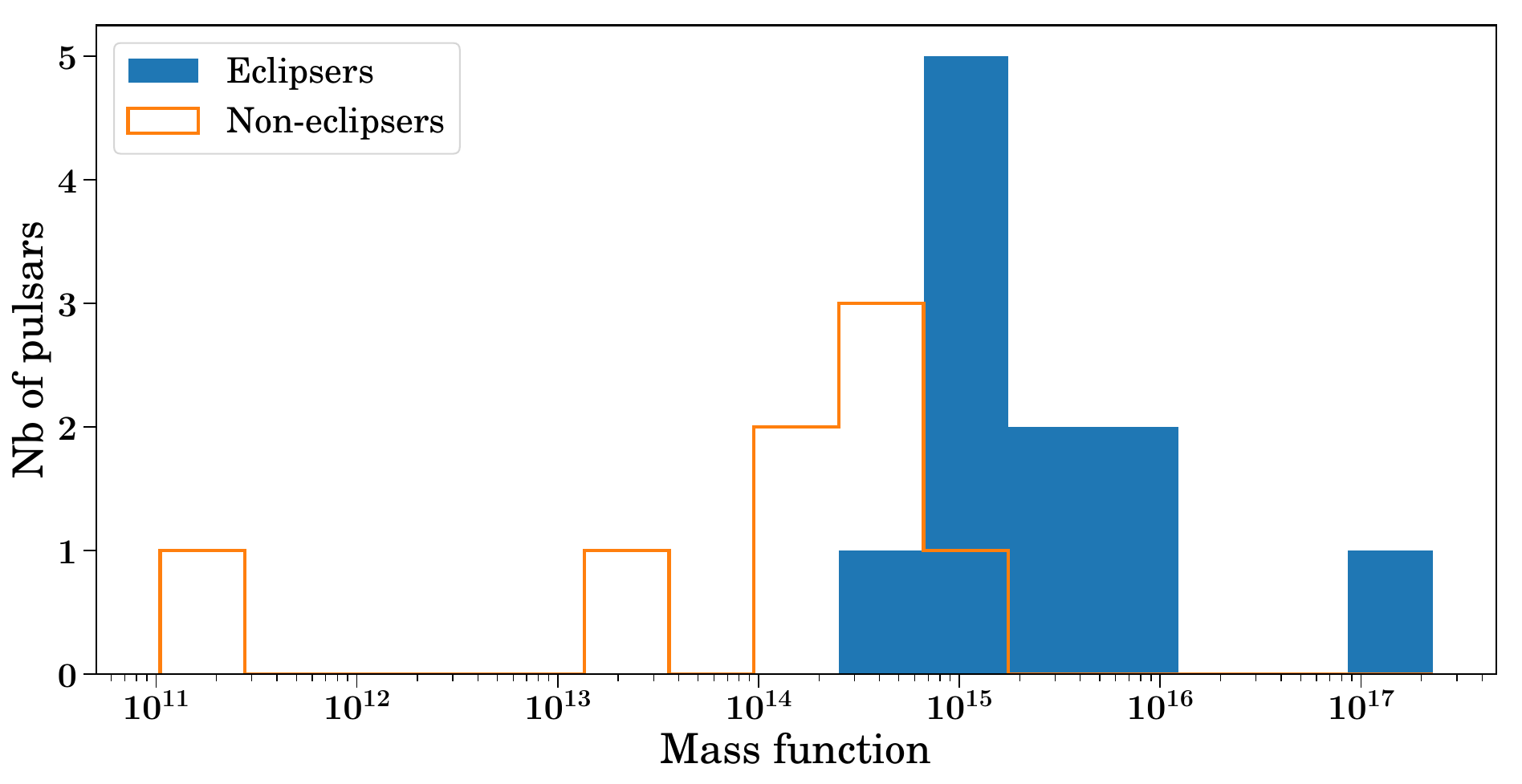}
      \caption{Mass function distributions for the pulsars in our sample.}
         \label{fig:hist_mass_f}
   \end{figure}

   \begin{figure*}[h!]
   \centering
   \includegraphics[width=\hsize]{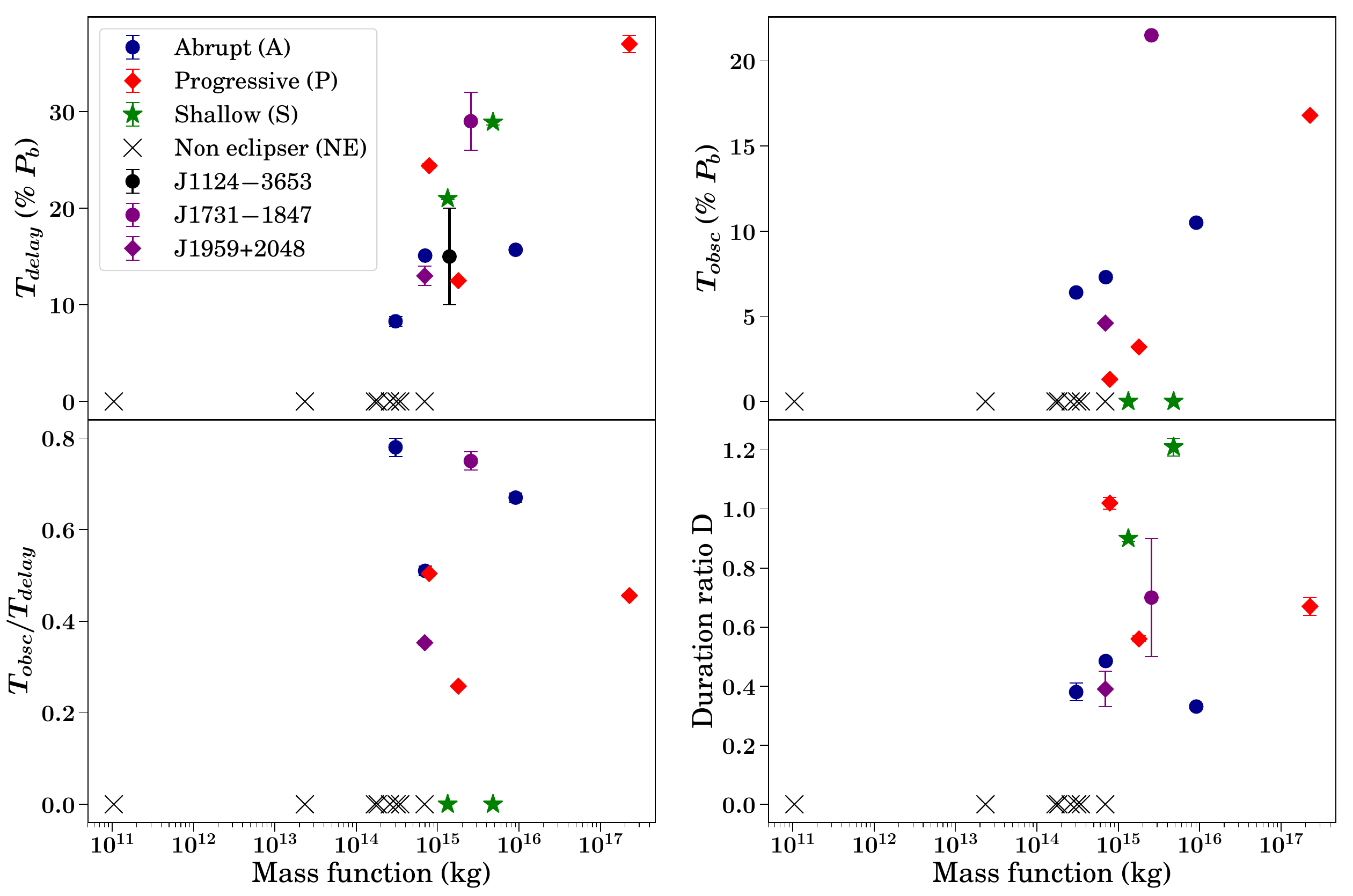}
      \caption{Eclipse parameters as determined from the phenomenological fits made in Sect.~\ref{sect:study} as a function of mass function values. As in Figs.~\ref{fig:TeVSTt} and~\ref{fig:ratioVSparams} we respectively represented ``abrupt'', ``progressive'' and ``shallow'' eclipsers with circles, diamonds and stars. Crosses represent non-eclipsing systems.}
         \label{fig:Td_vs_mass_f}
   \end{figure*}

With these caveats in mind, in Fig.~\ref{fig:hist_mass_f} we compare the mass functions of non-eclipsing pulsars with those of the eclipsing ones, for the pulsars in our sample. PSRs~J1719$-$1438 and J1628$-$3205 are located at the two extremes of the distribution plotted in Fig.~\ref{fig:hist_mass_f}. The one-dimensional Kolmogorov-Smirnov (KS) test \citep{Press1992} indicates a probability of $\sim$0.08\% for eclipsing and non-eclipsing pulsars, excluding PSRs~J1719$-$1438 and J1628$-$3205, to originate from the same parent distributions. Our results are thus in agreement with the conclusions of \citet{Freire_2005} and \citet{Guillemot_2019}.

\subsubsection{Dependence of eclipse parameters on mass function}
\label{subsec:ecl_massf}

In Sect.~\ref{subsec:mf_eclipsers_noneclipsers} we showed that the non-eclipsing pulsars in our sample tend to have lower mass functions than the eclipsing ones, suggesting that eclipsing pulsars are indeed seen with higher inclinations. Beyond the detection or non-detection of radio eclipses at superior conjunction, the coherent analysis of the eclipse properties of the pulsars in our sample, as presented in Sect.~\ref{sect:study}, provides an opportunity to address the question of the dependence of eclipse parameters on viewing geometry. More specifically, lower inclinations should result in less eclipsing material crossing the line-of-sight, leading to shorter eclipses, if any.

In Fig.~\ref{fig:Td_vs_mass_f} we plot a number of eclipse parameters determined from the analysis described in Sect.~\ref{sect:study}, against mass function. We observe no particular trends in the $T_\mathrm{obsc}$, $T_\mathrm{obsc} / T_\mathrm{delay}$, and duration ratio $D$ parameters as a function of mass function. This is confirmed by the Spearman correlation coefficients  \citep{Spearman} of 0.5 found for these three panels, indicating a low to moderate \citep{Hinkle_79} correlation. 
We also observe no particular differences in this figure between the three classes of eclipsing systems, other than those already mentioned in Sect.~\ref{subsect:eclipse_fit_results}.

However, the top-left panel of Fig.~\ref{fig:Td_vs_mass_f} hints at a linear correlation between $T_\mathrm{delay}$, \textit{i.e.}, the total duration of the eclipse, as a function of mass function. In this panel we included PSR~J1124$-$3653, which is seen to eclipse but for which, as mentioned in Sect.~\ref{sect:eclipses}, our phenomenological fits did not yield satisfactory results. From a visual inspection of Fig.~\ref{fig:SNR_drop_appendix1} we conservatively estimated $T_\mathrm{delay}$ to be about $15 \pm 5 \%$ of the orbital period. For the sub-sample of 10 eclipsing BWs (excluding the RB pulsar J1628$-$3205), we find a Spearman correlation coefficient \citep{Spearman} of 0.6. This coefficient indicates a moderate correlation, with a p-value of $\sim$10\% , the p-value in this test being the probability of an uncorrelated system producing datasets that have a Spearman correlation at least as high. With the entire sample of 17 objects (10 eclipsers and 7 non-eclipsers, excluding PSRs J1719$-$1438 and J1628$-$3205), and setting $T_\mathrm{delay}$ to 0 for non-eclipsing systems, the coefficient becomes 0.8, with a p-value of $\sim$0.04\%. This high correlation coefficient seems to suggest that eclipsing pulsars with lower mass function values have shorter eclipses, and vice versa, as could be expected under the hypotheses mentioned above. A larger pulsar sample is required to confirm this possible trend.

\begin{table*}
\caption[]{Published orbital inclinations and Roche-lobe filling factors, for the pulsars in our sample. }
\renewcommand{\arraystretch}{1.3}
\label{tab:optic_sample}
\centering

\begin{tabular}{ccccc}
\hline
\hline
Pulsar & Inclination, $i$ ($^\circ$) & Roche-lobe filling factor, $f$  & Reference & Note\\
\hline
J0023+0923 & $54\pm14$ & $0.3\pm0.3$ & \citet{Breton_2013} & $f=f_{\mathrm{eff}}$\\
 & & \underline{$0.2 \pm 0.2$} & & conversion to $f_{\mathrm{L1}}$\\
 & $77\pm13$ & \underline{$0.72\pm0.04$} & \citet{Draghis_2019} \\
 & $42\pm4$ & \underline{$0.36\pm0.18$} & \citet{MataSanchez_2023}\\
J0610$-$2100 & $ 73.2^{+15.6}_{-19.2} $ & $ \underline{0.86^{+0.08}_{-0.14}} $ & \citet{Wateren_2022} \\
J0636+5128 & $24.3\pm3.5$ & \underline{$0.75\pm0.20$ }& \citet{Kaplan_2018} & Values for $M_\mathrm{p} = 1.4 M_{\odot}$ \\
 & $23.3\pm0.3$ & \underline{$0.98\pm0.02$} & \citet{Draghis_2019} \\
 & $24.0\pm1.0$ & \underline{$f > 0.95 $} & \citet{MataSanchez_2023} \\
J1124$-$3653 & $44.9\pm0.4$ & \underline{$0.84\pm0.03$} & \citet{Draghis_2019} \\
*J1513$-$2550 & / & / & /  \\
J1544+4937 & $47^{+7}_{-4}$ & \underline{$f > 0.96$} & \citet{MataSanchez_2023} \\
*J1555$-$2908 & $ i > 75$ & \underline{$0.98\pm0.02$} & \citet{Kennedy_2022} \\
 & $i > 83$ & / & \citet{Clark_2023} & $\gamma$-ray analysis \\
*J1628$-$3205 & $i > 55$ & \underline{1} & \citet{Li_2014} & f is fixed and not fitted for \\
 & $i < 82.2$ & / & \citet{Clark_2023} & $\gamma$-ray analysis \\
*J1705$-$1903 & / & / & / \\
J1719$-$1438 & / & / & / \\
*J1731$-$1847 & / & / & / \\
J1745+1017 & / & / & / \\
*J1959+2048 & $65\pm2$ & \underline{$0.81 \leq f \leq 0.87$} & \citet{Reynolds_2007}\\
 & $62.5\pm1.3$ & \underline{$0.90\pm0.01$} & \citet{Draghis_2019} \\
 & $ i > 84$ & / & \citet{Clark_2023} & $\gamma$-ray analysis \\
 & $85.1\pm0.4$ & / & \citet{Du_2023} & Radio analysis \\
*J2051$-$0827 & $55.9^{+4.8}_{-4.1}$ & \underline{$0.88\pm0.02$} & \citet{Dhillon_2022} \\
 & $59.5\pm0.4$ & / & \citet{Du_2023} & Radio analysis \\
*J2055+3829 & $46.8\pm1.6$ & / & \citet{Du_2023} & Radio analysis \\
*J2115+5448 & / & / & / \\
J2214+3000 & / & / & / \\
J2234+0944 & / & / & / \\
*J2256$-$1024 & $68\pm11$ & $0.4\pm0.2$ & \citet{Breton_2013} & $f=f_{\mathrm{eff}}$ \\
 & & \underline{$0.3\pm0.1$}  & & conversion to $f_{\mathrm{L1}}$ \\
\hline
\end{tabular}
\tablefoot{Two different definitions exist for the filling factor $f$ in these articles. Most articles in this Table define $f$ as the ratio of the companion radius at the nose to the L1 radius. These filling factors, whose values are underlined in the table, are denoted by $f_{\rm L1}$ and are those used in the rest of the paper. In certain cases $f$ is defined as the ratio of the volume-averaged stellar radius to the volume-averaged Roche lobe radius. These values of $f$ are denoted by $f_\mathrm{eff}$. To facilitate comparison we converted $f_\mathrm{eff}$ estimates to $f_{\rm L1}$ estimates when necessary. See Appendix \ref{sec:filfac} for details on the conversion. Pulsars marked with a star are those that display eclipses. While most measurements are based on optical photometry of the companion objects, some of the orbital inclination measurements were obtained from radio or $\gamma$-ray observations. No constraints on the Roche-lobe filling factor could be obtained from these methods.}
\end{table*}

\subsection{Dependence of eclipse parameters on orbital inclination}
\label{sect:optical}

   \begin{figure*}[h]
   \centering
   \includegraphics[width=\hsize]{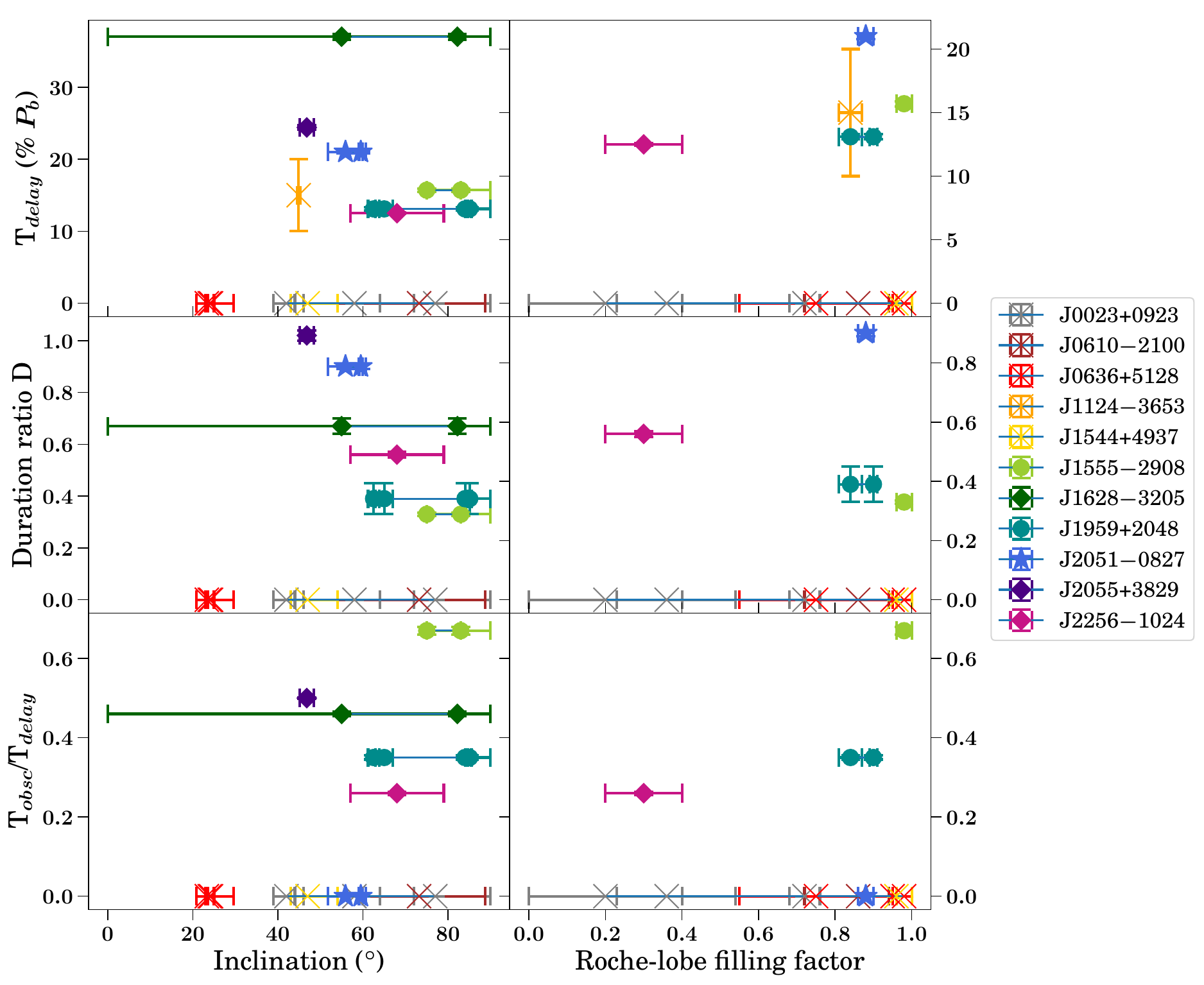}
      \caption{Comparison between the results of the phenomenological fits made in Sect.~\ref{sect:study}, summarized in Table~\ref{tab:params_fit}, and the published inclinations and Roche-lobe filling factors listed in Table~\ref{tab:optic_sample}. The marker shapes convention is the same as Fig~\ref{fig:TeVSTt} and Fig~\ref{fig:ratioVSparams}, circles correspond to the ``abrupt'' eclipsers, diamonds are the ``progressive'' eclipsers, stars are the ``shallow'' eclipsers with the addition of crosses for non-eclipsing systems.}
         \label{fig:inclination}
   \end{figure*}

In Sect.~\ref{subsec:mf_eclipsers_noneclipsers} and \ref{subsec:ecl_massf} we searched for correlations between eclipse properties and mass function values, arguing that since the mass function scales as $\sin^3 i$, it can be used as a proxy for the orbital inclination. However, for some of the pulsars in our sample, actual orbital inclination measurements have been published. 

We conducted a comprehensive review of published orbital inclination measurements for these objects. The results from our literature review are listed in Table~\ref{tab:optic_sample}. The majority of the orbital inclination measurements listed in this table have been obtained from optical measurements, fitting a heating model to the light curve of the companion object. Several parameters can be inferred from these analyses, including the inclination of the system, and the Roche-lobe filling factor (denoted $f$ in the Table) of the companion object. The other orbital inclination constraints listed in Table~\ref{tab:optic_sample} were obtained by fitting the eclipses seen in radio \citep{Du_2023} or $\gamma$ rays \citep{Clark_2023} or radio \citep{Du_2023}. As can be seen from the table, for some pulsars multiple $i$ and/or $f$ measurements were available in the literature; measurements that are in some cases inconsistent. For four of the eclipsing pulsars in our sample (PSRs~J1513$-$2550, J1705$-$1903, J1731$-$1847 and J2115+5448) no inclination or Roche-lobe filling factor constraints have been published, and for two objects (PSRs~J2055+3829 and J2256$-$1024) we found inclination measurements, but no measurements of $f$.

In Fig.~\ref{fig:inclination} we plot the eclipse parameters presented in Sect.~\ref{subsect:eclipse_fit_results} as a function of orbital inclinations $i$ and Roche-lobe filling factors $f$ found in the literature. Both eclipsers and non-eclipsers are shown in Fig.~\ref{fig:inclination}, with fit parameters set to 0 in the case of non-eclipsers. In cases where multiple $i$ and/or $f$ constraints were available, we plotted each individual measurement. We note that the inclinations and Roche-lobe filling factors of non-eclipsing pulsars almost span the entire allowed ranges, albeit with large uncertainties. 

Contrary to Fig.~\ref{fig:Td_vs_mass_f}, no obvious link can be found between the eclipse parameters we determined and the orbital inclinations and/or Roche-lobe filling factors. This is aggravated by the fact that few of the eclipsers in our sample have $i$ and $f$ measurements, with only six of the eclipsers having measured orbital inclinations and four having Roche-lobe filling factor estimates. The sample of four objects will both $i$ and $f$ estimates is too small to draw any conclusions at present, but can serve as a starting point for future studies, when more radio and/or optical studies of eclipsing spiders have been conducted.

\subsection{Correlations between mass functions and profile widths}
\label{subsect:width}

In the simple cone emission model \citep{Handbook}, the pulse width $W$ can be written as:
\begin{equation}
\sin^2 \left( \frac{W}{4} \right) = \frac{\sin^2(\rho/2) - \sin^2(\beta/2)}{\sin (\alpha) \sin (\zeta)} 
\label{eq:sin_w}
\end{equation} 

\noindent
where $\rho$ denotes the beam width, $\alpha$ is the angle between the magnetic axis and the spin axis, $\beta$ is the impact angle of the closest approach of the line-of-sight to the magnetic axis and $\zeta = \alpha + \beta$ is the angle between the spin axis and the line-of-sight. In this description, observable pulsars have $|\beta| < \rho$, indicating that $\beta$ needs to be small, and thus, the $\alpha$ and $\zeta$ angles need to be similar. Additionally, the long timescale of mass transfer in low mass X-ray binaries is expected to align the spin axis of recycled pulsars with the orbital angular momentum vector \citep{Hills_1983, Bhattacharya_1991}, leading to $\zeta \sim i$ for recycled pulsars. Furthermore, as mentioned above, the mass function scales as $\sin(i)^3$, and can thus serve as a proxy for the orbital inclination. As can be seen from Eq.\ref{eq:sin_w}, pulsars with small $\alpha$ values are expected to have broad pulses, under the simple cone model. Therefore, under the various hypotheses mentioned above, we might expect pulse profiles to be narrower for higher mass functions. An important caveat is, of course, the fact that although normal radio pulsars generally have profiles that are adequately represented with the simple cone model, pulse profiles of MSPs are often complex, as can be seen from Figs.~\ref{fig:profile_appendix1} to \ref{fig:profile_appendix3}.

   \begin{figure}[h]
   \centering
   \includegraphics[width=\hsize]{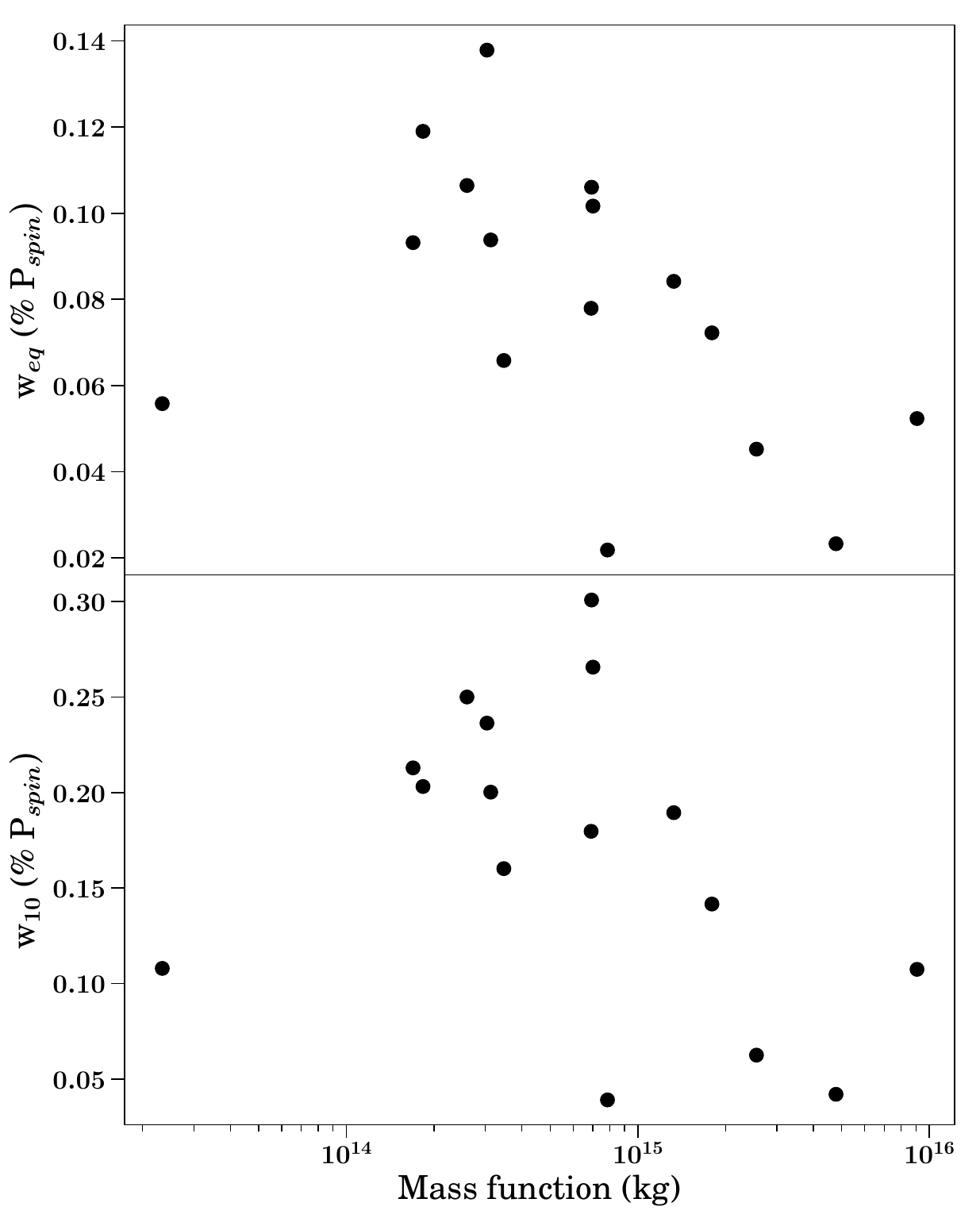}
    \caption{Equivalent width (top panel) and width at 10\% of the maximum (bottom panel) as function of the mass function. PSRs J1719$-$1438 and J1628$-$3205 are not shown, for the reasons stated in Sect.~\ref{subsect:mass_f}, and PSR~J1124$-$3653 is also not plotted as the noise baseline in its pulse profile is higher than 10\% of the maximum intensity (see Fig.~\ref{fig:profile_appendix1}).}
    \label{fig:pulse_width}
   \end{figure}

In Fig.~\ref{fig:pulse_width} we show the equivalent pulse profile widths, $w_\mathrm{eq}$, where $w_\mathrm{eq}$ corresponds to the width of a rectangle with a height equal to the profile maximum, and whose area is equal to that of the profile, and the profile width at 10\% of the maximum, $w_{10}$, as a function of mass functions. Values of $w_\mathrm{eq}$ and $w_{10}$ were determined using the profiles shown in Figs.~\ref{fig:profile_appendix1} to \ref{fig:profile_appendix3}. We find Spearman correlation coefficients for the data shown in the top and bottom panels of $-0.6$ and $-0.5$, corresponding to low p-values of chance occurrence of $\sim 2$\% and $\sim 5$\%, respectively. These marginally-significant correlations are to be confirmed with more pulsars, and need to be taken with a grain of salt, considering the number of hypotheses made and the caveats mentioned above. We finally searched for correlations between polarization position angle slopes and mass functions, using the polarimetric profiles shown in Figs.~\ref{fig:profile_appendix1} to \ref{fig:profile_appendix3}, and found none.


\section{Summary and conclusions}
\label{sect:conclusion}

In this article we presented an analysis of 19 spider-type pulsars observed using 1.4~GHz data taken with the NUPPI backend of the Nan\c{c}ay Radio Telescope. For the majority of the 19 objects, the datasets we analysed spanned more than 12 years, \textit{i.e.}, these datasets included NUPPI data taken since this instrument started operating in 2011. 

In order to search the data for the presence of eclipses at superior conjunction caused by outflowing material from the companion, we extracted TOAs from the NUPPI data using the 2D template matching technique implemented in \texttt{PulsePortraiture}, enabling us to obtain reliable timing data for all the considered pulsars, \textit{i.e.}, including for the pulsars with the lowest S/N detections. Additionally, the TOA datasets obtained using \texttt{PulsePortraiture} generally presented lower outlier rates than datasets constructed using the standard template-matching method and fully or partially frequency-scrunched observations. A timing analysis was then conducted for each of the considered pulsars, enabling us to properly determine the orbital phases associated with each of the TOAs, over the entire time spans of the datasets.

The analysis of the S/N values associated with the TOAs across the orbits revealed that eight of the 19 spider pulsars in the sample display ``total'' eclipses: namely, PSRs~J1513$-$2550, J1555$-$2908, J1628$-$3205, J1731$-$1847, J1959+2048, J2055+3829, J2115+5448 and J2256$-$1024. We also found that two pulsars display ``shallow'' eclipses, during which the pulsar does not get entirely obscured: PSRs~J1705$-$1903 and J2051$-$0827. Finally, we found that PSR~J1124$-$3653 also appears to display eclipses but is too faint for the eclipse properties to be constrained adequately with the NRT data. In total, we therefore found that 11 of the 19 spider pulsars in our sample display eclipses around superior conjunction at 1.4 GHz, while the remaining 8 pulsars do not appear to display eclipses at these frequencies. 

We conducted a phenomenological fit of the TOA residuals around superior conjunction for all eclipsing pulsars in our sample, except PSR~J1124$-$3653 as mentioned above. Eclipse properties appear to vary widely, with eclipse duty cycles ranging from 8 to 37\% of the orbit for the objects in our sample, and intervals of complete obscuration ranging from 0 to 22\%, with no apparent correlation between these two parameters. We further distinguished two categories of eclipsers, in addition to the ``shallow'' eclipsers mentioned above. On the one hand, eclipsing pulsars displaying no ingress phases: namely, PSRs~J1513$-$2550, J1555$-$2908 and J2115+5448, hence categorized as ``abrupt'' eclipsers. On the other hand, pulsars with more progressive ingress phases, namely, PSRs~J1628$-$3205, J2055+3829 and J2256$-$1024, categorized as ``progressive''. PSRs~J1731$-$1847 and J1959+2048 are more difficult to firmly categorize with the present data, as they display characteristics seen in the latter two categories. We found that most ingress/egress slope ratios $S$ and duration ratios $D$ are smaller than 1, meaning that eclipses generally have longer egress than ingress phases, and that ingress phases are steeper than the egress phases. We found indications for a negative correlation between $D$ (or $S$) and the fraction of the eclipse during which the beam is not detected; however, the pulsar sample is too small for the correlation to be firmly established. 

Eclipse properties are likely to depend on viewing geometry. We searched for correlations between some of the eclipse properties determined from the phenomenological fits and the mass functions of the systems, assuming that the mass function can be used as a proxy for the inclination. We found that the mass function distributions of eclipsing and non-eclipsing systems have a low probability of originating from the same parent distributions, in accordance with previous studies. We also found marginally-significant evidence for a positive correlation between eclipse durations and mass functions, suggesting that larger sections of the eclipsing material cross the line-of-sight in systems with high orbital inclinations.

To further the exploration of this correlation we also conducted a comprehensive review of published orbital inclinations and Roche lobe filling factors for the pulsars in our sample, and searched for trends between these parameters and the eclipse parameters. Contrary to the previous comparison, no correlations were found. Nevertheless, the limited pulsar sample, the paucity of available orbital inclination and Roche-lobe filling factor measurements for these pulsars, and the fact that the results for a given pulsar often contradict each other, hinder such correlation searches. 

In addition to determining the eclipse properties of the eclipsing systems, we constructed the polarimetric profiles of the 19 pulsars in our sample, and computed the polarized, linear, circular and absolute circular intensities. We put our sample in the broader context of a pulsar population study \citep{Wang_2023}. We also searched for correlations between pulse profile widths and mass functions, finding marginally-significant correlations.

We argue that a systematic orbital inclination and Roche-lobe filling factor measurement campaign, with a common analysis methodology, would be highly beneficial. Analyses of large samples of pulsars similar to the systematic study presented in our article would also be beneficial. In the future we will look into extending our study to data taken with other telescopes, using a similar methodology, with the goal of increasing the sample of pulsars with eclipse fits to confront with properties determined at other wavelengths. Additionally, in the future we will also use the timing datasets we produced by analyzing NUPPI data on the 19 spider pulsars in our sample, to conduct a systematic study of the timing properties of these pulsars. This systematic study will enable us to investigate a potential dependence of orbital instability on other parameters, such as the Roche-lobe filling factor.


\begin{acknowledgements}

The Nan\c{c}ay Radio Observatory is operated by the Paris Observatory, associated with the French Centre National de la Recherche Scientifique (CNRS). We acknowledge financial support from the ``Programme National de Cosmologie and Galaxies'' (PNCG), ``Programme National Hautes Energies'' (PNHE), and ``Programme National Gravitation, R\'ef\'erences, Astronomie, M\'etrologie'' (PNGRAM) of CNRS/INSU, France. 

\end{acknowledgements}


\bibliographystyle{aa}
\bibliography{refs.bib}

\begin{thebibliography}{87}
\expandafter\ifx\csname natexlab\endcsname\relax\def\natexlab#1{#1}\fi

\bibitem[{{Alpar} {et~al.}(1982){Alpar}, {Cheng}, {Ruderman}, \& {Shaham}}]{Alpar_1982}
{Alpar}, M.~A., {Cheng}, A.~F., {Ruderman}, M.~A., \& {Shaham}, J. 1982, \nat, 300, 728

\bibitem[{{Bailes} {et~al.}(2011){Bailes}, {Bates}, {Bhalerao}, {Bhat}, {Burgay}, {Burke-Spolaor}, {D'Amico}, {Johnston}, {Keith}, {Kramer}, {Kulkarni}, {Levin}, {Lyne}, {Milia}, {Possenti}, {Spitler}, {Stappers}, \& {van Straten}}]{Bailes_2011}
{Bailes}, M., {Bates}, S.~D., {Bhalerao}, V., {et~al.} 2011, Science, 333, 1717

\bibitem[{{Ballet} {et~al.}(2023){Ballet}, {Bruel}, {Burnett}, {Lott}, \& {The Fermi-LAT collaboration}}]{4FGL}
{Ballet}, J., {Bruel}, P., {Burnett}, T.~H., {Lott}, B., \& {The Fermi-LAT collaboration}. 2023, arXiv e-prints, arXiv:2307.12546

\bibitem[{{Bangale} {et~al.}(2024){Bangale}, {Bhattacharyya}, {Camilo}, {Clark}, {Cognard}, {DeCesar}, {Ferrara}, {Gentile}, {Guillemot}, {Hessels}, {et~al.}}]{Bangale_2024}
{Bangale}, P., {Bhattacharyya}, B., {Camilo}, F., {et~al.} 2024, \apj, 966, 161

\bibitem[{{Barr} {et~al.}(2013){Barr}, {Guillemot}, {Champion}, {Kramer}, {Eatough}, {Lee}, {Verbiest}, {Bassa}, {Camilo}, {{\c{C}}elik}, {Cognard}, {Ferrara}, {Freire}, {Janssen}, {Johnston}, {Keith}, {Lyne}, {Michelson}, {Parkinson}, {Ransom}, {Ray}, {Stappers}, \& {Wood}}]{Barr_2013}
{Barr}, E.~D., {Guillemot}, L., {Champion}, D.~J., {et~al.} 2013, \mnras, 429, 1633

\bibitem[{Bassa {et~al.}(2014)Bassa, Patruno, Hessels, Keane, Monard, Mahony, Bogdanov, Corbel, Edwards, Archibald, Janssen, Stappers, \& Tendulkar}]{Bassa_2014}
Bassa, C.~G., Patruno, A., Hessels, J. W.~T., {et~al.} 2014, Monthly Notices of the Royal Astronomical Society, 441, 1825

\bibitem[{{Bates} {et~al.}(2011){Bates}, {Bailes}, {Bhat}, {Burgay}, {Burke-Spolaor}, {D'Amico}, {Jameson}, {Johnston}, {Keith}, {Kramer}, {Levin}, {Lyne}, {Milia}, {Possenti}, {Stappers}, \& {van Straten}}]{Bates_2011}
{Bates}, S.~D., {Bailes}, M., {Bhat}, N.~D.~R., {et~al.} 2011, \mnras, 416, 2455

\bibitem[{{Bhattacharya} \& {van den Heuvel}(1991)}]{Bhattacharya_1991}
{Bhattacharya}, D. \& {van den Heuvel}, E.~P.~J. 1991, \physrep, 203, 1

\bibitem[{{Bhattacharyya} {et~al.}(2013){Bhattacharyya}, {Roy}, {Ray}, {Gupta}, {Bhattacharya}, {Romani}, {Ransom}, {Ferrara}, {Wolff}, {Camilo}, {Cognard}, {Harding}, {den Hartog}, {Johnston}, {Keith}, {Kerr}, {Michelson}, {Saz Parkinson}, {Wood}, \& {Wood}}]{Bhattacharyya_2013}
{Bhattacharyya}, B., {Roy}, J., {Ray}, P.~S., {et~al.} 2013, \apjl, 773, L12

\bibitem[{{Bisnovatyi-Kogan} \& {Komberg}(1974)}]{Bisnovatyi_1974}
{Bisnovatyi-Kogan}, G.~S. \& {Komberg}, B.~V. 1974, \sovast, 18, 217

\bibitem[{Breton {et~al.}(2012)Breton, Rappaport, van Kerkwijk, \& Carter}]{breton_koi_2012}
Breton, R.~P., Rappaport, S.~A., van Kerkwijk, M.~H., \& Carter, J.~A. 2012, 748, 115

\bibitem[{{Breton} {et~al.}(2013){Breton}, {van Kerkwijk}, {Roberts}, {Hessels}, {Camilo}, {McLaughlin}, {Ransom}, {Ray}, \& {Stairs}}]{Breton_2013}
{Breton}, R.~P., {van Kerkwijk}, M.~H., {Roberts}, M.~S.~E., {et~al.} 2013, \apj, 769, 108

\bibitem[{{Burgay} {et~al.}(2006){Burgay}, {Joshi}, {D'Amico}, {Possenti}, {Lyne}, {Manchester}, {McLaughlin}, {Kramer}, {Camilo}, \& {Freire}}]{Burgay_06}
{Burgay}, M., {Joshi}, B.~C., {D'Amico}, N., {et~al.} 2006, \mnras, 368, 283

\bibitem[{{Clark} {et~al.}(2023){Clark}, {Kerr}, {Barr}, {Bhattacharyya}, {Breton}, {Bruel}, {Camilo}, {Chen}, {Cognard}, {Cromartie}, {Deneva}, {Dhillon}, {Guillemot}, {Kennedy}, {Kramer}, {Lyne}, {Mata S{\'a}nchez}, {Nieder}, {Phillips}, {Ransom}, {Ray}, {Roberts}, {Roy}, {Smith}, {Spiewak}, {Stappers}, {Tabassum}, {Theureau}, \& {Voisin}}]{Clark_2023}
{Clark}, C.~J., {Kerr}, M., {Barr}, E.~D., {et~al.} 2023, Nature Astronomy, 7, 451

\bibitem[{{Crowter} {et~al.}(2020){Crowter}, {Stairs}, {McPhee}, {Archibald}, {Boyles}, {Hessels}, {Karako-Argaman}, {Lorimer}, {Lynch}, {McLaughlin}, {Ransom}, {Roberts}, {Stovall}, \& {van Leeuwen}}]{Crowter_2020}
{Crowter}, K., {Stairs}, I.~H., {McPhee}, C.~A., {et~al.} 2020, \mnras, 495, 3052

\bibitem[{{Desvignes} {et~al.}(2013){Desvignes}, {Cognard}, {Champion}, {Lazarus}, {Lespagnol}, {Smith}, \& {Theureau}}]{SPAN512}
{Desvignes}, G., {Cognard}, I., {Champion}, D., {et~al.} 2013, in Neutron Stars and Pulsars: Challenges and Opportunities after 80 years, ed. J.~{van Leeuwen}, Vol. 291, 375--377

\bibitem[{{Desvignes} {et~al.}(2022){Desvignes}, {Cognard}, {Smith}, {Champion}, {Guillemot}, {Kramer}, {Lespagnol}, {Octau}, \& {Theureau}}]{Desvignes_2022}
{Desvignes}, G., {Cognard}, I., {Smith}, D.~A., {et~al.} 2022, \aap, 667, A79

\bibitem[{{Dhillon} {et~al.}(2022){Dhillon}, {Kennedy}, {Breton}, {Clark}, {Mata S{\'a}nchez}, {Voisin}, {Breedt}, {Brown}, {Dyer}, {Green}, {Kerry}, {Littlefair}, {Marsh}, {Parsons}, {Pelisoli}, {Sahman}, {Wild}, {van Kerkwijk}, \& {Stappers}}]{Dhillon_2022}
{Dhillon}, V.~S., {Kennedy}, M.~R., {Breton}, R.~P., {et~al.} 2022, \mnras, 516, 2792

\bibitem[{{Draghis} \& {Romani}(2018)}]{Draghis_18}
{Draghis}, P. \& {Romani}, R.~W. 2018, \apjl, 862, L6

\bibitem[{{Draghis} {et~al.}(2019){Draghis}, {Romani}, {Filippenko}, {Brink}, {Zheng}, {Halpern}, \& {Camilo}}]{Draghis_2019}
{Draghis}, P., {Romani}, R.~W., {Filippenko}, A.~V., {et~al.} 2019, \apj, 883, 108

\bibitem[{{Du} {et~al.}(2023){Du}, {Yu}, {Chen}, {Wang}, {Zhou}, \& {Zheng}}]{Du_2023}
{Du}, Z.-X., {Yu}, Y.-W., {Chen}, A.~M., {et~al.} 2023, Research in Astronomy and Astrophysics, 23, 125024

\bibitem[{Eggleton(1983)}]{eggleton_aproximations_1983}
Eggleton, P.~P. 1983, 268, 368

\bibitem[{Espinoza {et~al.}(2013)Espinoza, Guillemot, Çelik, Weltevrede, Stappers, Smith, Kerr, Zavlin, Cognard, Eatough, Freire, Janssen, Camilo, Desvignes, Hewitt, Hou, Johnston, Keith, Kramer, Lyne, Manchester, Ransom, Ray, Shannon, Theureau, \& Webb}]{Espinoza_13}
Espinoza, C.~M., Guillemot, L., Çelik, O., {et~al.} 2013, Monthly Notices of the Royal Astronomical Society, 430, 571

\bibitem[{{Foreman-Mackey} {et~al.}(2013){Foreman-Mackey}, {Hogg}, {Lang}, \& {Goodman}}]{emcee}
{Foreman-Mackey}, D., {Hogg}, D.~W., {Lang}, D., \& {Goodman}, J. 2013, \pasp, 125, 306

\bibitem[{{Freire}(2005)}]{Freire_2005}
{Freire}, P.~C.~C. 2005, in Astronomical Society of the Pacific Conference Series, Vol. 328, Binary Radio Pulsars, ed. F.~A. {Rasio} \& I.~H. {Stairs}, 405

\bibitem[{{Fruchter} {et~al.}(1988){Fruchter}, {Stinebring}, \& {Taylor}}]{B1957}
{Fruchter}, A.~S., {Stinebring}, D.~R., \& {Taylor}, J.~H. 1988, \nat, 333, 237

\bibitem[{Gentile {et~al.}(2014)Gentile, Roberts, McLaughlin, Camilo, Hessels, Kerr, Ransom, Ray, \& Stairs}]{Gentile_2014}
Gentile, P.~A., Roberts, M. S.~E., McLaughlin, M.~A., {et~al.} 2014, The Astrophysical Journal, 783, 69

\bibitem[{{Guillemot} {et~al.}(2023){Guillemot}, {Cognard}, {van Straten}, {Theureau}, \& {G{\'e}rard}}]{Guillemot_2023}
{Guillemot}, L., {Cognard}, I., {van Straten}, W., {Theureau}, G., \& {G{\'e}rard}, E. 2023, \aap, 678, A79

\bibitem[{{Guillemot} {et~al.}(2012){Guillemot}, {Johnson}, {Venter}, {Kerr}, {Pancrazi}, {Livingstone}, {Janssen}, {Jaroenjittichai}, {Kramer}, {Cognard}, {Stappers}, {Harding}, {Camilo}, {Espinoza}, {Freire}, {Gargano}, {Grove}, {Johnston}, {Michelson}, {Noutsos}, {Parent}, {Ransom}, {Ray}, {Shannon}, {Smith}, {Theureau}, {Thorsett}, \& {Webb}}]{Guillemot_2012}
{Guillemot}, L., {Johnson}, T.~J., {Venter}, C., {et~al.} 2012, \apj, 744, 33

\bibitem[{{Guillemot} {et~al.}(2019){Guillemot}, {Octau}, {Cognard}, {Desvignes}, {Freire}, {Smith}, {Theureau}, \& {Burnett}}]{Guillemot_2019}
{Guillemot}, L., {Octau}, F., {Cognard}, I., {et~al.} 2019, \aap, 629, A92

\bibitem[{{Guo} {et~al.}(2022){Guo}, {Wang}, \& {Han}}]{Guo_2022}
{Guo}, Y., {Wang}, B., \& {Han}, Z. 2022, \mnras, 515, 2725

\bibitem[{{Han} {et~al.}(2018){Han}, {Manchester}, {van Straten}, \& {Demorest}}]{Han_2018}
{Han}, J.~L., {Manchester}, R.~N., {van Straten}, W., \& {Demorest}, P. 2018, \apjs, 234, 11

\bibitem[{{Hills}(1983)}]{Hills_1983}
{Hills}, J.~G. 1983, \apj, 267, 322

\bibitem[{Hinkle {et~al.}(1979)Hinkle, Wiersma, \& Jurs}]{Hinkle_79}
Hinkle, D.~E., Wiersma, W., \& Jurs, S.~G. 1979, Applied statistics for the behavioral sciences (Chicago : Rand McNally College Pub. Co), includes index

\bibitem[{{Hobbs} {et~al.}(2006){Hobbs}, {Edwards}, \& {Manchester}}]{Hobbs_2006}
{Hobbs}, G.~B., {Edwards}, R.~T., \& {Manchester}, R.~N. 2006, \mnras, 369, 655

\bibitem[{{Hotan} {et~al.}(2004){Hotan}, {van Straten}, \& {Manchester}}]{Hotan_2004}
{Hotan}, A.~W., {van Straten}, W., \& {Manchester}, R.~N. 2004, \pasa, 21, 302

\bibitem[{{Kansabanik} {et~al.}(2021){Kansabanik}, {Bhattacharyya}, {Roy}, \& {Stappers}}]{Kansabanik_2021}
{Kansabanik}, D., {Bhattacharyya}, B., {Roy}, J., \& {Stappers}, B. 2021, \apj, 920, 58

\bibitem[{{Kaplan} {et~al.}(2018){Kaplan}, {Stovall}, {van Kerkwijk}, {Fremling}, \& {Istrate}}]{Kaplan_2018}
{Kaplan}, D.~L., {Stovall}, K., {van Kerkwijk}, M.~H., {Fremling}, C., \& {Istrate}, A.~G. 2018, \apj, 864, 15

\bibitem[{{Keith} {et~al.}(2010){Keith}, {Jameson}, {van Straten}, {Bailes}, {Johnston}, {Kramer}, {Possenti}, {Bates}, {Bhat}, {Burgay}, {Burke-Spolaor}, {D'Amico}, {Levin}, {McMahon}, {Milia}, \& {Stappers}}]{Keith_2010}
{Keith}, M.~J., {Jameson}, A., {van Straten}, W., {et~al.} 2010, \mnras, 409, 619

\bibitem[{{Kennedy} {et~al.}(2022){Kennedy}, {Breton}, {Clark}, {Mata S{\'a}nchez}, {Voisin}, {Dhillon}, {Halpern}, {Marsh}, {Nieder}, {Ray}, \& {van Kerkwijk}}]{Kennedy_2022}
{Kennedy}, M.~R., {Breton}, R.~P., {Clark}, C.~J., {et~al.} 2022, \mnras, 512, 3001

\bibitem[{Kopal(1978)}]{kopal_dynamics_1978}
Kopal, Z. 1978, Dynamics of Close Binary Systems (Springer Netherlands), {OCLC}: 851368874

\bibitem[{{Kumari} {et~al.}(2024){Kumari}, {Bhattacharyya}, {Sharan}, {Kansabanik}, {Stappers}, \& {Roy}}]{Kumari_2024}
{Kumari}, S., {Bhattacharyya}, B., {Sharan}, R., {et~al.} 2024, \apj, 961, 155

\bibitem[{{Lazarus} {et~al.}(2016){Lazarus}, {Karuppusamy}, {Graikou}, {Caballero}, {Champion}, {Lee}, {Verbiest}, \& {Kramer}}]{Lazarus_2016}
{Lazarus}, P., {Karuppusamy}, R., {Graikou}, E., {et~al.} 2016, \mnras, 458, 868

\bibitem[{{Li} {et~al.}(2014){Li}, {Halpern}, \& {Thorstensen}}]{Li_2014}
{Li}, M., {Halpern}, J.~P., \& {Thorstensen}, J.~R. 2014, \apj, 795, 115

\bibitem[{{Lin} {et~al.}(2023){Lin}, {Main}, {Jow}, {Li}, {Pen}, \& {van Kerkwijk}}]{Lin_2023}
{Lin}, F.~X., {Main}, R.~A., {Jow}, D., {et~al.} 2023, \mnras, 519, 121

\bibitem[{{Lin} {et~al.}(2021){Lin}, {Main}, {Verbiest}, {Kramer}, \& {Shaifullah}}]{Lin_2021}
{Lin}, F.~X., {Main}, R.~A., {Verbiest}, J.~P.~W., {Kramer}, M., \& {Shaifullah}, G. 2021, \mnras, 506, 2824

\bibitem[{{Linares}(2020)}]{Linares_2020}
{Linares}, M. 2020, in Multifrequency Behaviour of High Energy Cosmic Sources - XIII. 3-8 June 2019. Palermo, 23

\bibitem[{{Lorimer} \& {Kramer}(2004)}]{Handbook}
{Lorimer}, D.~R. \& {Kramer}, M. 2004, {Handbook of Pulsar Astronomy}, Vol.~4

\bibitem[{{Main} {et~al.}(2018){Main}, {Yang}, {Chan}, {Li}, {Lin}, {Mahajan}, {Pen}, {Vanderlinde}, \& {van Kerkwijk}}]{Main_2018}
{Main}, R., {Yang}, I.~S., {Chan}, V., {et~al.} 2018, \nat, 557, 522

\bibitem[{{Manchester} {et~al.}(2005){Manchester}, {Hobbs}, {Teoh}, \& {Hobbs}}]{Manchester_2005}
{Manchester}, R.~N., {Hobbs}, G.~B., {Teoh}, A., \& {Hobbs}, M. 2005, \aj, 129, 1993

\bibitem[{{Mata S{\'a}nchez} {et~al.}(2023){Mata S{\'a}nchez}, {Kennedy}, {Clark}, {Breton}, {Dhillon}, {Voisin}, {Camilo}, {Littlefair}, {Marsh}, \& {Stringer}}]{MataSanchez_2023}
{Mata S{\'a}nchez}, D., {Kennedy}, M.~R., {Clark}, C.~J., {et~al.} 2023, \mnras, 520, 2217

\bibitem[{Morello {et~al.}(2018)Morello, Barr, Cooper, Bailes, Bates, Bhat, Burgay, Burke-Spolaor, Cameron, Champion, Eatough, Flynn, Jameson, Johnston, Keith, Keane, Kramer, Levin, Ng, Petroff, Possenti, Stappers, van Straten, \& Tiburzi}]{Morello_2018}
Morello, V., Barr, E.~D., Cooper, S., {et~al.} 2018, Monthly Notices of the Royal Astronomical Society, 483, 3673

\bibitem[{{Ng} {et~al.}(2022){Ng}, {Ray}, {Chakrabarty}, {Arzoumanian}, {Guillemot}, \& {Cognard}}]{Ng_2022}
{Ng}, M., {Ray}, P., {Chakrabarty}, D., {et~al.} 2022, in AAS/High Energy Astrophysics Division, Vol.~54, AAS/High Energy Astrophysics Division, 110.04

\bibitem[{{O'Sullivan} {et~al.}(2023){O'Sullivan}, {Shimwell}, {Hardcastle}, {Tasse}, {Heald}, {Carretti}, {Br{\"u}ggen}, {Vacca}, {Sobey}, {Van Eck}, {Horellou}, {Beck}, {Bilicki}, {Bourke}, {Botteon}, {Croston}, {Drabent}, {Duncan}, {Heesen}, {Ideguchi}, {Kirwan}, {Lawlor}, {Mingo}, {Nikiel-Wroczy{\'n}ski}, {Piotrowska}, {Scaife}, \& {van Weeren}}]{OSullivan_2023}
{O'Sullivan}, S.~P., {Shimwell}, T.~W., {Hardcastle}, M.~J., {et~al.} 2023, \mnras, 519, 5723

\bibitem[{{{\"O}zel} \& {Freire}(2016)}]{O&F_2016}
{{\"O}zel}, F. \& {Freire}, P. 2016, \araa, 54, 401

\bibitem[{Paczynski(1981)}]{paczynski_evolution_1981}
Paczynski, B. 1981, 31, 1

\bibitem[{Pallanca {et~al.}(2012)Pallanca, Mignani, Dalessandro, Ferraro, Lanzoni, Possenti, Burgay, \& Sabbi}]{Pallanca_2012}
Pallanca, C., Mignani, R.~P., Dalessandro, E., {et~al.} 2012, The Astrophysical Journal, 755, 180

\bibitem[{{Papitto} {et~al.}(2013){Papitto}, {Ferrigno}, {Bozzo}, {Rea}, {Pavan}, {Burderi}, {Burgay}, {Campana}, {di Salvo}, {Falanga}, {Filipovi{\'c}}, {Freire}, {Hessels}, {Possenti}, {Ransom}, {Riggio}, {Romano}, {Sarkissian}, {Stairs}, {Stella}, {Torres}, {Wieringa}, \& {Wong}}]{Papitto_2013}
{Papitto}, A., {Ferrigno}, C., {Bozzo}, E., {et~al.} 2013, \nat, 501, 517

\bibitem[{Pennucci {et~al.}(2014)Pennucci, Demorest, \& Ransom}]{Pennucci_2014}
Pennucci, T.~T., Demorest, P.~B., \& Ransom, S.~M. 2014, The Astrophysical Journal, 790, 93

\bibitem[{{Polzin} {et~al.}(2019){Polzin}, {Breton}, {Stappers}, {Bhattacharyya}, {Janssen}, {Os{\l}owski}, {Roberts}, \& {Sobey}}]{Polzin_2019}
{Polzin}, E.~J., {Breton}, R.~P., {Stappers}, B.~W., {et~al.} 2019, \mnras, 490, 889

\bibitem[{{Press} {et~al.}(1992){Press}, {Teukolsky}, {Vetterling}, \& {Flannery}}]{Press1992}
{Press}, W.~H., {Teukolsky}, S.~A., {Vetterling}, W.~T., \& {Flannery}, B.~P. 1992 (Cambridge University Press, Cambridge)

\bibitem[{{Ransom} {et~al.}(2011){Ransom}, {Ray}, {Camilo}, {Roberts}, {{\c{C}}elik}, {Wolff}, {Cheung}, {Kerr}, {Pennucci}, {DeCesar}, {Cognard}, {Lyne}, {Stappers}, {Freire}, {Grove}, {Abdo}, {Desvignes}, {Donato}, {Ferrara}, {Gehrels}, {Guillemot}, {Gwon}, {Harding}, {Johnston}, {Keith}, {Kramer}, {Michelson}, {Parent}, {Saz Parkinson}, {Romani}, {Smith}, {Theureau}, {Thompson}, {Weltevrede}, {Wood}, \& {Ziegler}}]{Ransom_2011}
{Ransom}, S.~M., {Ray}, P.~S., {Camilo}, F., {et~al.} 2011, \apjl, 727, L16

\bibitem[{{Ray} {et~al.}(2012){Ray}, {Abdo}, {Parent}, {Bhattacharya}, {Bhattacharyya}, {Camilo}, {Cognard}, {Theureau}, {Ferrara}, {Harding}, {Thompson}, {Freire}, {Guillemot}, {Gupta}, {Roy}, {Hessels}, {Johnston}, {Keith}, {Shannon}, {Kerr}, {Michelson}, {Romani}, {Kramer}, {McLaughlin}, {Ransom}, {Roberts}, {Saz Parkinson}, {Ziegler}, {Smith}, {Stappers}, {Weltevrede}, \& {Wood}}]{Ray_2012}
{Ray}, P.~S., {Abdo}, A.~A., {Parent}, D., {et~al.} 2012, arXiv e-prints, arXiv:1205.3089

\bibitem[{{Ray} {et~al.}(2022){Ray}, {Nieder}, {Clark}, {Ransom}, {Cromartie}, {Frail}, {Mooley}, {Intema}, {Jagannathan}, {Demorest}, {Stovall}, {Halpern}, {Deneva}, {Guillot}, {Kerr}, {Swihart}, {Bruel}, {Stappers}, {Lyne}, {Mickaliger}, {Camilo}, {Ferrara}, {Wolff}, \& {Michelson}}]{Ray_2022}
{Ray}, P.~S., {Nieder}, L., {Clark}, C.~J., {et~al.} 2022, \apj, 927, 216

\bibitem[{{Reynolds} {et~al.}(2007){Reynolds}, {Callanan}, {Fruchter}, {Torres}, {Beer}, \& {Gibbons}}]{Reynolds_2007}
{Reynolds}, M.~T., {Callanan}, P.~J., {Fruchter}, A.~S., {et~al.} 2007, \mnras, 379, 1117

\bibitem[{Roberts(2013)}]{Roberts_2012}
Roberts, M. S.~E. 2013, Proceedings of the International Astronomical Union, 8, 127–132

\bibitem[{{Roy} {et~al.}(2015){Roy}, {Ray}, {Bhattacharyya}, {Stappers}, {Chengalur}, {Deneva}, {Camilo}, {Johnson}, {Wolff}, {Hessels}, {Bassa}, {Keane}, {Ferrara}, {Harding}, \& {Wood}}]{Roy_2015}
{Roy}, J., {Ray}, P.~S., {Bhattacharyya}, B., {et~al.} 2015, \apjl, 800, L12

\bibitem[{{Ruderman} {et~al.}(1989){Ruderman}, {Shaham}, \& {Tavani}}]{Ruderman_1989}
{Ruderman}, M., {Shaham}, J., \& {Tavani}, M. 1989, \apj, 336, 507

\bibitem[{{Sanpa-Arsa}(2016)}]{SanpaArsa_PhD}
{Sanpa-Arsa}, S. 2016, PhD thesis, University of Virginia

\bibitem[{{Schroeder} \& {Halpern}(2014)}]{Schroeder_2014}
{Schroeder}, J. \& {Halpern}, J. 2014, \apj, 793, 78

\bibitem[{{Shaifullah} {et~al.}(2016){Shaifullah}, {Verbiest}, {Freire}, {Tauris}, {Wex}, {Os{\l}owski}, {Stappers}, {Bassa}, {Caballero}, {Champion}, {Cognard}, {Desvignes}, {Graikou}, {Guillemot}, {Janssen}, {Jessner}, {Jordan}, {Karuppusamy}, {Kramer}, {Lazaridis}, {Lazarus}, {Lyne}, {McKee}, {Perrodin}, {Possenti}, \& {Tiburzi}}]{Shaifullah_2016}
{Shaifullah}, G., {Verbiest}, J.~P.~W., {Freire}, P.~C.~C., {et~al.} 2016, \mnras, 462, 1029

\bibitem[{{Smith} {et~al.}(2023){Smith}, {Abdollahi}, {Ajello}, {Bailes}, {Baldini}, {Ballet}, {Baring}, {Bassa}, {Gonzalez}, {Bellazzini}, {Berretta}, {Bhattacharyya}, {Bissaldi}, {Bonino}, {Bottacini}, {Bregeon}, {Bruel}, {Burgay}, {Burnett}, {Cameron}, {Camilo}, {Caputo}, {Caraveo}, {Cavazzuti}, {Chiaro}, {Ciprini}, {Clark}, {Cognard}, {Corongiu}, {Orestano}, {Crnogorcevic}, {Cuoco}, {Cutini}, {D'Ammando}, {de Angelis}, {DeCesar}, {De Gaetano}, {de Menezes}, {Deneva}, {de Palma}, {Di Lalla}, {Dirirsa}, {Di Venere}, {Dom{\'\i}nguez}, {Dumora}, {Fegan}, {Ferrara}, {Fiori}, {Fleischhack}, {Flynn}, {Franckowiak}, {Freire}, {Fukazawa}, {Fusco}, {Galanti}, {Gammaldi}, {Gargano}, {Gasparrini}, {Giacchino}, {Giglietto}, {Giordano}, {Giroletti}, {Green}, {Grenier}, {Guillemot}, {Guiriec}, {Gustafsson}, {Harding}, {Hays}, {Hewitt}, {Horan}, {Hou}, {Jankowski}, {Johnson}, {Johnson}, {Johnston}, {Kataoka}, {Keith}, {Kerr}, {Kramer}, {Kuss}, {Latronico}, {Lee}, {Li}, {Li}, {Limyansky}, {Longo}, {Loparco}, {Lorusso},
  {Lovellette}, {Lower}, {Lubrano}, {Lyne}, {Maan}, {Maldera}, {Manchester}, {Manfreda}, {Marelli}, {Mart{\'\i}-Devesa}, {Mazziotta}, {McEnery}, {Mereu}, {Michelson}, {Mickaliger}, {Mitthumsiri}, {Mizuno}, {Moiseev}, {Monzani}, {Morselli}, {Negro}, {Nemmen}, {Nieder}, {Nuss}, {Omodei}, {Orienti}, {Orlando}, {Ormes}, {Palatiello}, {Paneque}, {Panzarini}, {Parthasarathy}, {Persic}, {Pesce-Rollins}, {Pillera}, {Poon}, {Porter}, {Possenti}, {Principe}, {Rain{\`o}}, {Rando}, {Ransom}, {Ray}, {Razzano}, {Razzaque}, {Reimer}, {Reimer}, {Renault-Tinacci}, {Romani}, {S{\'a}nchez-Conde}, {Parkinson}, {Scotton}, {Serini}, {Sgr{\`o}}, {Shannon}, {Sharma}, {Shen}, {Siskind}, {Spandre}, {Spinelli}, {Stappers}, {Stephens}, {Suson}, {Tabassum}, {Tajima}, {Tak}, {Theureau}, {Thompson}, {Tibolla}, {Torres}, {Valverde}, {Venter}, {Wadiasingh}, {Wang}, {Wang}, {Wang}, {Weltevrede}, {Wood}, {Yan}, {Zaharijas}, {Zhang}, \& {Zhu}}]{Smith_3PC}
{Smith}, D.~A., {Abdollahi}, S., {Ajello}, M., {et~al.} 2023, \apj, 958, 191

\bibitem[{{Smith} {et~al.}(2019){Smith}, {Bruel}, {Cognard}, {Cameron}, {Camilo}, {Dai}, {Guillemot}, {Johnson}, {Johnston}, {Keith}, {Kerr}, {Kramer}, {Lyne}, {Manchester}, {Shannon}, {Sobey}, {Stappers}, \& {Weltevrede}}]{Smith_2019}
{Smith}, D.~A., {Bruel}, P., {Cognard}, I., {et~al.} 2019, \apj, 871, 78

\bibitem[{Spearman(1904)}]{Spearman}
Spearman, C. 1904, The American Journal of Psychology, 15, 72

\bibitem[{{Spiewak} {et~al.}(2022){Spiewak}, {Bailes}, {Miles}, {Parthasarathy}, {Reardon}, {Shamohammadi}, {Shannon}, {Bhat}, {Buchner}, {Cameron}, {Camilo}, {Geyer}, {Johnston}, {Karastergiou}, {Keith}, {Kramer}, {Serylak}, {van Straten}, {Theureau}, \& {Venkatraman Krishnan}}]{Spiewak_2022}
{Spiewak}, R., {Bailes}, M., {Miles}, M.~T., {et~al.} 2022, \pasa, 39, e027

\bibitem[{Stappers {et~al.}(1996)Stappers, Bailes, Lyne, Manchester, D'Amico, Tauris, Lorimer, Johnston, \& Sandhu}]{Stappers_1996}
Stappers, B.~W., Bailes, M., Lyne, A.~G., {et~al.} 1996, The Astrophysical Journal, 465, L119

\bibitem[{{Stovall} {et~al.}(2014){Stovall}, {Lynch}, {Ransom}, {Archibald}, {Banaszak}, {Biwer}, {Boyles}, {Dartez}, {Day}, {Ford}, {Flanigan}, {Garcia}, {Hessels}, {Hinojosa}, {Jenet}, {Kaplan}, {Karako-Argaman}, {Kaspi}, {Kondratiev}, {Leake}, {Lorimer}, {Lunsford}, {Martinez}, {Mata}, {McLaughlin}, {Roberts}, {Rohr}, {Siemens}, {Stairs}, {van Leeuwen}, {Walker}, \& {Wells}}]{Stovall_14}
{Stovall}, K., {Lynch}, R.~S., {Ransom}, S.~M., {et~al.} 2014, \apj, 791, 67

\bibitem[{{Tang} {et~al.}(2014){Tang}, {Kaplan}, {Phinney}, {Prince}, {Breton}, {Bellm}, {Bildsten}, {Cao}, {Kong}, {Perley}, {Sesar}, {Wolf}, \& {Yen}}]{Tang_2014}
{Tang}, S., {Kaplan}, D.~L., {Phinney}, E.~S., {et~al.} 2014, \apjl, 791, L5

\bibitem[{{Thompson} {et~al.}(1994){Thompson}, {Blandford}, {Evans}, \& {Phinney}}]{Thompson_1994}
{Thompson}, C., {Blandford}, R.~D., {Evans}, C.~R., \& {Phinney}, E.~S. 1994, \apj, 422, 304

\bibitem[{{van der Wateren} {et~al.}(2022){van der Wateren}, {Bassa}, {Clark}, {Breton}, {Cognard}, {Guillemot}, {Janssen}, {Lyne}, {Stappers}, \& {Theureau}}]{Wateren_2022}
{van der Wateren}, E., {Bassa}, C.~G., {Clark}, C.~J., {et~al.} 2022, \aap, 661, A57

\bibitem[{{van Haaften} {et~al.}(2013){van Haaften}, {Nelemans}, {Voss}, \& {Jonker}}]{Haaften_2013}
{van Haaften}, L.~M., {Nelemans}, G., {Voss}, R., \& {Jonker}, P.~G. 2013, in IAU Symposium, Vol. 291, Neutron Stars and Pulsars: Challenges and Opportunities after 80 years, ed. J.~{van Leeuwen}, 133--136

\bibitem[{{van Kerkwijk} {et~al.}(2011){van Kerkwijk}, {Breton}, \& {Kulkarni}}]{Kerkwijk_2011}
{van Kerkwijk}, M.~H., {Breton}, R.~P., \& {Kulkarni}, S.~R. 2011, \apj, 728, 95

\bibitem[{Voisin {et~al.}(2020)Voisin, Clark, Breton, Dhillon, Kennedy, \& Mata-Sánchez}]{Voisin_2020}
Voisin, G., Clark, C.~J., Breton, R.~P., {et~al.} 2020, Monthly Notices of the Royal Astronomical Society, 494, 4448

\bibitem[{{Wahl} {et~al.}(2022){Wahl}, {McLaughlin}, {Gentile}, {Jones}, {Spiewak}, {Arzoumanian}, {Crowter}, {Demorest}, {DeCesar}, {Dolch}, {Ellis}, {Ferdman}, {Ferrara}, {Fonseca}, {Garver-Daniels}, {Jones}, {Lam}, {Levin}, {Lewandowska}, {Lorimer}, {Lynch}, {Madison}, {Ng}, {Nice}, {Pennucci}, {Ransom}, {Ray}, {Stairs}, {Stovall}, {Swiggum}, \& {Zhu}}]{Wahl_2022}
{Wahl}, H.~M., {McLaughlin}, M.~A., {Gentile}, P.~A., {et~al.} 2022, \apj, 926, 168

\bibitem[{{Wang} {et~al.}(2023){Wang}, {Wang}, {Li}, {Yao}, {Manchester}, {Hobbs}, {Wang}, {Dai}, {Xu}, {Luo}, {Feng}, {Wang}, {Li}, {Yu}, {Du}, {Niu}, {Zhang}, \& {Zhang}}]{Wang_2023}
{Wang}, S.~Q., {Wang}, J.~B., {Li}, D.~Z., {et~al.} 2023, \apj, 955, 36

\bibitem[{{Wu} {et~al.}(2012){Wu}, {Kong}, {Huang}, {Takata}, {Tam}, {Wu}, \& {Cheng}}]{Wu_2012}
{Wu}, J.~H.~K., {Kong}, A.~K.~H., {Huang}, R.~H.~H., {et~al.} 2012, \apj, 748, 141

\bibitem[{{Yao} {et~al.}(2017){Yao}, {Manchester}, \& {Wang}}]{Yao_2017}
{Yao}, J.~M., {Manchester}, R.~N., \& {Wang}, N. 2017, \apj, 835, 29

\end{thebibliography}


\begin{appendix}
\onecolumn

\section{Timing residuals and S/N histograms for the pulsar sample}

\begin{figure*}[h]
\centering
\includegraphics[width=18cm]{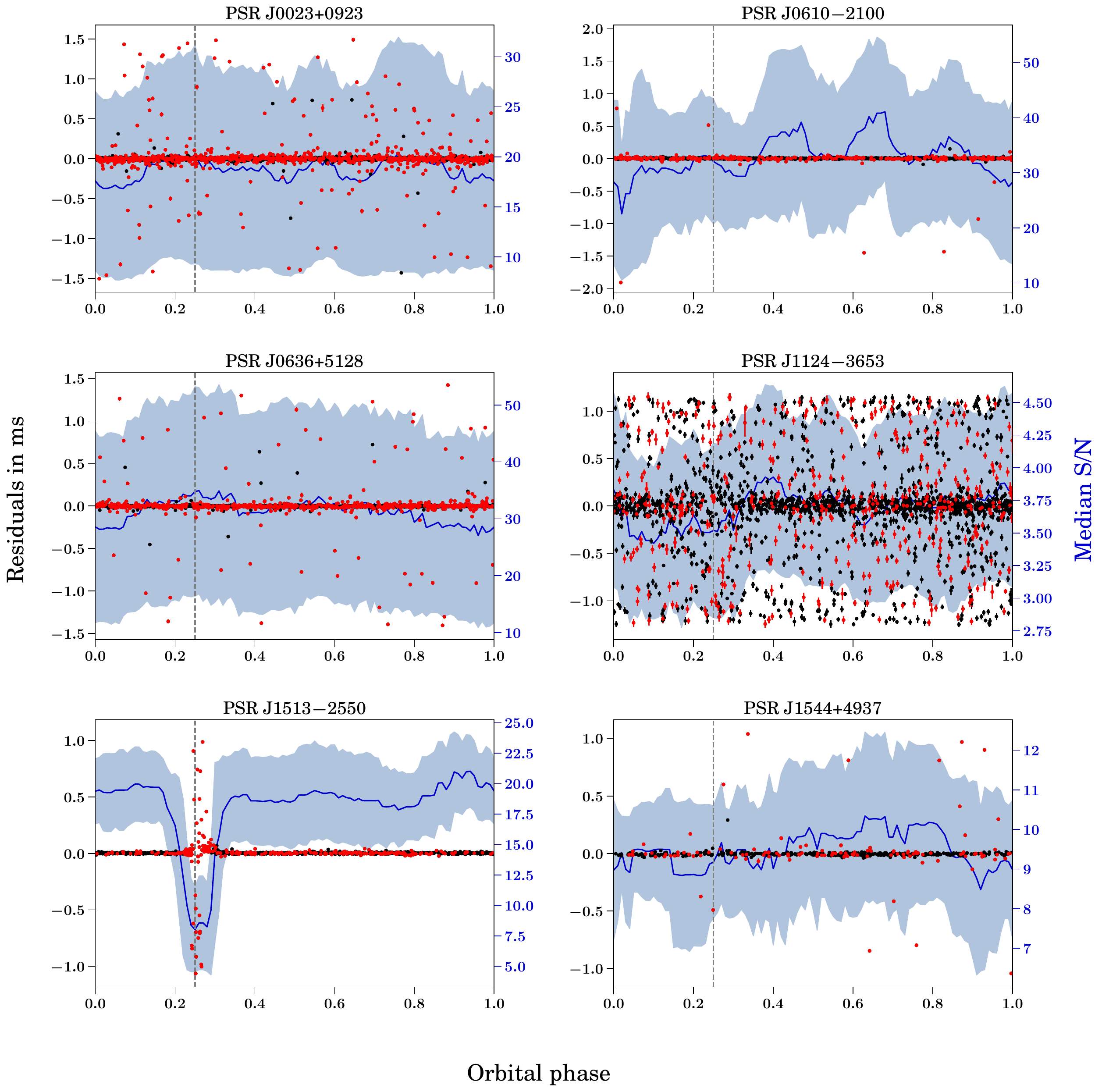}
    \caption{Same as Fig.~\ref{fig:SNR_drop}, for PSRs J0023+0923, J0610$-$2100, J0636+5128, J1124$-$3653, J1513$-$2550 and J1544+4937. Note that in this figure S/N histograms and corresponding axes are shown in blue. TOA residuals are shown as red and black dots. Red dots correspond to observations with S/N values among the 25\% lowest. Dashed vertical lines indicate superior conjunction, corresponding to orbital phase 0.25. In some plots the error bars are too small to be apparent.}
    \label{fig:SNR_drop_appendix1}
\end{figure*}

\begin{figure*}[h]
\centering
\includegraphics[width=18cm]{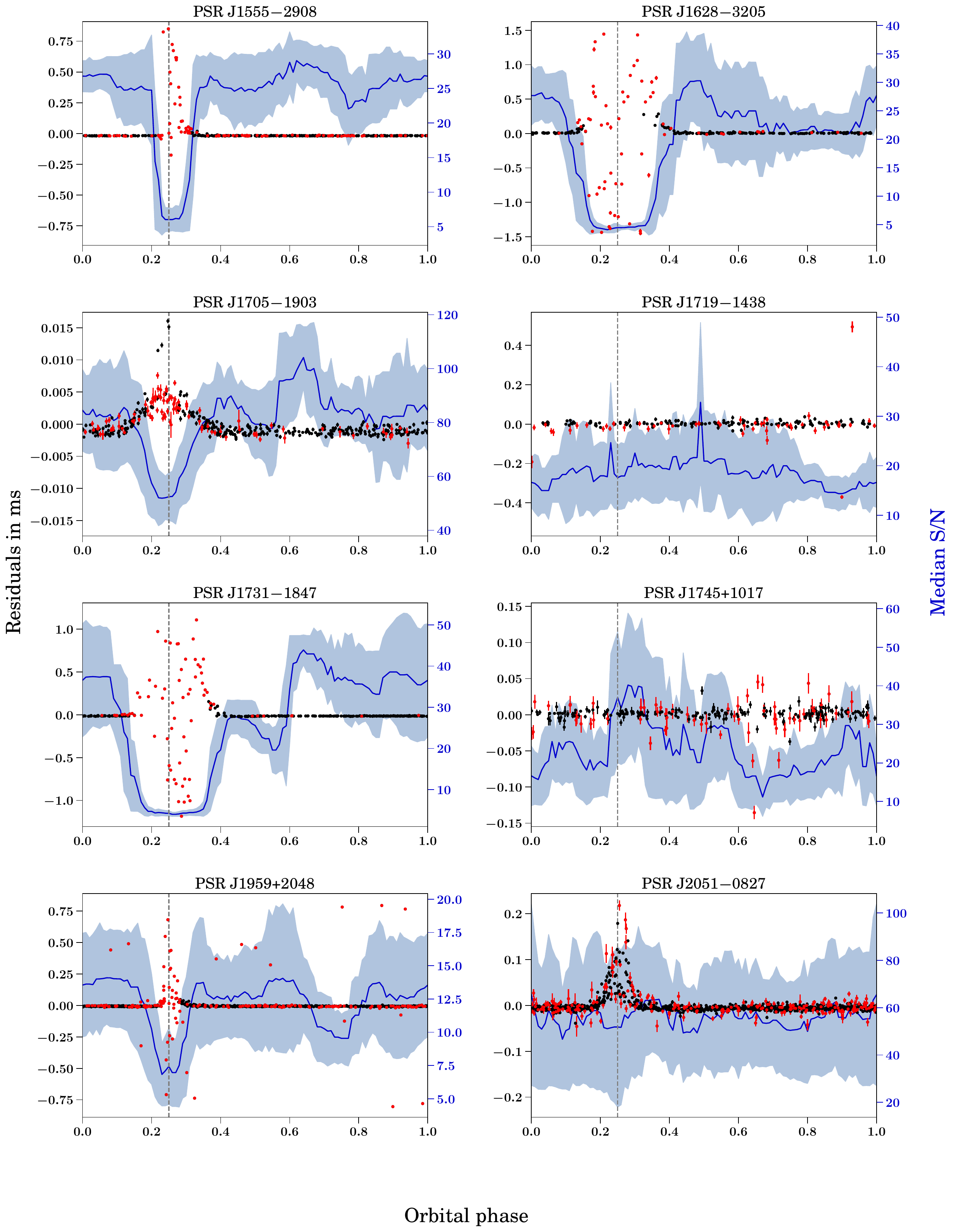}
    \caption{Same as Fig.~\ref{fig:SNR_drop_appendix1}, for PSRs J1555$-$2908, J1628$-$3205, J1705$-$1903, J1719$-$1438, J1731$-$1847, J1745+1017, J1959+2048 and J2051$-$0827.}
    \label{fig:SNR_drop_appendix2}
\end{figure*}

\begin{figure*}[h]
\centering
\includegraphics[width=18cm]{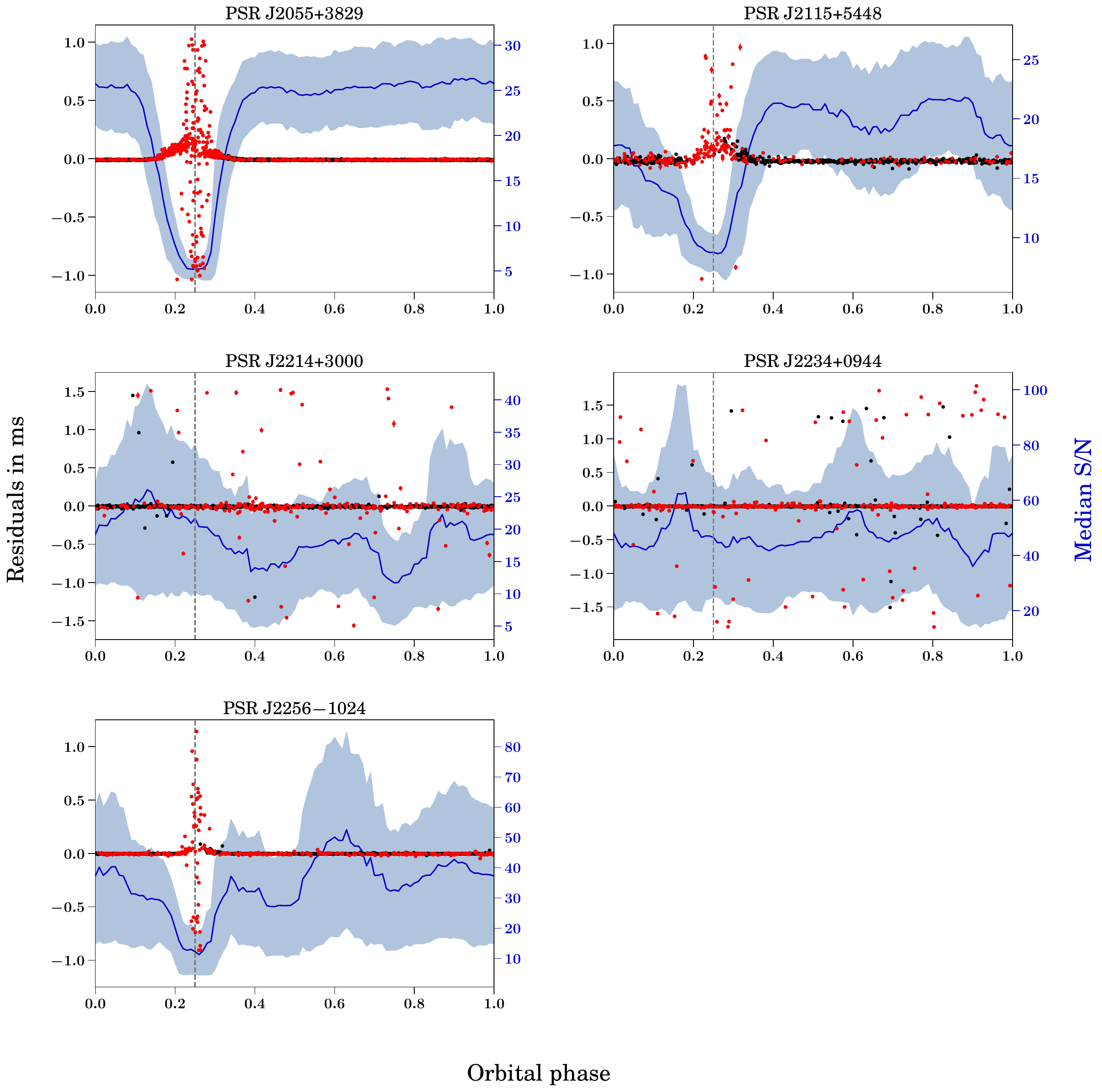}
    \caption{Same as Fig.~\ref{fig:SNR_drop_appendix1}, for PSRs J2055+3829, J2115+5448, J2214+3000, J2234+0944 and J2256$-$1024.}
    \label{fig:SNR_drop_appendix3}
\end{figure*}

In Fig.~\ref{fig:SNR_drop} we show an example of a plot of timing residuals and S/N values as a function of orbital phase, for PSR~J1731$-$1847. In Figs.~\ref{fig:SNR_drop_appendix1}, \ref{fig:SNR_drop_appendix2} and \ref{fig:SNR_drop_appendix3} we show the same plots for all the pulsars in our sample.

\clearpage
\section{Eclipse residuals fits}

This appendix refers to Sect.~\ref{sect:study}. 

\begin{figure*}[h]
\centering
\includegraphics[width=18cm]{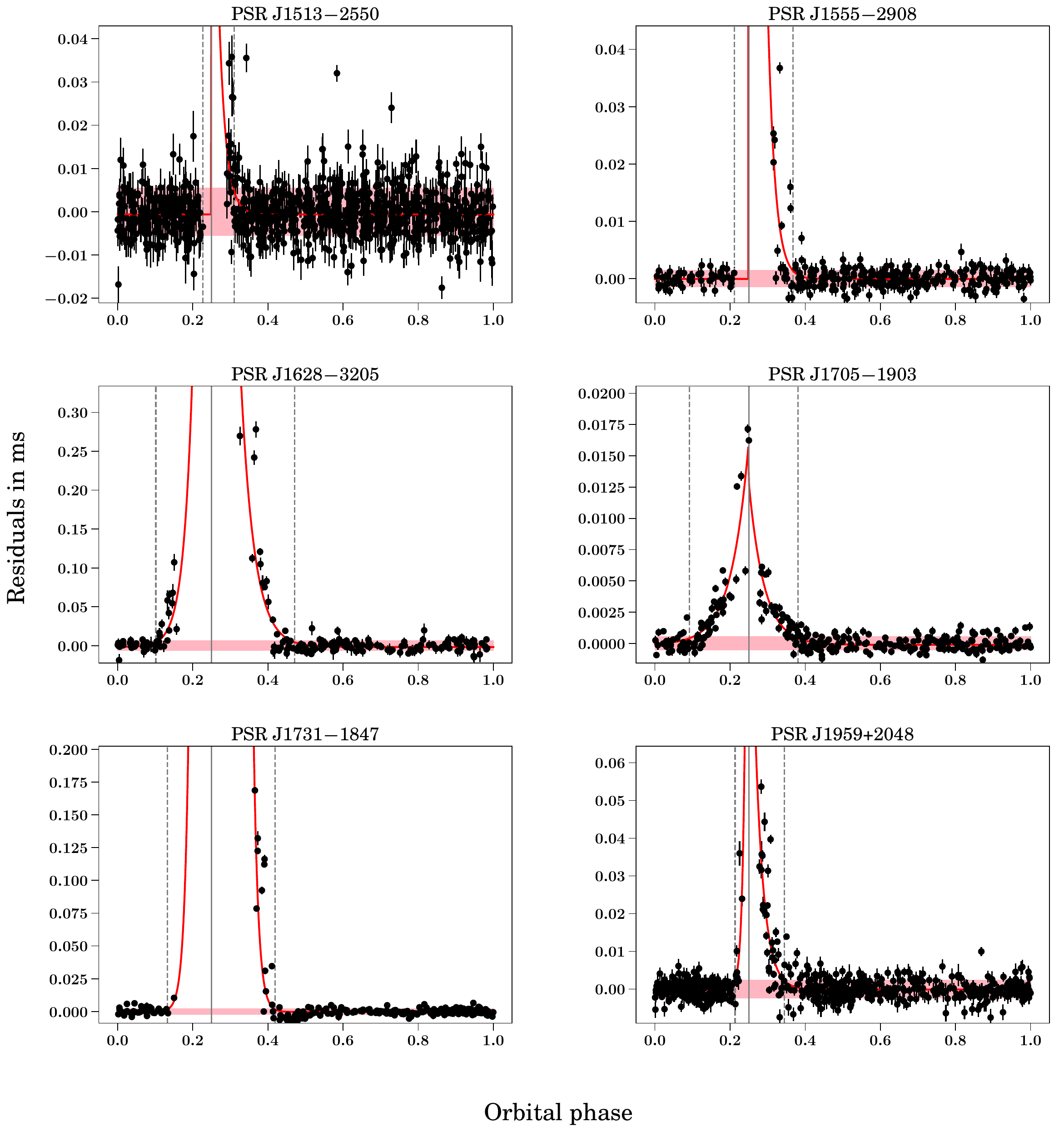}
    \caption{Same as Fig.~\ref{fig:J2256_fit}, for PSRs J1513$-$2550, J1555$-$2908, J1628$-$3205, J1705$-$1903, J1731$-$1847 and J1959+2048. The median-fit function obtained from the analysis presented in Sect.~\ref{sect:study} is shown in red. The light red zone represents the RMS of the timing residuals for orbital phases $\phi > 0.5 $, and the dashed vertical lines correspond to the values of $\Phi_\mathrm{start}$ and $\Phi_\mathrm{end}$ as inferred from the fit.}
    \label{fig:fit_appendix1}
\end{figure*}

\begin{figure*}[h]
\centering
\includegraphics[width=18cm]{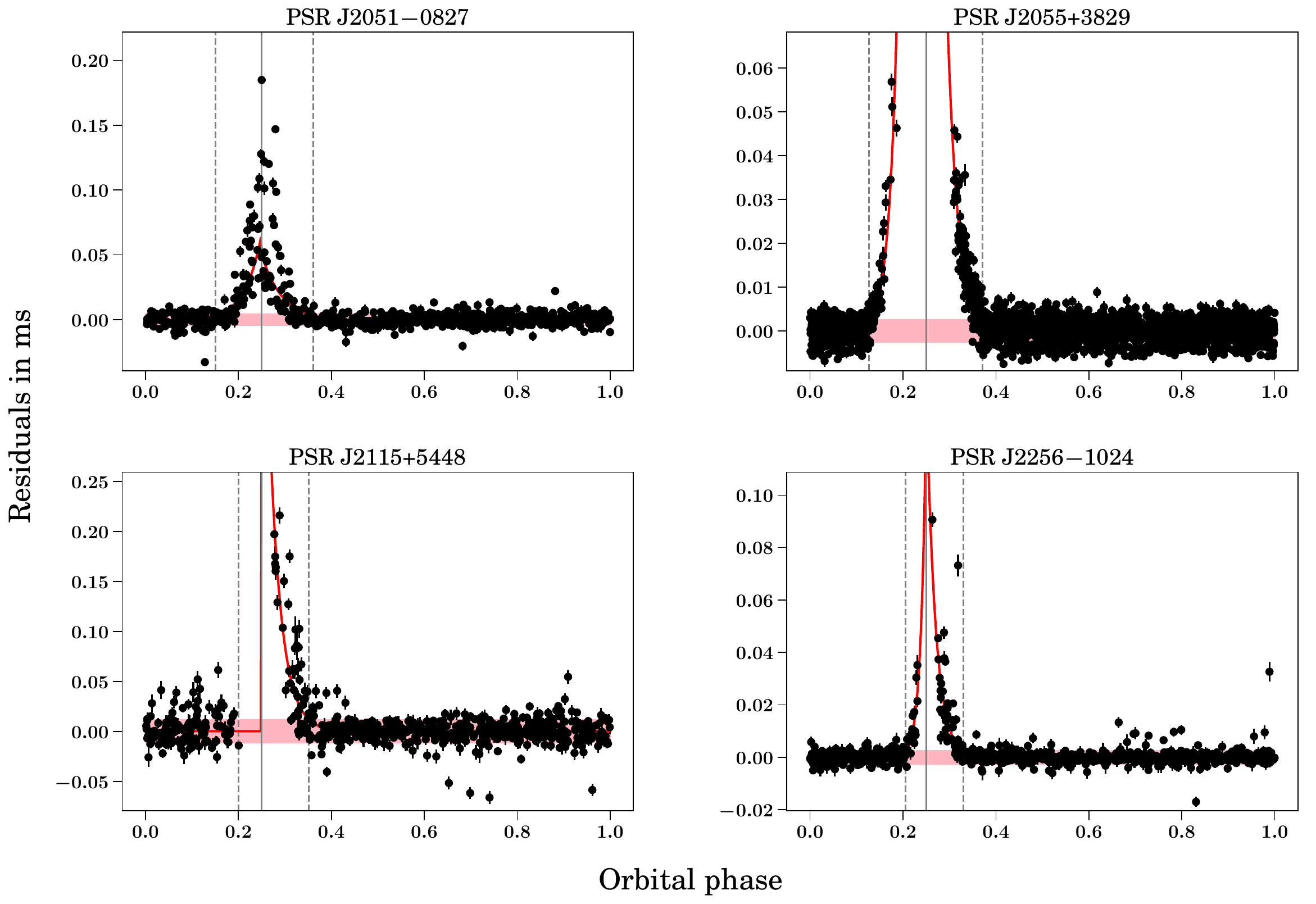}
    \caption{Same as Fig.~\ref{fig:fit_appendix1}, for PSRs J2051$-$0827, J2055+3829, J2115+5448, and J2256$-$1024.}
    \label{fig:fit_appendix2}
\end{figure*}

Figure~\ref{fig:J2256_fit} shows an example of eclipse residuals plot, for PSR~J2256$-$1024. In Figs.~\ref{fig:fit_appendix1} and \ref{fig:fit_appendix2} we show the same plot for all eclipsing pulsars in our sample. Best-fit parameters from this analysis are listed in Table~\ref{tab:results_fit}.

\clearpage
\section{Polarimetric profiles for the 19 spider pulsars}

This appendix refers to Sect.~\ref{sect:polar}. 

The 1.4~GHz composite profiles shown below were constructedby summing the profiles from individual NUPPI observations conducted after MJD 58800, \textit{i.e.}, NUPPI observations that were calibrated using the improved polarization calibration scheme presented in \citet{Guillemot_2023}. For polarimetric profiles with S/N values larger than 1000, we kept the original number of phase bins of 2048. We reduced the number of phase bins by a factor of 2 for S/N values under 1000, by a factor of 4 under 300, by a factor of 8 under 100 and by a factor of 16 under 30.

\begin{figure*}[h]
\centering
\includegraphics[width=18cm]{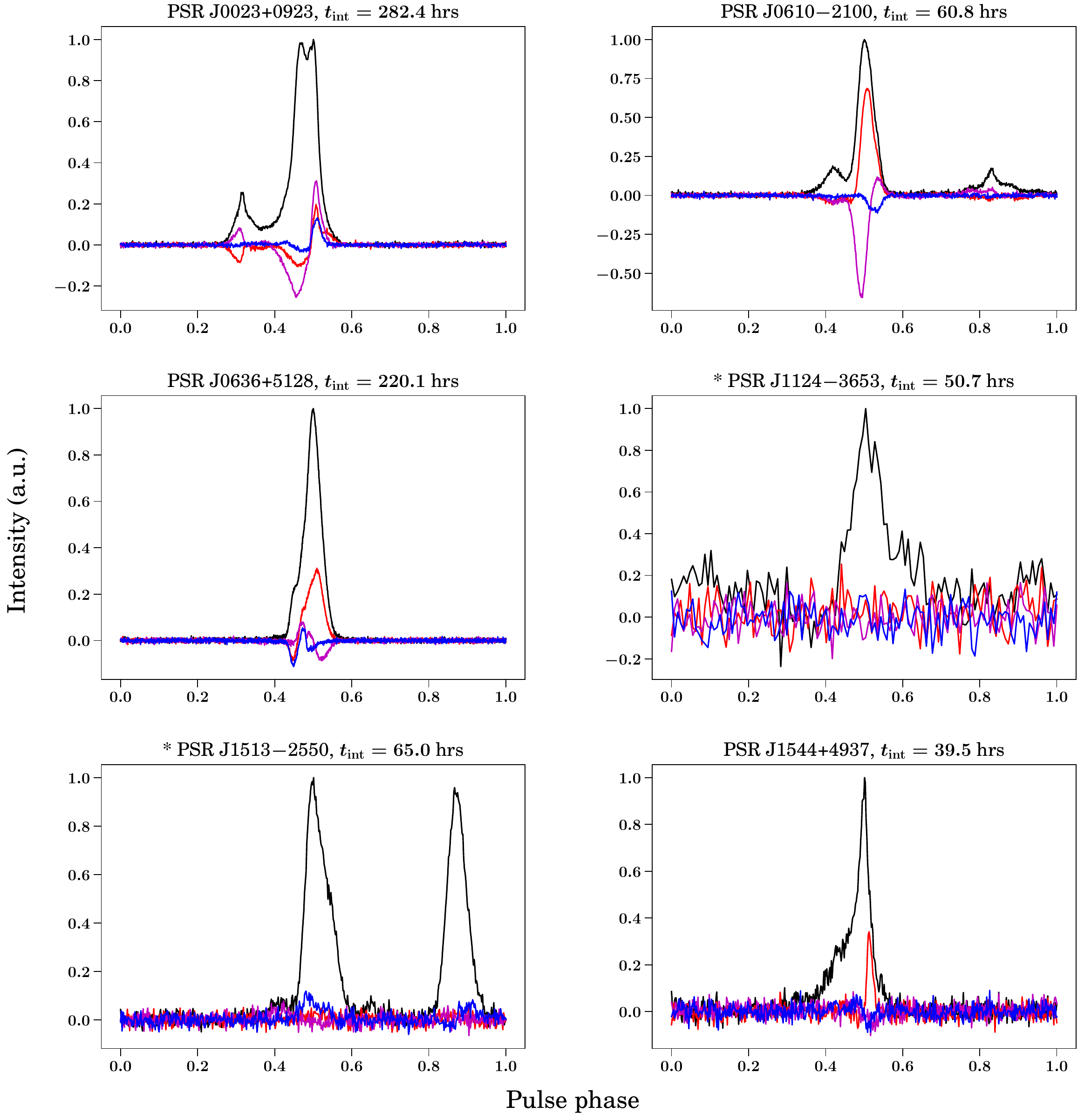}
    \caption{Polarimetric profiles for PSRs~J0023+0923, J0610$-$2100, J0636+5128, J1124$-$3653, J1513$-$2550 and J1544+4937. In each panel, the black line represents the total intensity (Stokes parameter I), the red and purple lines correspond to the Stokes parameters Q and U describing the linear polarization, and the blue line is the circular polarization (Stokes parameter V). Pulse profiles were rotated in order for the total intensity maxima to be at phase 0.5. For pulsars with no published RM value we did not correct the Stokes parameters Q and U for Faraday rotation. These pulsars are indicated with an asterisk.}
    \label{fig:profile_appendix1}
\end{figure*}

\begin{figure*}[h]
\centering
\includegraphics[width=18cm]{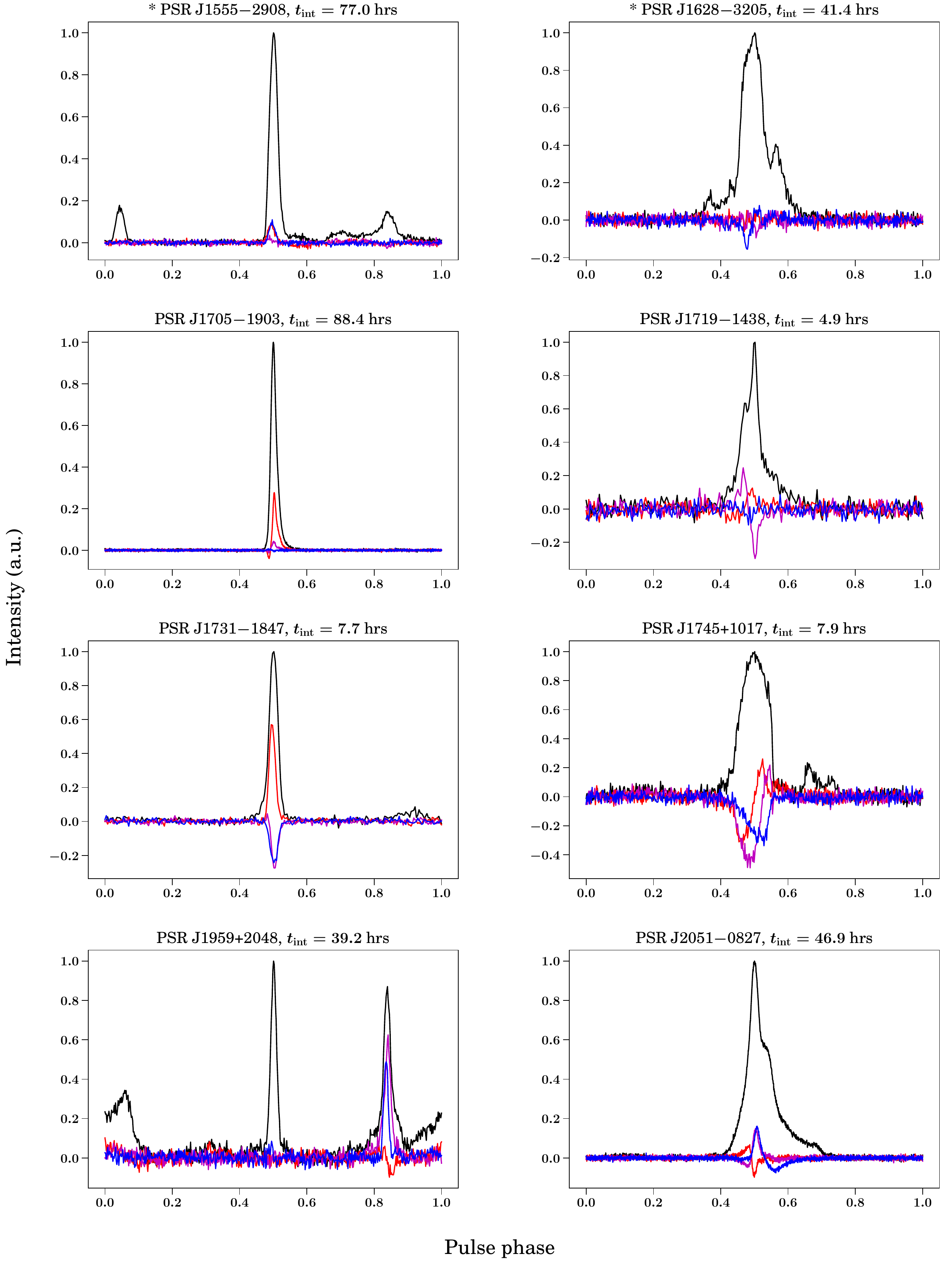}
    \caption{Same as Fig.~\ref{fig:profile_appendix1}, for PSRs J1555$-$2908, J1628$-$3205, J1705$-$1903, J1719$-$1438, J1731$-$1847, J1745+1017, J1959+2048 and J2051$-$0827.}
    \label{fig:profile_appendix2}
\end{figure*}

\begin{figure*}[h]
\centering
\includegraphics[width=18cm]{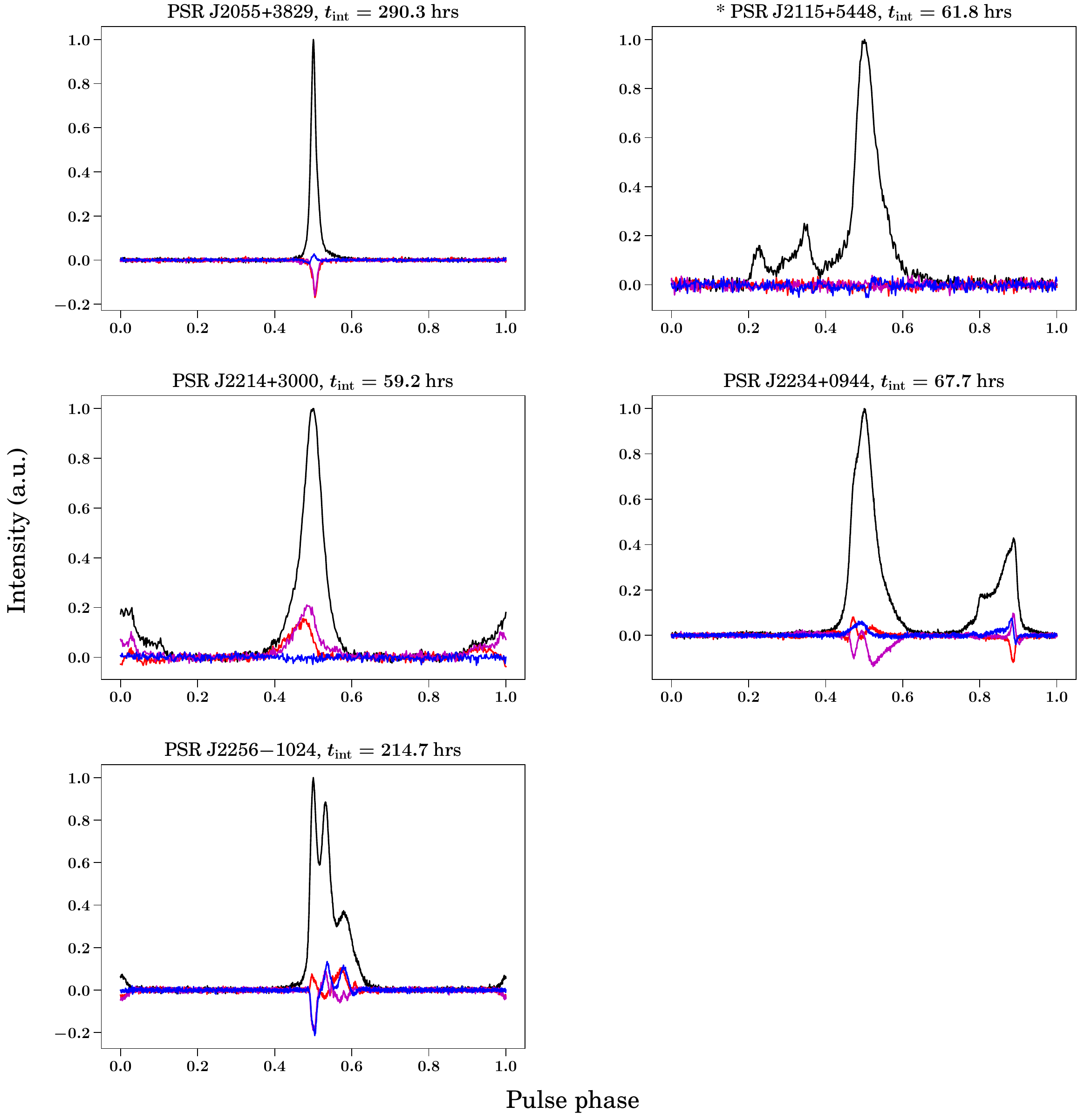}
    \caption{Same as Fig.~\ref{fig:profile_appendix1}, for PSRs J2055+3829, J2115+5448, J2214+3000, J2234+0944 and J2256$-$1024.}
    \label{fig:profile_appendix3}
\end{figure*}

\clearpage
\section{Relationship between effective and L1 filling factors
\label{sec:filfac}}

This appendix refers to Table~\ref{tab:optic_sample}.

Roche-lobe geometry is determined by only two parameters: the binary mass ratio $q$ and the binary separation $a$. The latter acts as a scaling factor. 
The stellar surface follows an equipotential surface, the largest one being by definition the boundary of the Roche lobe which goes through the L1 Lagrange point. 

There is not a unique way of defining the filling factor. The L1 filling factor is defined as 
\begin{equation}    
f_{\rm L1} = \frac{r_{\rm s}}{r_{\rm L1}},
\end{equation}
where $r_{\rm s}$ is the distance from the center to the surface of the star along the line connecting the two components of the binary system (the intra-binary line), and $r_{\rm L1}$ is the distance of the Lagrange L1 point. 
On the other hand the effective-radius filling factor is defined as 
\begin{equation}
f_{\rm eff} = \frac{\bar R_{\rm s}}{\bar R_{\rm L1}},
\end{equation}
where $\bar R_{\rm s}$ and $\bar R_{\rm L1}$ are the volume-averaged radii corresponding to the stellar surface and the Roche lobe, that is the radii of spheres with the same volumes as those enclosed in the stellar surface equipotential and Roche lobe, respectively \citep{paczynski_evolution_1981, kopal_dynamics_1978}. The effective Roche lobe radius $\bar R_{\rm L1}$ can be conveniently calculated using an approximation due to \citet{eggleton_aproximations_1983}.

In this paper we have generated a numerical interpolation of the functions $f_{\rm eff}(f_{\rm L1}, q)$, and inversely  $f_{\rm L1}(f_{\rm eff}, q)$, where $q = m_{p}/m_{c}$ is the mass ratio between the pulsar and its companion. The separation $a$ does not appear since it cancels out in the definition of the filling factors. 

To do so, we used the Icarus software \citep{breton_koi_2012} which is able to compute numerically $f_{\rm eff}(f_{\rm L1}, q)$. We computed $f_{\rm eff}$ on a grid of 500 mass ratios $1.4\leq q \leq 250$, and 100 L1 filling factors $0.1 \leq f_{\rm L1} \leq 1$. This grid can then be interpolated (we used basic linear interpolation). In order to invert the relation we used this interpolation coupled with a root finding algorithm to solve $f_{\rm eff}(f_{\rm L1}, q) = y$ for $f_{\rm L1}(y,q)$ \footnote{We used the \texttt{Scipy} library for both interpolation and root finding.}. The difference between the interpolated results and direct calculation with Icarus was not above $0.1\%$ in our validations. The file containing the interpolation grid as well as the \texttt{python} routines converting both ways are available at \url{Add_zenodo_repository_upon_acceptance}.

\begin{figure}[h]
    \centering
    \includegraphics[width=0.5\linewidth]{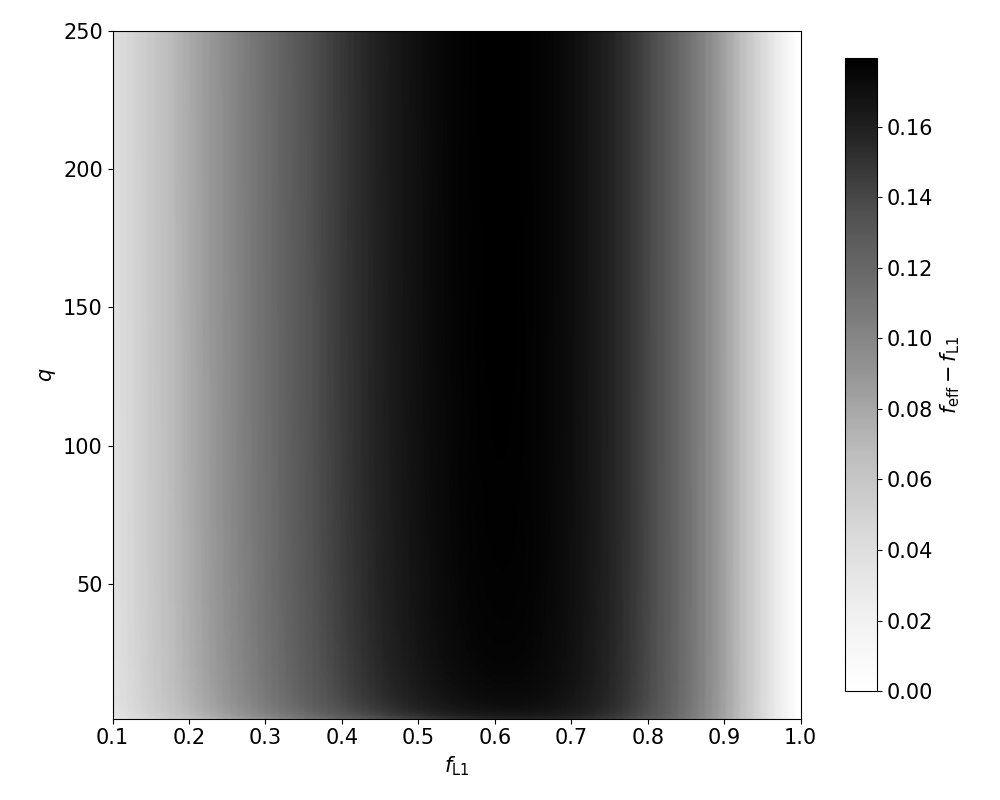}
    \caption{Difference between effective radius filling $f_{\rm eff}$ factor and L1 filling factor $f_{\rm L1}$ as a function $ (f_{\rm L1}, q)$, where $q$ is the mass ratio.}
    \label{fig:fdiff}
\end{figure}

In Fig. \ref{fig:fdiff} we show the difference between the two definitions on our interpolation grid. We see that the difference depends weakly on $q$ as it varies only by $\sim 10\%$ on the whole range for a fixed $f_{\rm L1}$. The largest difference between the two definitions is around $f_{\rm L1} \simeq 0.6$ where $f_{\rm eff} \simeq 0.75$.

\end{appendix}

\end{document}